\documentclass[aps,prd,10pt,showpacs,amsmath,amssymb,twocolumn,nofootinbib,nobibnotes,preprintnumbers
]{revtex4-1}
\usepackage{float}
\usepackage{mathrsfs,amsfonts}
\usepackage{mathtools}
\usepackage{tensor}
\usepackage[T1]{fontenc} 
\usepackage{xcolor}
\usepackage{url}
\usepackage[hyperindex,breaklinks]{hyperref}
\usepackage{graphicx}
\pdfoutput=1
\newcommand{\sig}{\ensuremath{{\hat{\sigma}}}}
\newcommand{\s}{\ensuremath{{\hat{s}}}}
\begin{document}
\allowdisplaybreaks[1]
\title{Primordial black hole dark matter in dilaton-extended two-field Starobinsky inflation}
\author{Anirudh Gundhi~${}^{a,b}$}
\author{Sergei V. Ketov~${}^{c,d,e}$}
\author{Christian F. Steinwachs~${}^{f}$}
\email{anirudh.gundhi@phd.units.it}
\email{ketov@tmu.ac.jp}
\email{christian.steinwachs@physik.uni-freiburg.de}
\affiliation{${}^{a}$Dipartimento di Fisica, Universit\`a degli Studi di Trieste, Strada Costiera 11, 34151 Miramare-Trieste, Italy\\
${}^{b}$ Istituto Nazionale di Fisica Nucleare, Trieste Section, Via Valerio 2, 34127 Trieste, Italy}
\affiliation{${}^c$~Department of Physics, Tokyo Metropolitan University,
1-1 Minami-ohsawa, Hachioji-shi, Tokyo 192-0397, Japan \\
${}^d$~Research School of High-Energy Physics, Tomsk Polytechnic University, 
2a Lenin Avenue, Tomsk 634028, Russian Federation\\
${}^e$~Kavli Institute for the Physics and Mathematics of the Universe (WPI),
The University of Tokyo Institutes for Advanced Study, Kashiwa 277-8583, Japan\\}
\affiliation{${}^{f}$Physikalisches Institut, Albert-Ludwigs-Universit\"at Freiburg,\\
Hermann-Herder-Str.~3, 79104 Freiburg, Germany}
%
%
\begin{abstract}
We investigate the production of primordial black holes and their contribution to the presently observed dark matter in a dilaton two-field extension of Starobinsky's quadratic $f(R)$ model of inflation. The model features a multi-field amplification mechanism which leads to the generation of a sharp peak in the inflationary power spectrum at small wavelengths responsible for the production of primordial black holes. This mechanism is significantly different from single-field models and requires a stochastic treatment during an intermediate phase of the inflationary dynamics. We find that the model leads to a successful phase of effective single-field Starobinsky inflation for wavelengths probed by the cosmic microwave background radiation and explains the observed cold dark matter content in the Universe by the formation of primordial black holes.    
\end{abstract}
%
%
\maketitle
%
%
\section{Introduction}
\label{Into}
%
%
\label{Sec:Introduction}
Primordial Black Holes (PBHs) could provide an explanation of the origin of Cold Dark Matter (CDM) without assuming new particles, generate the seeds of Large Scale Structure (LSS), and probe very high energy physics including quantum gravity, see e.g.~\cite{Carr:2003bj,M3,Ketov:2019mfc,Carr:2020gox,Carr:2020xqk} for a review, cosmological and astrophysical constraints on PBHs, and references therein to original papers. 

The phenomenologically most relevant PBH formation scenario is due to large density fluctuations generated during inflation. Aside from the relevance of PBHs in explaining (part of) the currently observed CDM, they offer a unique opportunity to probe the power spectrum of perturbations at smaller wavelengths. Therefore, they provide constraints complementary to those from the observations of the Cosmic Microwave Background (CMB) at large wavelengths and have the potential to limit the number of phenomenologically viable inflationary models.  
Among the theoretically best motivated models, whose predictions for the inflationary spectral observables are in perfect agreement with recent PLANCK data \cite{Akrami:2018odb}, are Starobinsky's quadratic model of $f(R)$ gravity \cite{Starobinsky1980} and the model of Higgs inflation \cite{Bezrukov2008}, see \cite{Steinwachs:2019hdr} for a recent review. A combination of the two individual models leads to the unified scalaron-Higgs two-field model of inflation considered in \cite{Salvio2015,Kaneda2016,Calmet:2016fsr,Ema2017a,Wang2017,He:2018gyf,Gundhi:2018wyz}. 
Another multi-field extension of Higgs inflation and Starobinsky's $R+R^2$ model is based on a coupling to a dilaton field considered in \cite{Shaposhnikov:2008xi,GarciaBellido:2011de}. 

The formation of PBHs in the context of multi-field models of inflation has recently gained much attention, see e.g.~\cite{Pi:2017gih,Inomata:2017vxo,Canko:2019mud,Fumagalli:2020adf,Braglia:2020eai,Palma:2020ejf,Zhou:2020kkf}.
In contrast to the PBH formation in single-field models of inflation \cite{Garcia-Bellido:2017mdw,Kannike:2017bxn,Ezquiaga:2017fvi,Motohashi:2017kbs,Rasanen:2018fom,Mishra:2019pzq,Cheong:2019vzl}, the PBH formation in multi-field models relies on the multidimensional potential landscape, the curved field space geometry and the isocurvature sourcing of adiabatic modes.

In this paper we investigate the formation of PBHs in a two-field dilaton extension of Starobinsky's model.
Although the structure is similar to the scalaron-Higgs model \cite{Ema2017a,Wang2017,He:2018gyf,Gundhi:2018wyz}, regarding the coupling of the dilaton field, we consider the operators $R$ and $R^2$ on equal footing. The non-minimal coupling to the linear Einstein-Hilbert term leads to an effective dilaton-dependent Planck mass, while the coupling to the quadratic $R^2$ term leads to an effective dilaton-dependent scalaron mass. The latter is a key element in the successful realization of the multi-field amplification mechanism that leads to an enhancement of the power spectrum of scalar perturbations crucial for the formation of PBHs. 
 
Our paper is organized as follows. We formulate our model in the Jordan frame and perform the transition to the two-field scalar-tensor formulation in the Einstein frame in Sect.~\ref{Model}. We summarize the formalism required for the covariant description of the background dynamics and the evolution of the perturbations in Sect.~\ref{CovMultiFieldFormalism}. The properties of the Einstein frame two-field potential landscape are discussed in Sect.~\ref{PotLandscape}. In Sect.~\ref{Properties}, we explain how inflation proceeds in three subsequent stages involving an intermediate stochastic phase connecting two effective single-field phases of slow-roll inflation. Sect.~\ref{PeakFormation} is devoted to the details of the peak formation mechanism, while we present numerical results of the inflationary dynamics and observables in Sect.~\ref{NumericalTreatment}. In Sect.~\ref{PBHDM}, we summarize the PBH formation mechanism and the formalism required for the calculation of the PBH mass distribution and the total PBH fraction of the currently observed CDM.
Finally, in Sect.~\ref{NumResults}, we present our numerical results for the PBH mass distribution in the three mass windows which permit a significant PBH fraction of CDM. We summarize our main results and discuss various future applications of our model in Sect.~\ref{Conclusions}. A derivation of the PBH mass distribution function is presented in Appendix \ref{Appendix1}. An approximate analytical estimate for the power spectrum peak amplitude required for a significant PBH production is provided in Appendix \ref{Analytic}.
%
%
\section{The Model}
\label{Model}

Starobinsky's model is defined by the action \cite{Starobinsky1980}
\begin{align} \label{star}
S_{\mathrm{Star}}[g] = \frac{M^2_{\mathrm{P}}}{2}\int \mathrm{d}^4x\sqrt{-g} \left( R +\frac{1}{6m_0^2}R^2\right)~.
\end{align}
The higher derivatives entering \eqref{star} via the marginal $R^2$ operator lead to an additional scalar propagating degree of freedom, the scalaron \cite{Stelle1978,Starobinsky1980}, which becomes manifest in the scalar-tensor representation of \eqref{star}.\footnote{Curved spacetime is described by a four-dimensional pseudo-Riemannian manifold with local coordinates $\mu,\nu,\ldots=0,1,2,3$, metric field ${g}_{\mu\nu}(x)$, and metric-compatible connection $\nabla_{\mu}$. Our sign conventions are ${\mathrm{sig}(g)=(-,+,+,+,)}$, $\tensor{R}{^{\rho}_{\mu\sigma\nu}}=\partial_{\sigma}\Gamma^{\rho}_{\mu\nu}-\partial_{\nu}\Gamma^{\rho}_{\mu\sigma}+\ldots$, and $R_{\mu\nu}=\tensor{R}{^{\rho}_{\mu\rho\nu}}$.}
The action \eqref{star} is characterized by two dimensional parameters: the reduced Planck mass
 ${M_{\mathrm{P}}=1/\sqrt{8\pi G_{\mathrm{N}}}\approx 2.4\times 10^{18}\;\mathrm{GeV}}$, with Newton's constant $G_{\mathrm{N}}$ and the scalaron mass $m_0$. In the regime of large curvatures ${R/m_0^2\gg1}$, the marginal $R^2$ operator in \eqref{star} dominates and an inflationary quasi de Sitter phase and an almost scale invariant spectrum of perturbations is naturally realized due to the asymptotic scale invariance. In contrast, in the regime of small curvatures $R/m_0^2\ll1$, the linear Einstein-Hilbert term in \eqref{star} dominates and naturally realizes a graceful exit from inflation.
The particular degeneracy structure of $f(R)$ theories ensures that the scalaron  is neither a tachyon nor a ghost, provided $M_{\mathrm{P}}^2$ and $m_0^2$ are both positive \cite{Woodard2007,Ruf2018}.

Starobinsky's geometric single-field model of inflation \eqref{star} is in excellent agreement with recent Planck data \cite{Akrami:2018odb}. The main inflationary predictions are the power spectra of tensor perturbations $h_{ij}$ and scalar curvature perturbation $\mathcal{R}$. Due to their weak logarithmic dependence on the wave number $k$, they are parametrized (neglecting the higher order $k$-dependent terms) by the power laws
\begin{align}
 \mathcal{P}_{h}&\approx {}A_{h}\left(\frac{k}{k^{*}}\right)^{n_{h}},\qquad
 \mathcal{P}_{\mathcal{R}} \approx A_{\mathcal{R}}\left(\frac{k}{k^{*}}\right)^{n_{\mathcal{R}}-1}.\label{PowerLawPh}
\end{align} 
The power spectra \eqref{PowerLawPh} are fully characterized by the two amplitudes $A_{h}$ and $A_{\mathcal{R}}$, the two spectral indices $n_h$ and $n_{\mathcal{R}}$, and the pivot scale $k^{*}$ chosen within the CMB window of scales accessible to PLANCK \cite{Aghanim:2018eyx},
\begin{align}
2\times10^{-4}\mathrm{Mpc}^{-1}\lesssim k_{\mathrm{CMB}}\lesssim 2\,\mathrm{Mpc}^{-1}.\label{CMBModes}
\end{align}
The moment when $k^{*}$ first crosses the Hubble horizon is chosen to correspond to $50\leq N\leq 60$ efolds. Since no primordial gravitational waves have been measured yet, it is more convenient to express the amplitude $A_{h}$ in terms of the tensor-to-scalar ratio $r=A_{h}/A_{\mathcal{R}}$. Due to the single-field consistency condition $r=-8\,n_{h}$, the parametrization of the two spectra \eqref{PowerLawPh} is effectively determined only by the three parameters $A_{\mathcal{R}}$, $n_{\mathcal{R}}$ and $r$. PLANCK data \cite{Akrami:2018odb} constrains the values for $A_{\mathcal{R}}$ and $n_{\mathcal{R}}$ at ${k^{*}= 0.05\; \text{Mpc}^{-1}}$,
\begin{align}
A_{\mathcal{R}}^{*}={}&\left(2.099\pm0.014\right)\times 10^{-9}&& (68\%\;\mathrm{CL}),\label{AsCMB}\\
n_{\mathcal{R}}^{*} ={}& 0.9649 \pm 0.0042&& (68\%\;\mathrm{CL}),\label{Planckns}
\end{align}
and for $r$ at ${k^{*}= 0.002\; \text{Mpc}^{-1}}$,
\begin{align}
r^{*}<0.064.
\end{align}  
Starobinsky's model \eqref{star} predicts a scalar amplitude
\begin{align}
A_{\mathcal{R}}^{*}={}&\frac{N^2}{24 \pi^2 }\frac{m_0^2 }{M_{\mathrm{P}}^2}.\label{AsStar}
\end{align}
The normalization condition \eqref{AsCMB} fixes the only free parameter, the scalaron mass as
\begin{align}\label{starm}
m_0\approx1.18\times10^{-5}~M_{\mathrm{P}}=2.8\times 10^{13}~\mathrm{GeV}.
\end{align}
Since the spectral observables are independent of the model parameter, they provide predictions of \eqref{star},
\begin{align}
n_{\mathcal{R}}^{*}\approx {}&1-\frac{2}{N}\approx0.9667,\qquad r^{*}\approx 
\frac{12}{N^2}\approx 0.0033.\label{ObsStar}
\end{align}
The numerical values in \eqref{starm} and \eqref{ObsStar} have been obtained for ${N=60}$.

Even though Starobinsky's inflationary model \eqref{star} has maximal predictive power, it does not exclude interactions with other fields. In particular, in string theory inspired cosmological models, as well as in modified supergravity, the coupling 
of gravity to a dilaton field $\varphi$ arises naturally, see e.g.~\cite{Callan:1985ia,Alexandrov:2016plh,Aldabergenov:2020bpt,Aldabergenov:2020yok}.
A generic coupling of the action \eqref{star} to a canonically normalized dilaton field $\varphi$ with generic potential $V(\varphi)$ can be written in the form
\begin{align}\label{fg}
S[g,\varphi] = \int \mathrm{d}^4x\sqrt{-g} &\left\{f(R,\varphi) 
-\frac{1}{2}g^{\mu\nu}\partial_{\mu}\varphi\partial_{\nu}\varphi \right\},
\end{align}
with the function
\begin{align}\label{smodf}
f(R,\varphi) = 
\frac{U(\varphi)}{2}\left(R + \frac{1}{6\,M^2(\varphi)}R^2\right)-V(\varphi).
\end{align}
Compared to \eqref{star}, the dimensional parameters in \eqref{smodf} have been promoted to generic functions of $\varphi$. The non-minimal coupling function $U(\varphi)$ replaces the constant Planck mass $M_{\mathrm{P}}$, while the the mass function $M(\varphi)$ replaces the scalaron mass $m_0$.

Following the general derivation in \cite{Gundhi:2018wyz}, the Jordan frame (JF) action \eqref{fg} can be written as a two-field model in the classically equivalent\footnote{The equivalence of scalar-tensor theories formulated in different frames, as well as the equivalence between $f(R)$ gravity theories and scalar-tensor theories, also hold at the one-loop quantum level for on-shell field configurations  \cite{Kamenshchik2015,Ruf2018a}.} Einstein frame (EF) by performing the non-linear field redefinitions
\begin{align}
g_{\mu\nu}=\frac{1}{2}\,\frac{M_{\mathrm{P}}^2}{\chi^2}\hat{g}_{\mu\nu},\qquad
\chi=\frac{M_{\mathrm{P}}}{\sqrt{2}}\exp\left(\frac{\hat{\chi}}{\sqrt{6}M_{\mathrm{P}}}\right).\label{conftraf}
\end{align}
In terms of the field covariant formulation,
the EF two-field action resembles a non-linear sigma-model \cite{Gundhi:2018wyz},
\begin{align}
S[\hat{g},\Phi]={}
\int\mathrm{d}^4 x\sqrt{-\hat{g}}&\left[\frac{M_{\mathrm{P}}^2}{2}\hat{R}-\frac{\hat{g}^{\mu\nu}}{2}G_{IJ}\Phi^{I}_{,\mu}\Phi^{J}_{,\nu}-\hat{W}\right].\label{ActScal}
\end{align}
The scalars $\Phi^{I}(x)$ are the local coordinates of the scalar field space manifold 
with the target metric $G_{IJ}$,
\begin{align}
\Phi^{I}=\left(\begin{array}{c}\hat{\chi}\\\varphi\end{array}\right),\qquad G_{IJ}(\Phi)=\left(
\begin{array}{cc}
1&0\\
0&F^{-1}\left(\hat{\chi}\right)
\end{array}
\right)\,. \label{MetricG}
\end{align}
In terms of the parametrizations
\begin{align}
F (\hat{\chi}):={}& \exp\left( \sqrt{\frac{2}{3}} \frac{\hat{\chi}}{M_{\mathrm{P}}}\right),\\ m^2(\varphi):={}&M^2(\varphi)\frac{M_{\mathrm{P}}^2}{U(\varphi)} ~\label{ffactor},
\end{align}
the scalar two-field potential $\hat{W}(\Phi)$ reads
\begin{align} \label{spot1}
\hat{W}(\varphi,\hat{\chi}) = \frac{V}{F^{2}}+ \frac{3}{4} m^2\, M^2_{\mathrm{P}} \left( 1 -\frac{U}{M^2_{\mathrm{P}} F}\right)^2.  
\end{align}
In order to specify a concrete dilaton extension of Starobinsky's geometric model, we assume the following explicit form of the functions $U(\varphi)$, $m^2(\varphi)$, and $V(\varphi)$,
\begin{align}
U(\varphi)&= M^2_{\mathrm{P}}+\xi\varphi^2, \label{coup31}\\
m^2(\varphi)&= m_0^2+\zeta\varphi^2, \label{coup32}\\ 
V(\varphi)  &=\frac{\lambda}{4}\varphi^4. \label{coup33}
\end{align}
The choices \eqref{coup31}-\eqref{coup33} are motivated by several considerations: First, we demand that Starobinsky's model \eqref{star} is  recovered in the limit $\varphi\to0$, justifying the presence of the field independent constants $M_{\mathrm{P}}$ and $m_0^2$ in \eqref{coup31} and \eqref{coup33} and the absence of a field independent contribution (a cosmological constant $\Lambda$) to \eqref{coup33}. Second, we assume an additional internal $\mathbb{Z}_{2}$ symmetry $\varphi\to-\varphi$ leaving only $\varphi$-even operators. Third, we assume that the invariance under the global scale transformations $g_{\mu\nu}\to\alpha^{-2}g_{\mu\nu}$ and $\varphi\to\alpha\varphi$ with constant parameter $\alpha$ is asymptotically realized for large field values 
${\varphi/M_{\mathrm{P}}\gg1}$, justifying the absence of higher order $\varphi$ monomials in 
\eqref{coup31}-\eqref{coup33}.
For \eqref{coup31}-\eqref{coup33}, $\hat{W}$ defined in \eqref{spot1} reduces to
\begin{align}\label{TwoFieldPotential}
\hat{W}(\hat{\chi},\varphi) = \frac{\lambda\varphi^4+ 3 M^2_{\mathrm{P}}\left(m_0^2+\zeta\varphi^2\right) \left( 1+ \xi\frac{\varphi^2}{M^2_{\mathrm{P}}}-F\right)^2}{4F^2}. 
\end{align}
In addition to the two mass parameters $M_{\mathrm{P}}$ and $m_0$, present in the Starobinsky model \eqref{star}, the EF two-field potential \eqref{TwoFieldPotential} is characterized by the three dimensionless parameters
$\xi$, $\zeta$, and $\lambda$.
We do not include a dilaton mass term $m^2_{\mathrm{D}}\,\varphi^2$ in \eqref{coup33}, which in view of the EF potential \eqref{TwoFieldPotential}, can be safely neglected as long as $m^2_{\mathrm{D}}\ll\zeta M_{\mathrm{P}}^2$ or $m^2_{\mathrm{D}}\ll\xi m_{0}^2$. Moreover, since in Sect.~\ref{NumResults} we find that a significant PBH production requires $\xi\gg1$, a dilaton mass term can be neglected as long as $m_{\mathrm{D}}^2\lesssim m_0^2$. It would be interesting to study the impact of a large dilaton mass $m_{\mathrm{D}}^2\gtrsim m_0^2$, which, however, goes beyond the scope of our present work.

\section{Covariant multi-field formalism and inflationary observables} 
\label{CovMultiFieldFormalism}
Following the general treatment in \cite{Gundhi:2018wyz}, we use the covariant multi-field formalism\footnote{A completely covariant treatment of all field variables, including the metric field, was proposed in \cite{Vilkovisky1984} in the general field theoretical context and in \cite{Steinwachs2013, Steinwachs2014,Kamenshchik2015} in the context of cosmological scalar-tensor theories.} to formulate the inflationary dynamics of the background and perturbations.
%
%
\subsection{Background dynamics}
The homogeneous and isotropic background dynamics of the metric is determined by the flat Friedmann-Lema\^itre-Robertson-Walker (FLRW) line element 
\begin{align}
\mathrm{d}s^2 = -\mathrm{d}t^2+a^2 \delta_{ij}\mathrm{d}x^i\mathrm{d}x^j\,.\label{FLRW}
\end{align}
Here, $t$ is the cosmic Friedmann time, $a(t)$ is the scale factor, $i,\,j,\ldots=1,2,3$ are spatial indices and ${\delta_{ij}=\mathrm{diag}(1,1,1)}$ is the spatial metric.
The Friedmann equations and the Klein-Gordon equations for the homogeneous scalar fields $\Phi^{I}(t)$ read
\begin{align}
H^2 ={}& \frac{M^{-2}_{\mathrm{P}}}{3}\left[\frac{1}{2}G_{IJ}\dot{\Phi}^I\dot{\Phi}^J + \hat{W}(\Phi)\right],\label{Friedmann1}\\
\dot{H} ={}& -\frac{M^{-2}_{\mathrm{P}}}{2}G_{IJ}\dot{\Phi}^I\dot{\Phi}^J,\label{Friedmann2}\\
D_{t}\dot{\Phi}^I={}& - 3H\dot{\Phi}^I - G^{IJ}\hat{W},_{J}.\label{KleinGordon}
\end{align}
Here the dots denote the derivatives with respect to the cosmic time $t$.  The Hubble parameter $H(t)$ and the covariant time derivative $D_t$ are defined as
\begin{align}
H(t):=\frac{\dot{a}(t)}{a(t)},\qquad D_{t}V^{I}:=\dot{V}^I+\dot{\Phi}^{J}\Gamma^{I}_{JK}(\Phi)V^{K}.
\end{align}
The Christoffel connection $\Gamma^{I}_{JK}$ is defined with respect to the field space metric \eqref{MetricG}.
The unit vector tangential to the inflationary trajectory reads
\begin{align}
\hat{\sigma}^I:={}& \frac{\dot{\Phi}^I}{\dot{\sigma}},\qquad G_{IJ}\hat{\sigma}^{I}\hat{\sigma}^{J}={}1,\qquad\dot{\sigma}:= \sqrt{G_{IJ}\dot{\Phi}^I\dot{\Phi}^J}.\label{dsigma}
\end{align}
The two-field background dynamics is decomposed into a direction along $\hat{\sigma}^I$ and a direction along the unit vector $\hat{s}^{I}$ orthogonal to $\hat{\sigma}^I$,
\begin{align}
\label{norms}
G_{IJ}\hat{s}^I\hat{s}^J = 1\,,\qquad G_{IJ}\hat{s}^I\sig^J = 0.
\end{align} 
The unit vector $\hat{s}^{I}$ is proportional to the acceleration vector $\omega^I$ which defines the turn-rate $\omega$,
\begin{align}
 \omega^{I}=D_{t}\hat{\sigma}^{I}\,,\qquad \hat{s}^{I}:=\frac{\omega^{I}}{\omega}\,,\qquad  \omega:={}\sqrt{G_{IJ}\omega^I\omega^J}.\label{TurnVector}
\end{align}
Projecting \eqref{Friedmann1}-\eqref{KleinGordon} along $\hat{\sigma}^{I}$ and $\hat{s}^{I}$ leads to the set of background equations
\begin{align}
H^2 ={}& \frac{M_{\mathrm{P}}^{-2}}{3}\left(\frac{1}{2}{\dot{\sigma}}^2 + \hat{W}\right),&
\dot{H} ={}& -\frac{M^{-2}_{\mathrm{P}}}{2}\dot{\sigma}^2,\label{Friedmann2Sig}\\
\ddot{\sigma}={}&-3H\dot{\sigma}- \hat{W}_{,\sigma},& \omega ={}& -\frac{\hat{W},_s}{{\dot{\sigma}}}\label{TurnRateIsoDerive}.
\end{align}
The derivatives of $\hat{W}$ along $\hat{\sigma}^{I}$ and $\hat{s}^{I}$ are defined by 
\begin{align}
\hat{W}_{,\sigma}:=\frac{\partial \hat{W}}{\partial\Phi^I}\hat{\sigma}^{I},\qquad \hat{W}_{,s}:=\frac{\partial \hat{W}}{\partial\Phi^I}\s^{I}.
\end{align}
%
%
\subsection{Perturbations}\label{SecIsoSourcing}

The perturbed FLRW line element reads
\begin{align}
\mathrm{d}s^2=&-\left(1+2A\right)\mathrm{d}t^2+2a B_{,i}\mathrm{d}x^i\mathrm{d}t\nonumber
\\&+a^2\left(\delta_{ij}+2E_{ij}\right)\mathrm{d}x^i\mathrm{d}x^{j}.
\end{align}
Here ${E_{ij}:=\psi\delta_{ij}+E_{,ij}}$ and the scalar metric perturbations $A(t,\mathbf{x})$, $B(t,\mathbf{x})$, $\psi(t,\mathbf{x})$, and $E(t,\mathbf{x})$ combine with the perturbation of the scalar fields $\delta\Phi^{I}(t,\mathbf{x})$.
Instead of $\delta\Phi^{I}(t,\mathbf{x})$, we work with the gauge-invariant multi-field Mukhanov-Sasaki variables \cite{Mukhanov1988,Sasaki1986,Greenwood2013},
\begin{align}
\delta\Phi^{I}_{\mathrm{g}}:=\delta\Phi^I+\frac{\dot{\Phi}^{I}}{H}\psi.\label{MukSasPhi}
\end{align}
The equation for the Fourier modes of the perturbation $\delta\Phi^{I}_{\mathrm{g}}(t,\mathbf{k})$ is found to be \cite{Sasaki1996,Nakamura1996,Greenwood2013},
\begin{align}
D^2_t\delta\Phi^I_\mathrm{g}+ 3HD_t \delta\Phi^I_\mathrm{g}
+\left(\frac{k^2}{a^2}\delta^I_J+\tensor{\Omega}{^{I}_{J}}\right)\delta\Phi^J_\mathrm{g}=0.\label{DynEQPertPhi}
\end{align}
Following the conventions introduced in \cite{Gundhi:2018wyz}, $\tensor{\Omega}{^{I}_{J}}$ and the effective mass tensor $\tensor{M}{^{I}_{J}}$ are defined by
\begin{align}
\tensor{\Omega}{^{I}_{J}}&:={}\tensor{M}{^I_J}-M^{-2}_\mathrm{P}a^{-3}D_t\left(\frac{a^3}{H}{\dot{\Phi}}^I{\dot{\Phi}}_J\right),\label{Omega}\\
M_{IJ} &:={} \nabla_I\nabla_J\hat{W} + R_{IKJL}\dot{\Phi}^K\dot{\Phi}^L.\label{EffMassMatrix}
\end{align}
The effective mass tensor $M_{IJ}$ includes the Riemannian curvature tensor $R_{IJKL}$ associated with the curved scalar field space manifold, as well as the curvature of the multi-field potential $\hat{W}$. 
The tensor modes $h(t,\mathbf{k})$ (suppressing tensor indices) satisfy the simple mode equation,
\begin{align}
\ddot{h}+3H\dot{h}+\frac{k^2}{a^2}h=0.\label{Tensor equation}
\end{align} 
Projecting \eqref{MukSasPhi} along $\hat{\sigma}^{I}$ and $\hat{s}^{I}$ defines the adiabatic and isocurvature perturbations
\begin{align}
Q_{\sigma}=\hat{\sigma}^{J}G_{IJ}\delta\Phi^{I}_{\mathrm{g}},\qquad Q_{\mathrm{s}}=\hat{s}^{J}G_{IJ}\delta\Phi^{I}_{\mathrm{g}}.\label{PertDecomp}
\end{align}
Inserting  $\delta\Phi_{\mathrm{g}}^{I}=Q_{\sigma}\hat{\sigma}^{I}+Q_{\mathrm{s}}\hat{s}^{I}$
into \eqref{DynEQPertPhi}, the dynamical equations for the Fourier modes $Q_{\sigma}(t,\mathbf{k})$, $Q_{\mathrm{s}}(t,\mathbf{k})$ and $h(t,\mathbf{k})$ in the large wavelength limit $k\ll aH$ read
\begin{align}
\ddot{Q}_{\sigma} + 3H\dot{Q}_{\sigma} + \Omega_{\sigma\sigma}Q_{\sigma}={}& f(\mathrm{d}/\mathrm{d}t)(\omega Q_{\mathrm{s}}),\label{EQPertQsig} \\
\ddot{Q}_\mathrm{s} + 3H\dot{Q}_\mathrm{s} + m_\mathrm{s}^2Q_\mathrm{s}={}&0\label{EQPertQs},\\
\ddot{h}+3H\dot{h}={}&0.\label{TensorPert}
\end{align}
The projections of \eqref{Omega} and \eqref{EffMassMatrix} include the additional effective contributions of the turn rate,
\begin{align}
\Omega_{\sigma\sigma}=\hat{\sigma}^I\hat{\sigma}^J\Omega_{IJ}-w^2,\qquad
m^2_{\mathrm{s}}= \s^I\s^J M_{IJ} + 3w^2\label{IsoMassDef}.
\end{align}
The operator $f(\mathrm{d}/\mathrm{d}t)$ in \eqref{EQPertQsig}, acting on the product $\omega Q_{\mathrm{s}}$, is defined as \cite{Gundhi:2018wyz}
\begin{align}
f(\mathrm{d}/\mathrm{d}t)=2\left[\frac{\mathrm{d}}{\mathrm{d}t}-\left(\frac{W,_{\sigma}}{\dot{\sigma}} + \frac{\dot{H}}{H}\right)\right].
\end{align}
Equation \eqref{EQPertQsig} shows that the adiabatic mode $Q_{\sigma}$ is sourced by the product $\omega Q_{\mathrm{s}}$. The turn rate $\omega$ is determined by the background dynamics \eqref{TurnRateIsoDerive}, while the isocurvature mode $Q_{\mathrm{s}}$ is obtained by solving the homogeneous equation \eqref{EQPertQs}. Only if the combination of $\omega$ and $Q_{\mathrm{s}}$ is sufficiently large, $Q_{\sigma}$ is sourced by the isocurvature mode, leading to the amplification of the adiabatic power spectrum.
This ``isocurvature pumping'' amplification mechanism of $Q_{\sigma}$ was already described in \cite{Gundhi:2018wyz} and, as we describe in detail in Sect.~\ref{PeakFormation}, is crucial for the production of PBHs and the numerical results obtained in Sect.~\ref{NumResults}.

As in single-field models of inflation, the deviation from de Sitter space $\dot{H}\neq0$ is quantified  by the Hubble slow-roll parameters
\begin{align}
\varepsilon_{\mathrm{H}}=-\frac{1}{H}\frac{\mathrm{d}\ln H}{\mathrm{d}t},\qquad \eta_{\mathrm{H}}=\frac{1}{H}\frac{\mathrm{d}\ln\varepsilon_{\mathrm{H}}}{\mathrm{d}t}.
\end{align}
The power spectra of the scalar perturbations
\begin{align}
\mathcal{R}(t,\mathbf{k})=\frac{H}{\dot{\sigma}}Q_{\sigma}(t,\mathbf{k}),\label{CurvIsoPert}\qquad \mathcal{S}(t,\mathbf{k})=\frac{H}{\dot{\sigma}}Q_{\mathrm{s}}(t,\mathbf{k}),
\end{align}
and the tensor perturbation $h(t,\mathbf{k})$ read
\begin{align}
\mathcal{P}_{\mathcal{R}}={}&\frac{k^3}{2\pi^2}\left(\frac{H}{\dot{\sigma}}\right)^2\left|Q_{\sigma}\right|^2=\frac{k^3}{4\pi^2M_{\mathrm{P}}^2\varepsilon_{\mathrm{H}}}\left|Q_{\sigma}\right|^2,\label{PowerR}\\ \mathcal{P}_{\mathcal{S}}={}&\frac{k^3}{2\pi^2}\left(\frac{H}{\dot{\sigma}}\right)^2\left|Q_{\mathrm{s}}\right|^2=\frac{k^3}{4\pi^2M_{\mathrm{P}}^2\varepsilon_{\mathrm{H}}}\left|Q_{\mathrm{s}}\right|^2,\label{PowerS}\\
\mathcal{P}_{h}={}&8\frac{k^3}{2\pi^2}\left|h\right|^2.\label{Powerh}
\end{align} 
Note that a simple power law ansatz for the power spectra \eqref{PowerR}-\eqref{Powerh}, as in \eqref{PowerLawPh}, is no longer adequate when the power spectra feature peaks at small wavelengths.
%
%
\section{Two-field potential landscape}
\label{PotLandscape}

\begin{figure}[!ht]
	\centering
	\begin{tabular}{cc}
		&\includegraphics[width=0.45\linewidth]{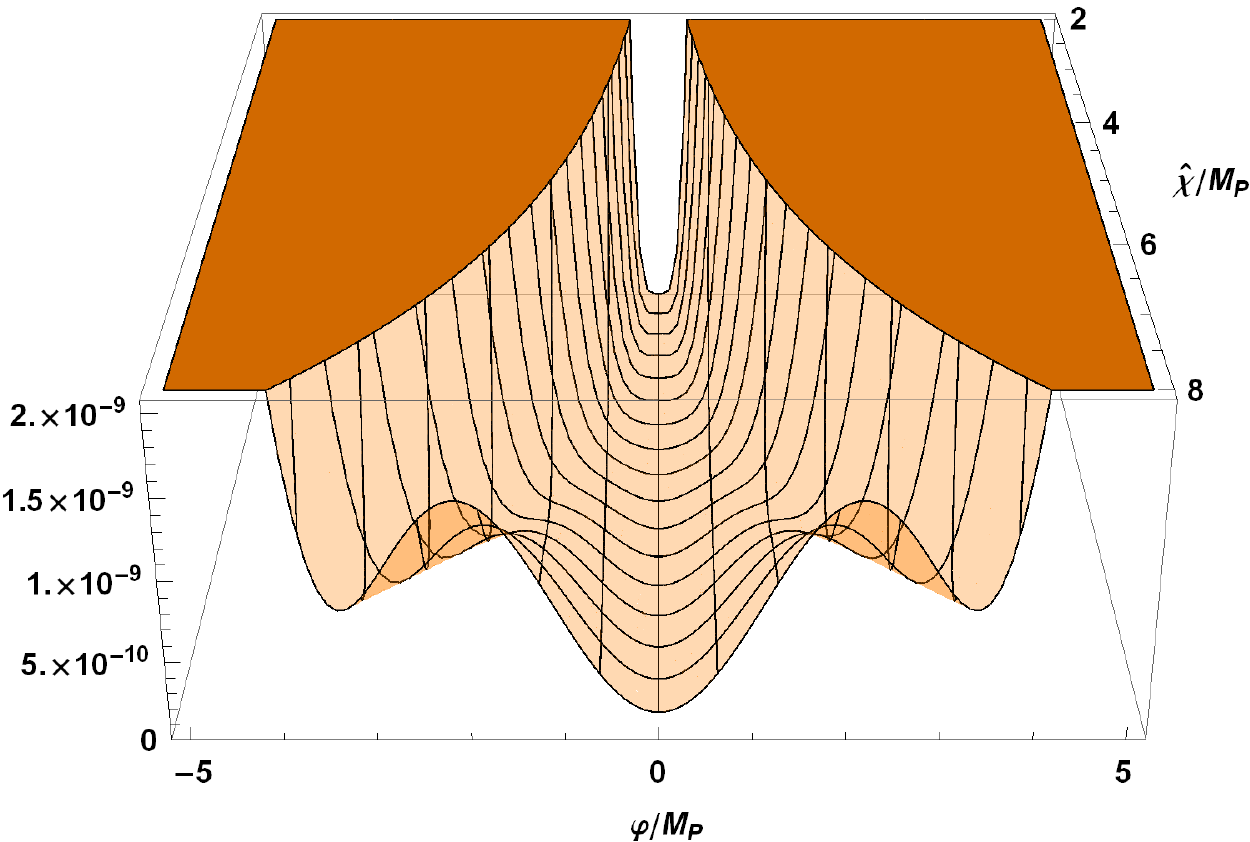}
		\includegraphics[width=0.45\linewidth]{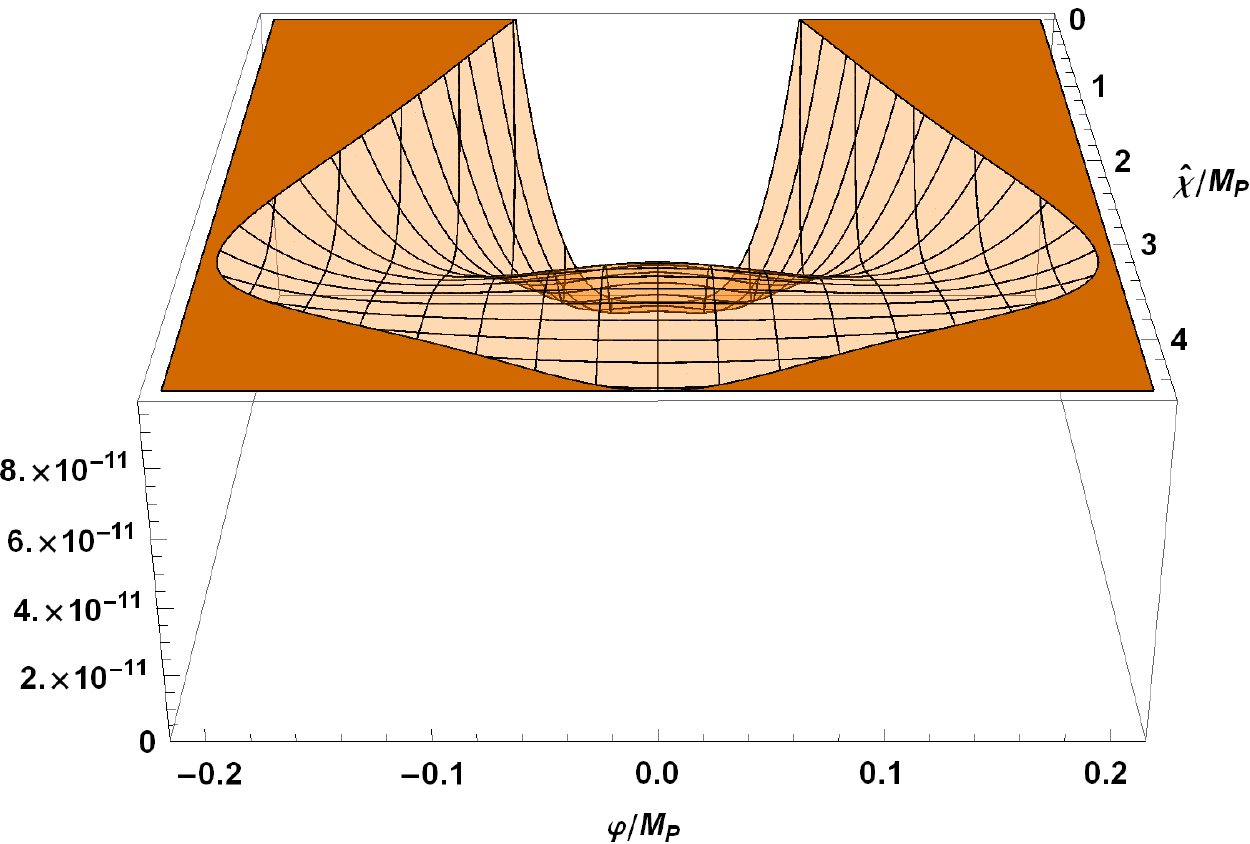}\\
		&\includegraphics[width=0.45\linewidth]{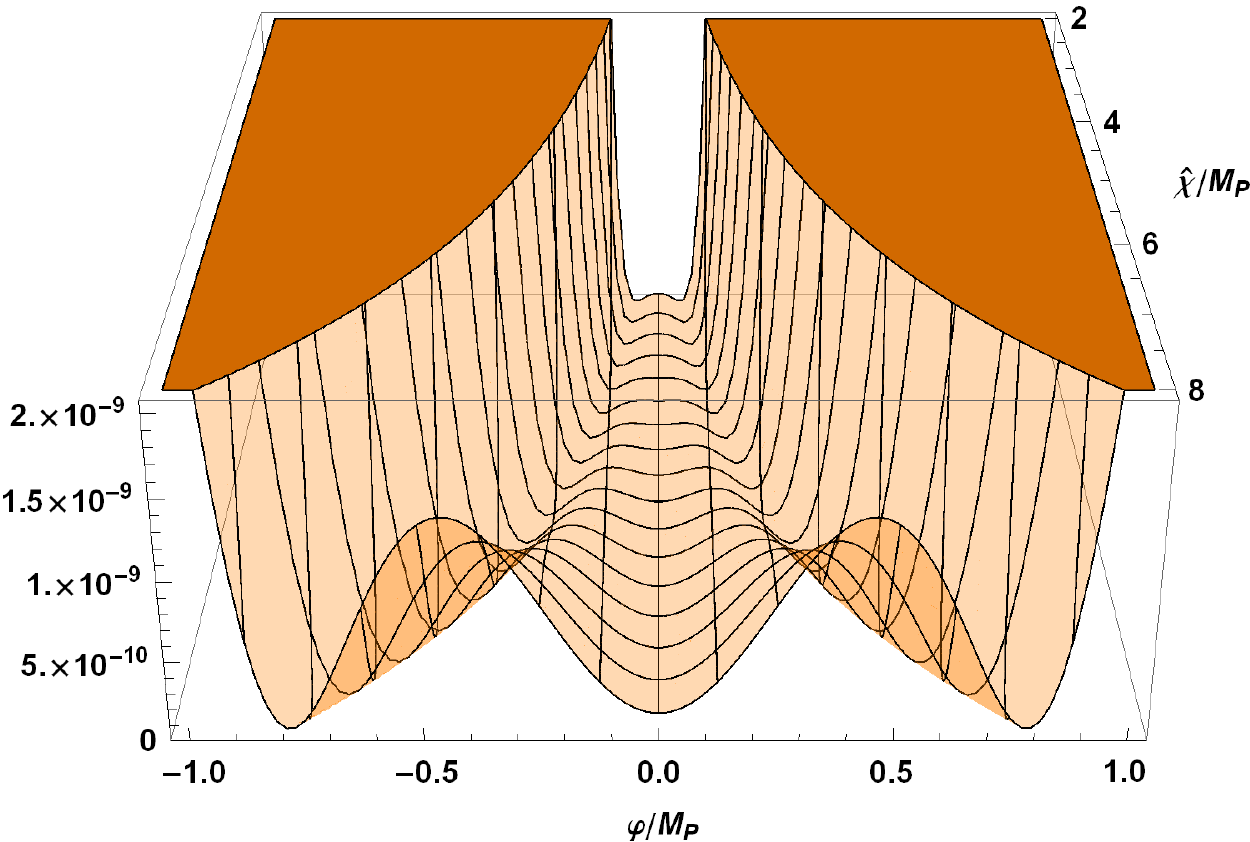}
		\includegraphics[width=0.45\linewidth]{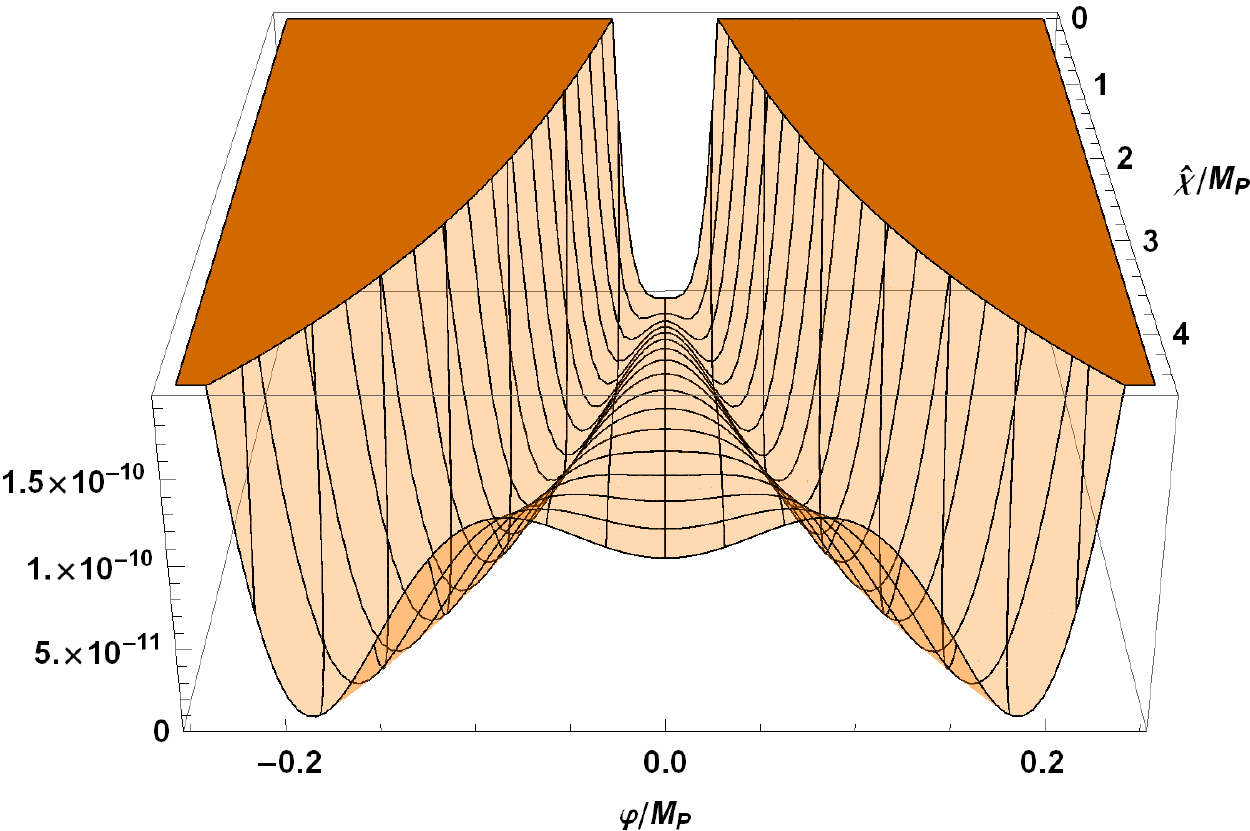}
	\end{tabular}
	\caption{The landscape of the two-field potential \eqref{TwoFieldPotential} (left column) and the vicinity of $\hat{\chi}_{\mathrm{c}}$ (right column) in \mbox{Scenario I} (top row) and \mbox{Scenario II} (bottom row), respectively. }	\label{PotentialLandscape}
\end{figure}

As is shown in Fig.~\ref{PotentialLandscape}, the landscape of the two-field potential \eqref{TwoFieldPotential} is dominated by three valleys separated by two hills symmetrically located around $\varphi=0$ . 
The action \eqref{fg} with the potential \eqref{TwoFieldPotential} can also be viewed as an extension of the 
scalaron-Higgs potential having one additional parameter $\zeta$.

The location of the three valleys and the two hills of the potential \eqref{TwoFieldPotential} are determined by the five roots of the valley equation
\begin{align}
\hat{W}_{,\varphi}=0.\label{ExtrW}
\end{align}
The condition \eqref{ExtrW} follows from the background dynamics of $\varphi$ and $\hat{\chi}$,
\begin{align}
\hat{\chi}^{\prime\prime}+\left(\varepsilon_{\mathrm{H}}-3\right)\left(\hat{\chi}^{\prime}-M_{\mathrm{P}}^2\frac{\hat{W}_{,\hat{\chi}}}{\hat{W}}\right)+\frac{1}{2}F_{,\hat{\chi}}\left(\frac{\varphi^{\prime}}{F}\right)^2 ={}&0,\label{EOMphasespaceChidot}\\
\varphi^{\prime\prime}+\left(\varepsilon_{\mathrm{H}}-3\right)F\left(\frac{\varphi^{\prime}}{F}-M_{\mathrm{P}}^2\frac{\hat{W}_{,\varphi}}{\hat{W}}\right)-F_{,\hat{\chi}}\hat{\chi}^{\prime}\frac{\varphi^{\prime}}{F}={}&0.\label{EOMphasespacePhidot}
\end{align} 
The equations \eqref{EOMphasespaceChidot} and \eqref{EOMphasespacePhidot} in turn follow from \eqref{KleinGordon} with $\hat{\chi}$ and $\varphi$ being functions of the number of efolds $N$ defined by $\mathrm{d}N=-H\mathrm{d}t$, i.e.~we count $N$ backwards with $N=0$ at the end of inflation. Primes denote derivatives with respect to $N$.
For fixed $\hat{\chi}$, a necessary condition to reach a stationary point in phase space $(\varphi,\varphi^{\prime})$ is $\hat{W},_{\varphi}|_{\hat{\chi}}=0$. As $\hat{\chi}$ takes different values during the background evolution, the classical trajectory is obtained by solving \eqref{ExtrW} for $\varphi(\hat{\chi})$. Equation \eqref{ExtrW} has five solutions 
\begin{align}
\varphi_{0}(\hat{\chi})={}&0,\qquad \varphi_{\mathrm{v}}^{\pm}(\hat{\chi}),\qquad\varphi_{\mathrm{h}}^{\pm}(\hat{\chi}).\label{ExtrSolW}
\end{align}
The solutions \eqref{ExtrSolW} correspond to the central valley at $\varphi_0$, the two outer valleys at $\varphi_{\mathrm{v}}^{\pm}$, and the two hills at $\varphi_{\mathrm{h}}^{\pm}$.

At the onset of the inflationary dynamics the initial value $\hat{\chi}_{\mathrm{i}}$ must be sufficiently large $\hat{\chi}_{\mathrm{i}}/M_{\mathrm{P}}\gg 1$, in order to guarantee that inflation lasts at least $N\approx 60$ efolds. Inflation ends in one of the outer valleys $\varphi_{\mathrm{v}}^{\pm}$ close to the global minimum at $\left(\hat{\chi},\varphi\right)=\left(0,0\right)$ when $\varepsilon_{\mathrm{H}}(\varphi_{\mathrm{v}}^{\pm},\hat{\chi}_{\mathrm{f}})=1$.

For inflationary background trajectories, which run along the $\varphi_0$ valley, there is a field value $\hat{\chi}_{\mathrm{c}}$ at which the local $\varphi_0$ minimum turns into an unstable maximum. The critical point $\hat{\chi}_{\mathrm{c}}$ is determined by the condition
\begin{align}
\left.\hat{W}_{,\varphi\varphi}(\hat{\chi},\varphi)\right|_{\varphi=0}=0.\label{critchicon}
\end{align}
The solution of \eqref{critchicon} only depends on $m_0/M_{\mathrm{P}}$ and $\xi/\zeta$,
\begin{align}\label{ChiCrit}
\hat{\chi}_{\mathrm{c}}=M_{\mathrm{p}}\sqrt{\frac{3}{2}} \ln \left[1+2\frac{\xi }{\zeta }\left(\frac{ m_0 }{M_{\mathrm{P}}}\right)^2\right].
\end{align}
For a fixed $m_0^2$ and a suitable ratio $\xi/\zeta$, the value of $\hat{\chi}_{\mathrm{c}}$ can be made sufficiently small, such that all CMB modes \eqref{CMBModes} cross the horizon before the inflationary trajectory crosses the critical point \eqref{ChiCrit}.

\noindent Depending on the parameters of the model, two qualitatively different scenarios are possible. They are shown in the top and bottom rows of Fig.~\ref{PotentialLandscape}, respectively.
In \mbox{Scenario I}, the two outer $\varphi_{\mathrm{v}}^{\pm}$ valleys merge with the central $\varphi_0$ valley before the critical point is reached $\hat{\chi}>\hat{\chi}_{\mathrm{c}}$ and the resulting landscape is that of a single global attractor at $\varphi=0$ (top left plot in Fig.~\ref{PotentialLandscape}). At $\hat{\chi}=\hat{\chi}_{\mathrm{c}}$, when the second derivative $W,_{\varphi\varphi}$ turns negative, the local minimum along the $\varphi$ direction turns into an unstable local maximum and the two valleys symmetrically located around $\varphi_0$ emerge again (top right plot in Fig.~\ref{PotentialLandscape}). 

In \mbox{Scenario II}, the two $\varphi_{\mathrm{v}}^{\pm}$ valleys always run parallel to the central $\varphi_0$ valley and at no stage merge with it into a single global $\varphi_0$ attractor. Nevertheless, 
for initial conditions $\varphi_{\mathrm{i}}/M_{\mathrm{P}}\ll 1$, at $\hat{\chi}_{\mathrm{i}}\gg\hat{\chi}_{\mathrm{c}}$, the local minimum at $\varphi=0$ keeps the inflationary trajectory trapped along the $\varphi_0$ solution until $\hat{\chi}$ reaches $\hat{\chi}_{\mathrm{c}}$. 
As in \mbox{Scenario I}, after the stable local minimum in the $\varphi$ direction turns into an unstable local maximum, the trajectory falls into one of the two lower-lying adjacent $\varphi_{\mathrm{v}}^{\pm}$ valleys.
\begin{figure}[!ht]
	\centering
	\begin{tabular}{cc}
		&\includegraphics[width=0.45\linewidth]{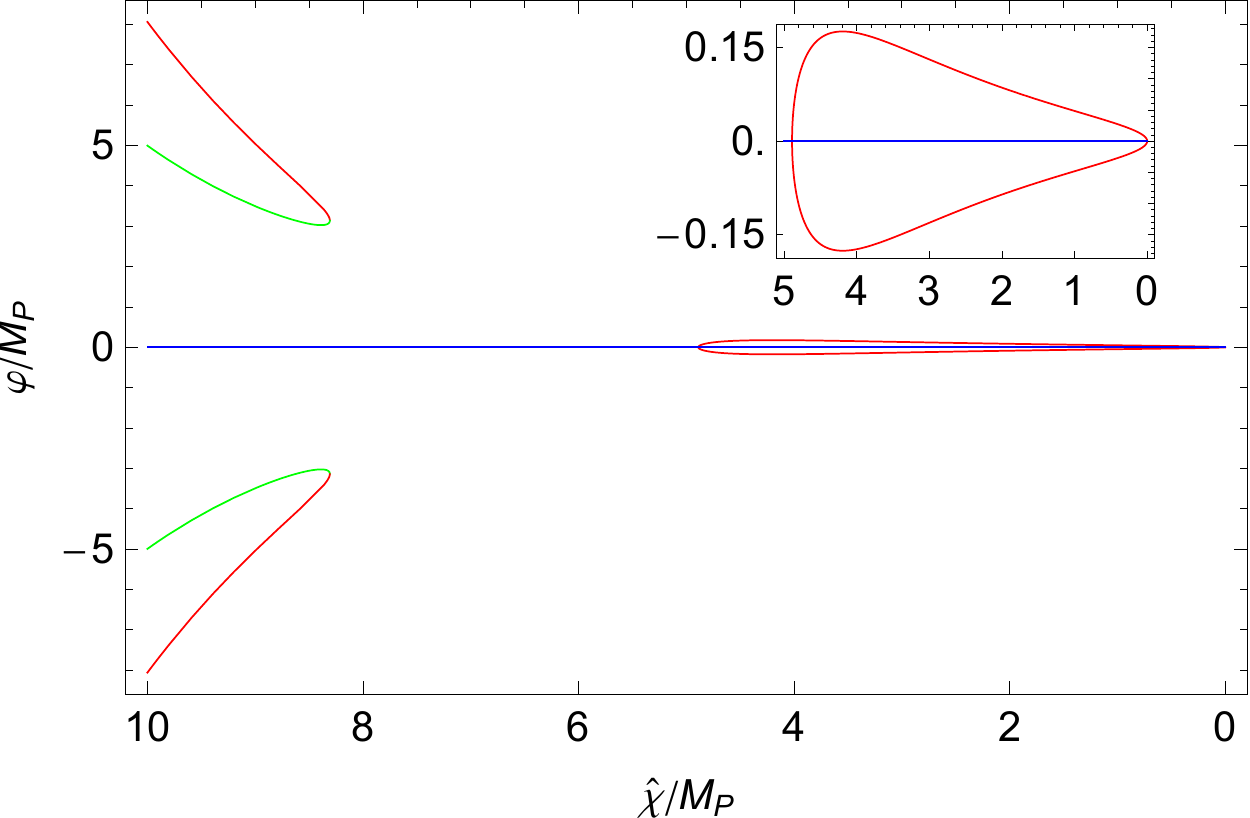}
		\includegraphics[width=0.45\linewidth]{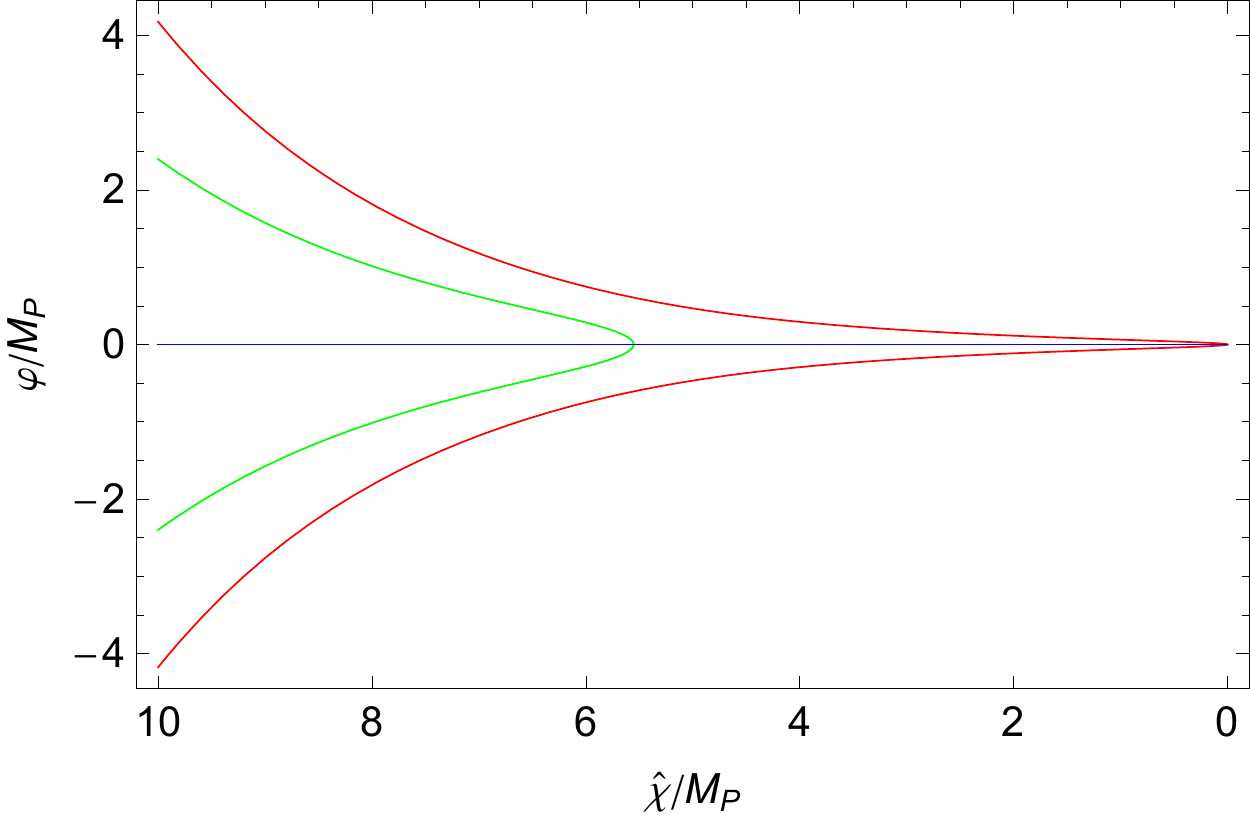}\\
	\end{tabular}
	\caption{Analytic solutions \eqref{ExtrSolW} in \mbox{Scenario I} for $\lambda=10^{-5}$, $\xi=50$, $\zeta=2.6\times 10^{-10}$ (left). The $\varphi^{\pm}_{\mathrm{v}}$ solutions are depicted in red, $\varphi^{\pm}_{\mathrm{h}}$ in green, and $\varphi_0$ in blue. Analytic solutions \eqref{ExtrSolW} in \mbox{Scenario II} for $\lambda=10^{-5}$, $\xi=200$, $\zeta=6\times 10^{-10}$ with the same color coding (right).}
	\label{Fig:ScOne_V_ScTwo}
\end{figure}

\noindent The parameter combination, which distinguishes between the two scenarios, is obtained by observing that in \mbox{Scenario I} the $\varphi_{\mathrm{v}}^{\pm}$ valleys merge with $\varphi^{\pm}_{\mathrm{h}}$ for $\hat{\chi}>\hat{\chi}_{\mathrm{c}}$ and re-emerge for $\hat{\chi}<\hat{\chi}_{\mathrm{c}}$.  In \mbox{Scenario II}, the valleys $\varphi_{0}$ and $\varphi_{\mathrm{v}}^{\pm}$ run parallel to each other for the entire inflationary dynamics. 

As is shown in Fig.~\ref{Fig:ScOne_V_ScTwo}, mathematically this observation is reflected by the fact that in \mbox{Scenario I}, during a certain period before reaching $\hat{\chi}_{\mathrm{c}}$, the valley solutions $\varphi_{\mathrm{v}}^{\pm}$ and $\varphi^{\pm}_{\mathrm{h}}$ become complex. In contrast, in \mbox{Scenario II}, the solutions $\varphi^{\pm}_{\mathrm{v}}$ are real for all values of $\hat{\chi}$. Only the two maxima $\varphi^{\pm}_{\mathrm{h}}$, which earlier separated the $\varphi_0$ valley from the $\varphi^{\pm}_{\mathrm{v}}$ valleys, disappear (or become complex) for $\hat{\chi}<\hat{\chi}_{\mathrm{c}}$.
Demanding that the $\varphi^{\pm}_{\mathrm{v}}$ valley solutions are real for all values of $\hat{\chi}$ leads to the constraint $\lambda\leq 6 \xi^2m_{0}^2/M_{\mathrm{P}}^2$. Hence, \mbox{Scenario I} and \mbox{Scenario II} are obtained for the parameter combinations 
\begin{align}
\lambda> 6 \xi^2{}\frac{m_{0}^2}{M_{\mathrm{P}}^2}&\quad \textrm{Scenario I}\label{ScenarioI},\\
\lambda\leq 6\xi^2\frac{m_{0}^2}{M_{\mathrm{P}}^2}{}&\quad \textrm{Scenario II}.\label{ScenarioII}
\end{align}
As shown in the top right plot in Fig.~\ref{PotentialLandscape}, the merger of the $\varphi^{\pm}_{\mathrm{v}}$ valleys with the $\varphi^{\pm}_{\mathrm{h}}$ hills in \mbox{Scenario I} implies that the potential landscape reduces to a broad global attractor around $\varphi_{0}$ for some $\hat{\chi}>\hat{\chi}_{\mathrm{c}}$. In general this would make the predictions of this scenario independent of the initial field values $\varphi_\mathrm{i}$\footnote{Strictly speaking, for the predictions to be independent of $\varphi_{\mathrm{i}}$,  the valleys must merge well before the CMB modes cross the horizon. Otherwise, for $\varphi_{\mathrm{i}}=\varphi^{\pm}_{\mathrm{v}}(\hat{\chi}_{\mathrm{i}})$, one could have CMB predictions which are different from those of Starobinsky inflation. However, the merger typically happens well before the largest scales of interest cross the horizon and thus makes the whole inflationary dynamics independent of the initial conditions.}. In contrast, in \mbox{Scenario II} the background trajectory always remains in one of the two $\varphi_{\mathrm{v}}^{\pm}$ valleys over the course of the entire inflationary dynamics for initial values $\varphi_{\mathrm{i}}=\varphi^{\pm}_{v}(\hat{\chi_{\mathrm{i}}})$. 

In both scenarios, the dynamics in the vicinity of  $\hat{\chi}_{\mathrm{c}}$ is important due to a rapid growth of isocurvature modes which, in combination with a non-zero turn rate, leads to a strong amplification of the adiabatic modes. This amplification mechanism is described in more detail in Sect.~\ref{PeakFormation}. As explained in Sect.~\ref{Stochastic}, it requires a careful stochastic treatment including diffusive quantum effects. 

Finally, let us compare the potential landscape \eqref{TwoFieldPotential} to that in the model of scalaron-Higgs inflation \cite{Gundhi:2018wyz}, which is recovered from \eqref{TwoFieldPotential} in the limit $\zeta\to0$. In this limit, instead of the central valley, the potential features a central hill at $\varphi=0$, where quantum diffusive effects almost immediately push any background trajectory running along the $\varphi=0$ line into one of the outer valleys at the very onset of inflation. This prevents multi-field effects to have any observable consequences, see \cite{Gundhi:2018wyz} for a detailed discussion.
%
%
\section{Three stages of the inflationary background dynamics}\label{Properties}
The dynamics is naturally divided into three stages:\\

\noindent \textbf{\mbox{Stage 1}}:~Effective single-field slow-roll Starobinsky inflation and CMB observables at large wavelengths.\label{P1}\\

\noindent \textbf{\mbox{Stage 2}}:~Stochastic transition regime with dominant quantum diffusion and tachyonic isocurvature mass.\label{P2}\\

\noindent \textbf{\mbox{Stage 3}}:~PBH peak formation at small wavelengths and effective singe-field slow-roll dynamics in a $\varphi_{\mathrm{v}}^{\pm}$ valley.\label{P3}
%
%
\subsection{\mbox{Stage 1}: Starobinsky inflation and CMB}

During the first stage $\hat{\chi}_{\mathrm{i}}\geq\hat{\chi}>\hat{\chi}_{c}$, the inflationary dynamics effectively reduces to a single-field slow-roll phase along the $\varphi_0$ valley.
For $\hat{\chi}>\hat{\chi}_{\mathrm{c}}$, due to the large and positive value of the isocurvature mass (directly related to the curvature of the potential in the $\varphi$ direction), any deviation of the background trajectory from $\varphi=0$ and any growth of the $\delta\varphi$ perturbation will be immediately suppressed. 

Along the $\varphi_{0}$ valley, the two-field potential \eqref{TwoFieldPotential} reduces to the EF potential of the Starobinsky model \eqref{star},
\begin{align}\label{StarobinskyPotential}
\hat{W}_{\mathrm{Star}}(\hat{\chi}):=\hat{W}(\varphi,\hat{\chi})|_{\varphi=0}=\frac{3}{4}m^2_0M_{\mathrm{P}}^2  \left(1-F^{-1}\right)^2.
\end{align}
Thus, the inflationary predictions are identical to those of the Starobinsky model \eqref{star} for wavelengths that can be probed by CMB measurements, provided the scalaron mass is fixed to the value in \eqref{starm}, and the ratio $\xi/\zeta$ in \eqref{ChiCrit} is chosen such that $\hat{\chi}_{c}$ is sufficiently small to ensure that all CMB modes \eqref{CMBModes} cross the horizon during \mbox{Stage 1}. In particular, $m_0$ is no longer a free parameter but is fixed at this stage, in order  to ensure consistency with CMB measurements.
%
%
\subsection{\mbox{Stage 2}: Stochastic dynamics and tachyonic isocurvature mass}\label{Stochastic}
The formalism of linear perturbation theory, in which the field $\varphi(N,\mathrm{x})$ is decomposed into a homogeneous background field $\bar{\varphi}(N)$ and a perturbation $\delta\varphi(N,\mathrm{x})$, can be safely applied to situations in which $\delta\varphi\ll\bar{\varphi}$.

Any linear perturbation $\delta\varphi$ around the classical solution $\varphi_0$ is strongly suppressed by a large and positive curvature of the potential $\hat{W}_{,\varphi\varphi}$. Along $\varphi_0$ the unit vector tangential to the inflationary trajectory $\hat{\sigma}^{I}$ points in $\hat{\chi}$ direction. Consequently, in view of \eqref{PertDecomp}, along $\varphi_0$ the perturbation $\delta\varphi$ is directly related to the isocurvature perturbation $Q_{\mathrm{s}}$ and $\hat{W}_{,\varphi\varphi}$ to the effective isocurvature mass $m_{\mathrm{s}}^2$ defined in \eqref{IsoMassDef}.
 
While the inflationary trajectory along $\bar{\varphi}=\varphi_0=0$ is classically stable for $\hat{\chi}\gg\hat{\chi}_{\mathrm{c}}$, in the vicinity of the critical point $\hat{\chi}_{\mathrm{c}}$, the restoring classical force, which keeps the trajectory focused to the $\varphi_0$ attractor, is no longer sufficiently strong to counteract the diffusive force that originates from the unavoidable quantum zero-point fluctuations the trajectory experiences in the $\varphi$ direction. 
For $\hat{\chi}<\hat{\chi}_{\mathrm{c}}$, the isocurvature mass turns negative and the solution $\varphi_0$ becomes unstable. At the same time, the perturbation $\delta\varphi$ starts to grow and even dominates over the classical solution $\varphi_0$.

In situations where the quantum diffusive force dominates the classical background drift, the formalism, in which the background dynamics of the scalar fields is considered to be independent of the time evolution of the quantum fluctuations, breaks down. Instead, the application of the stochastic formalism introduced in \cite{Starobinsky:1986fx}, which properly takes into account the back reaction of quantum fluctuations on the coarse grained classical background dynamics, is required. 
In the stochastic formalism, the dynamics of $\varphi$ during the transition stage around $\hat{\chi}_{\mathrm{c}}$ is determined by a probability density function (PDF) $P(\varphi ,N)$ that specifies the probability of the field having the value $\varphi$ at a time $N$.

The time evolution of $P(\varphi ,N)$ is described by the Fokker-Planck equation, see e.g.~\cite{Vennin:2020kng},
\begin{align}
\frac{\partial P}{\partial N}=-\frac{\partial}{\partial \varphi }\left[\mathcal{D}P\right]+\frac{1}{2} \frac{\partial ^2}{\partial \varphi^2}\left[\mathcal{F}P\right].\label{FP}
\end{align}
The right-hand-side of the Fokker-Planck equation \eqref{FP} is characterized by two terms: a classical drift term with  the coefficient $\mathcal{D}(\varphi,N)$ and a quantum diffusion, or ``fluctuation'', term with the coefficient $\mathcal{F}(\varphi,N)$. 

In the context of the inflationary dynamics in the vicinity of $\hat{\chi}_{\mathrm{c}}$, the drift coefficient $\mathcal{D}$ in \eqref{FP} corresponds to the rate of change of the classical (averaged) field, while the fluctuation coefficient $\mathcal{F}$ corresponds to the rate of change of the variance. 
For the decomposition ${\varphi(N,\mathbf{x})=\bar{\varphi}(N)+\delta\varphi(N,\mathbf{x})}$ into a homogeneous background $\bar{\varphi}(N)$ and a Gaussian random fluctuation $\delta\varphi(N,\mathbf{x})$ with $\langle\delta\varphi\rangle=0$, the coefficients are obtained as
\begin{align}
\mathcal{D}=\frac{\mathrm{d}\langle\varphi\rangle}{\mathrm{d}N},\qquad\mathcal{F}=\frac{\mathrm{d}\langle\delta\varphi^2\rangle}{\mathrm{d}N}.
\end{align}
In the following we omit the bar over a background quantity.
Assuming slow-roll in the $\hat{\chi}$ and $\varphi$ directions\footnote{In fact, slow-roll is automatically realized by the phase of Starobinsky inflation in \mbox{Stage 1} which sets the initial conditions for the subsequent evolution in \mbox{Stage 2}. Within the slow-roll approximation $\varphi^{\prime\prime}\approx0$, $\varepsilon_{\mathrm{H}}\ll3$ and the term proportional to $F_{,\hat{\chi}}\hat{\chi}^{\prime}\varphi^{\prime}/F$ can be neglected in \eqref{EOMphasespacePhidot}}, the equation of motion \eqref{EOMphasespacePhidot} reduces to the single-field dynamics of $\varphi$ depending only parametrically on $\hat{\chi}(N)$,
\begin{align}
\varphi^{\prime}\approx \frac{F\hat{W},_{\varphi}}{3H^2}.\label{slp}
\end{align}
Since we are interested in the dynamics around ${\varphi=0}$,
Taylor expansion of $\hat{W},_{\varphi}$ yields the linearized equation which determines the  drift coefficient
\begin{align}
\mathcal{D}(\varphi,N)=\varphi^{\prime}\approx \frac{m^2_{\varphi}}{3H^2}\varphi,\label{Drift}
\end{align}
with the effective $N$-dependent mass
\begin{align}
m^2_{\varphi}(N):=F\hat{W},_{\varphi\varphi}|_{\varphi=0}.
\end{align}
The variance $\left\langle\delta\varphi^2\right\rangle$ after coarse graining over the horizon scale is given in terms of the power spectrum ${\mathcal{P}_{\varphi}(k)\equiv k^3|\delta\varphi_k|^2/(2\pi^2)}$ of  the scalar perturbation $\delta\varphi$,
\begin{align}\label{VariancePhi}
\left\langle\delta\varphi^2\right\rangle=\int_{0}^{k\lesssim aH}\mathcal{P}_{\varphi}(k)d\ln k.
\end{align}
The modes $k$ crossing the Hubble horizon in the time interval $-\Delta N$ contribute to the increment of the integral \eqref{VariancePhi} by increasing its upper bound and satisfy $k=aH=a_*e^{(N-N_*)} H$ at their respective moments of horizon crossing. Here $a_*$ is the value of the scale factor at some earlier time, $N>N_*$. Treating $H$ as a constant, we obtain $d\ln k\approx - dN$, which results in the diffusive random ``quantum kicks'' the coarse-grained field experiences from the Fourier modes $\delta\varphi_{\mathrm{k}}$ continuously crossing the Hubble horizon, 
\begin{align}
\mathcal{F}(N)=-\left.\mathcal{P}_{\varphi}(k)\right|_{k\lesssim aH}\approx-\left(\frac{H}{2\pi}\right)^2.\label{Fluc}
\end{align}  
Inserting \eqref{Drift} and \eqref{Fluc} into \eqref{FP}, $P(\varphi ,N)$ satisfies the Fokker-Planck equation
\begin{align}\label{FockerPlanck}
\frac{\partial P}{\partial N}=-\frac{m^2_{\varphi}}{3 H^2} \frac{\partial (\varphi  P)}{\partial \varphi }-\frac{H^2}{8 \pi ^2} \frac{\partial ^2P}{{\partial \varphi} ^2}.
\end{align}  
The Fokker Planck equation \eqref{FockerPlanck} is solved by a Gaussian ansatz with a time dependent variance $S(N)=\langle\varphi^2\rangle(N)$,
\begin{align}\label{GaussianGuess}
P(\varphi,N)=\frac{1}{\sqrt{2\pi S(N)}}\exp\left(-\frac{\varphi^2}{2S(N)}\right).
\end{align} 
Inserting \eqref{GaussianGuess} into \eqref{FockerPlanck} yields an equation for $S$,
\begin{align}\label{VarianceTimeEv}
\frac{d S}{dN}=\frac{2}{3}\frac{m^2_{\varphi}}{H^2}S -\frac{H^2}{4\pi^2}.
\end{align} 
Hence, we may interpret the equation \eqref{VarianceTimeEv} for the variance $S$ as determining the time evolution of effective amplitude of the scalar field by identifying \cite{Randall:1995dj},
\begin{align}
\varphi(N)\equiv\sqrt{S(N)}.
\end{align}
We apply the stochastic formalism for a period in which the quantum diffusive term  $H^2/4\pi^2$ dominates the classical term $2 m^2_{\varphi}S/(3H^2)$.
The stochastic stage starts at some point $\hat{\chi}>\hat{\chi}_{\mathrm{c}}$ at which $m^2_{\varphi}$ falls from large positive values towards zero, and ends when it takes negative values for some $\hat{\chi}<\hat{\chi}_{\mathrm{c}}$. The onset of the stochastic stage can be determined by the condition $2m^2_{\varphi}/3H^2<1$. This ensures that for any inevitable initial increment of $S(N)\sim H^2/4\pi^2$ the diffusive term dominates the classical one. Similarly, the end of stochastic phase can be roughly estimated by the condition $2m^2_{\varphi}/3H^2\approx-1$.\footnote{A more precise discussion is presented in Sect.~\ref{NumericalTreatment}} If the isocurvature mass becomes more negative, the classical dynamics will dominate again and we resort back to the original equation \eqref{EOMphasespacePhidot}. Thus, the duration $\Delta N$ of \mbox{Stage 2}, can be estimated by calculating the time it takes $2m^2_{\varphi}/3H^2$ to change from $1$ to $-1$. For the potential \eqref{TwoFieldPotential}, the ratio  $2m^2_{\varphi}/3H^2$ is given by
\begin{align}
\frac{2m^2_{\varphi}}{3H^2}=4 \zeta F \frac{M_{\mathrm{P}}^2}{m^2_0}-8\xi\frac{F }{F-1}
\end{align}
that, in turn, provides the estimate
\begin{align}
\Delta F(\hat{\chi})\approx\frac{1}{\zeta}\frac{m^2_0}{M_{\mathrm{P}}^2}.
\end{align} 
Along $\varphi_0$, the difference $\Delta F$ can be estimated from Starobinsky inflation $\Delta F\approx \Delta N$, leading to
\begin{align}
\Delta N\approx \frac{1}{\zeta}\frac{m^2_0}{M_{\mathrm{P}}^2}.\label{DeltaN}
\end{align}
%
%
\subsection{\mbox{\mbox{Stage 3}}: Peak formation and slow-roll }
After the isocurvature mass first crosses zero and becomes tachyonic at $\hat{\chi}_{\mathrm{c}}$, it grows again (but negatively) while the trajectory still follows the $\varphi_0$ solution (which is now a hill). Eventually its magnitude becomes again comparable with the magnitude of the quantum diffusive kicks \eqref{Fluc} and the classical dynamics takes over again. At this point \mbox{Stage 3} starts, the trajectory turns/falls into one of the $\varphi_{\mathrm{v}}^{\pm}$ valleys and the growing isocurvature modes source the adiabatic modes, ultimately leading to a peak in the adiabatic power spectrum. 

The wavenumber $k_{\mathrm{p}}$, at which the power spectrum features a peak, crosses the horizon shortly after the trajectory passes $\hat{\chi}_{\mathrm{c}}$. Therefore, a peak at a fixed $k_{\mathrm{p}}$ requires that $\hat{\chi_{\mathrm{c}}}\neq0$. In addition, a narrow peak in the power spectrum centered at $k_{\mathrm{p}}$ requires that stage two lasts for less than one efold $\Delta N\leq 1$, which according to \eqref{DeltaN} implies $\zeta\geq m^2_0/M_{\mathrm{P}}^2$. Hence, according to \eqref{ChiCrit}, a non-zero $\hat{\chi}_{\mathrm{c}}$ implies $\xi\gg1$. As we will see in Sect.~\ref{NumericalTreatment}, the numerical results confirm that the parameter combinations, which lead to a suitable peak in the power spectrum, respect these expectations.

Although, \mbox{Stage 2} lasts for less than one efold, the stochastic treatment is essential as it provides the initial conditions for the dynamics in \mbox{Stage 3} during which the power spectrum is amplified.
The remaining inflationary dynamics until the end of inflation takes place inside the $\varphi_{\mathrm{v}}^{\pm}$ valley and is again that of an effective single-field model with a slow-roll potential.
%
%
\section{Peak formation mechanism}\label{PeakFormation}
The peak formation mechanism during \mbox{Stage 3} is based on the ``isocurvature pumping'' effect described in \cite{Sfakianakis:2014xxe} and explained in detail in the context of scalaron-Higgs inflation in \cite{Gundhi:2018wyz}.
Qualitatively, this mechanism can be easily understood from the dynamical equations for the scalar perturbations \eqref{EQPertQsig} and \eqref{EQPertQs}. 
Although the exact dynamics of the isocurvature modes \eqref{EQPertQs} is more complicated in general, let's assume for purely illustrative purposes that $\ddot{Q}_{\mathrm{s}}\approx0$. In this case the solution to \eqref{EQPertQs} would read
\begin{align}
Q_{\mathrm{s}}\approx \exp\left(-\int_{N_1}^{N_2}\mathrm{d}N\frac{m_{\mathrm{s}}^2(N)}{3H^2(N)}\right).\label{IsoExp}
\end{align}  
For a positive isocurvature mass $m_{\mathrm{s}}^2>0$, the amplitude of the isocurvature modes $Q_{\mathrm{s}}$ is exponentially damped. Conversely, for a tachyonic isocurvature mass $m_{\mathrm{s}}^2<0$, the amplitude of the isocurvature modes $Q_{\mathrm{s}}$ is exponentially amplified. The total amplification depends on the amplification factor $m_{\mathrm{s}}^2/(3H^2)$ and the duration $\Delta N=N_2-N_1$.

Next, let us look at the equation for the adiabatic modes \eqref{EQPertQsig}. They are sourced by the product of the turn rate $\omega$ and the isocurvature modes $Q_{\mathrm{s}}$. Hence, only if $\omega\neq0$ and the amplitude of $Q_{\mathrm{s}}$ is sufficiently large, the source term will have a sizeable impact on the adiabatic modes.\footnote{The case in which the effective isocurvature mass in \eqref{IsoMassDef} is dominated by the (positive) contribution of the turn rate (assuming a Gaussian profile function for  $\eta_{\perp}(N)=\omega(N)/H(N)$) arising from the curvature of the scalar field space manifold was studied in \cite{Fumagalli:2020adf}. Even if under such assumption the analysis can be carried out in a model-independent way, it does not seem to be applicable to most realistic models in which the model-dependent potential dominates the effective isocurvature mass and is responsible for its tachyonic instability.} 
Thus, in order for this multi-field amplification mechanism to be realized in a concrete model, several factors have to work together in a synchronized way. First, a sufficiently long inflationary phase with a tachyonic effective isocurvature mass is required for the isocurvature modes to grow. Next, this amplification must be transmitted to the adiabatic modes, which requires that the inflationary trajectory follows a curved path in the field space to yield a non-zero turn rate. To avoid an over-amplification, the sourcing must terminate at some point, either via the vanishing turn rate or via the exponential suppression of the isocurvature modes which requires the effective isocurvature mass to turn positive again.

\begin{figure}[!ht]
\centering
\includegraphics[width=0.65\linewidth]{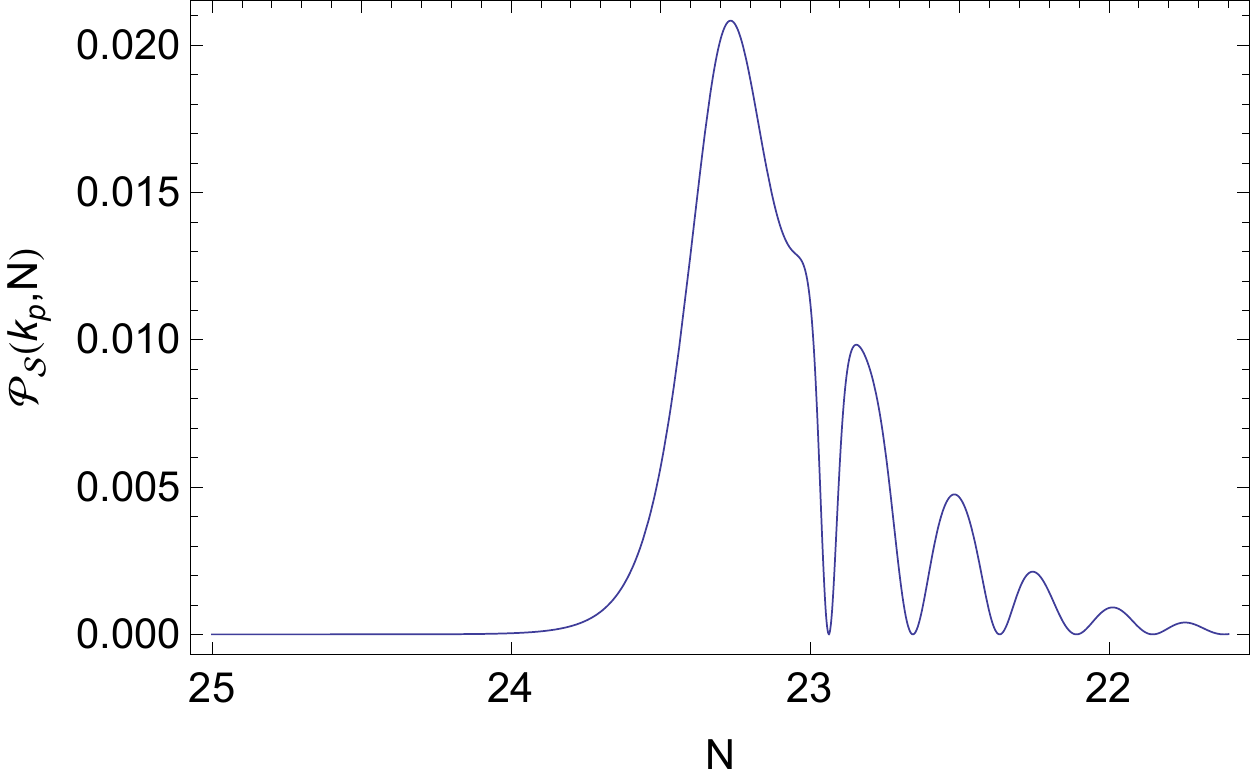}
\caption{Time evolution of the isocurvature power spectrum $\mathcal{P}_{\mathcal{S}}$ for the mode $k_{\mathrm{p}}=2.06\times10^{12} \mathrm{Mpc}^{-1}$ at which the power spectrum in Fig.~\ref{Scenario1Fig2} peaks. Comparing it with the bottom left plot in Fig.~\ref{Scenario1Fig1} shows that the isocurvature perturbation $Q_s$ starts growing when the isocurvature mass turns negative. }
\label{Fig:IsocurvaturePumping}
\end{figure}

The two-field potential \eqref{TwoFieldPotential} satisfies all these requirements. During the transition from \mbox{Stage 1} to \mbox{Stage 2}, the effective isocurvature mass turns tachyonic and the isocurvature modes start growing. At the same time, the stable $\varphi_0$ valley turns into an unstable hill, such that the trajectory eventually turns/falls into one of the $\varphi_{\mathrm{v}}^{\pm}$ valleys, and thereby permits a sourcing of adiabatic modes and ultimately leads to the formation of a peak in the adiabatic power spectrum.
During the fall the effective isocurvature mass turns positive again and leads to an exponential suppression of isocurvature modes during the subsequent slow-roll phase of inflation in $\varphi_{\mathrm{v}}^{\pm}$. This growth and subsequent damping of the isocurvature modes for \mbox{Scenario I} is shown in Fig.~\ref{Fig:IsocurvaturePumping}. It is obtained for the same parameter values as for those in Fig.~\ref{Scenario1Fig1} and Fig.~\ref{Scenario1Fig2} obtained in Sect.~\ref{NumericalTreatment}, so that its relation to the temporary tachyonic nature of the effective isocurvature mass and the peak in the power spectrum can be directly compared.

\noindent The efficiency and magnitude of the amplification is highly sensitive to the time the trajectory spends on the hill during its unstable dynamics in the vicinity of the critical point $\hat{\chi}_c$. Around $\hat{\chi}_{\mathrm{c}}$, the inflationary trajectory is no longer protected by the strong restoring classical force in $\varphi$ direction and, therefore, is  exposed to the diffusive quantum kicks driving it away from $\varphi_0=0$. Although this phase of unstable motion along the $\varphi_0$ direction is very short, the magnitude of the quantum kicks it gives to the background trajectory is crucial for determining the time for which the isocurvature pumping lasts. Hence, the stochastic treatment is crucial for the quantitative treatment of the amplification mechanism, and is described in detail in Sect.~\ref{Stochastic}.  
Note that this peak formation mechanism is very different from that in single-field models of inflation, see e.g.~\cite{Garcia-Bellido:2017aan,Ezquiaga:2017fvi,Garcia-Bellido:2017mdw} for models in which the amplification is based on an intermediate ultra slow-roll phase resulting from a potential featuring an inflection point.
%
%
\section{Numerical treatment}\label{NumericalTreatment}

As described in Sect.~\ref{Properties} and Sect.~\ref{PeakFormation}, the individual stages of the background dynamics can be easily understood qualitatively. However, a precise calculation including the dynamics of the perturbations has to be carried out numerically. The entire background dynamics is obtained by patching the numerical solutions of the equations in the individual stages in such a way that the preceding stages provide the initial conditions for the subsequent stages.

During \mbox{Stage 1} and \mbox{Stage 3}, we numerically solve the exact equations of motion \eqref{EOMphasespacePhidot} and \eqref{EOMphasespaceChidot} for both scalar fields. In \mbox{Stage 2}, in which the stochastic formalism is used to describe the $\varphi$ dynamics, we numerically solve the equations of motion
\begin{align}
\frac{d S}{dN}={}&\frac{2}{3} \frac{m^2_{\varphi}}{H^2}S-\frac{H^2}{4\pi^2},\label{stage2Stoch}\\
\frac{d^2\hat{\chi}}{dN^2}={}&(3-\varepsilon_{H})\left( \frac{d\hat{\chi}}{dN}-\frac{ \hat{W},_{\hat{\chi}}}{\hat{W}}M_{\mathrm{P}}^2\right).\label{stage2Chi}
\end{align} 
Since the actual fall into the valley occurs during \mbox{Stage 3}, during \mbox{Stage 2} both scalar fields $\hat{\chi}$ and $\varphi$ can be safely  considered as slowly rolling with the inflaton unit vector $\hat{\sigma}$ still pointing in the $\hat{\chi}$ direction.
Consequently, we neglect the ${\varphi'}^2$ term in \eqref{EOMphasespaceChidot} during \mbox{Stage 2}. This can also be seen from 
Fig.~\ref{Scenario1Fig1} where the spike in the slow roll parameter induced by the fall into the $\varphi^{+}_{\mathrm{v}}$ valley occurs well to the right of the red vertical dashed lines, marking the beginning and end of \mbox{Stage 2}. 
 
In order to patch the numerical solutions obtained in the different stages, we have to find the transition moments between them.
During the first phase along $\varphi_0$, the steep positive curvature of the potential along the $\varphi$ direction provides a strong restoring force which immediately erases the effect of the continuous quantum kicks trying to drive $\varphi$ away from $\varphi=0$. 
The moment $N_1$ of the transition between \mbox{Stage 1} and \mbox{Stage 2} can, therefore, be inferred from the moment at which for the first time the effect of a quantum kick $H/(2\pi)$ on $\varphi=\sqrt{S}$ will not be erased, i.e.~when the drift term in \eqref{stage2Stoch} becomes comparable to the diffusive term for $S(N_1)=H^2(N_1)/(4\pi^2)$. The resulting condition is solved numerically for $N_1$ as
\begin{align}
m^2_{\varphi}(N_1)=\frac{3}{2}H^2(N_1).
\end{align} 
Since $S=\varphi^2$ is effectively zero before $N_1$, the complete set of initial conditions which result from patching
\mbox{Stage 1} and \mbox{Stage 2} read 
\begin{align}\label{InCondStage2}
S(N_1)={}&0,\\ 
\left.\hat{\chi}\right|^{\mathrm{s2}}_{ N_1}={}&\left.\hat{\chi}\right|^{\mathrm{s1}}_{N_1},\qquad \left.\frac{\mathrm{d}\hat{\chi}}{\mathrm{d}N}\right|^{\mathrm{s2}}_{ N_1}=\left.\frac{\mathrm{d}\hat{\chi}}{\mathrm{d}N}\right|^{\mathrm{s1}}_{N_1}.
\end{align}
\mbox{Stage 2} lasts until the curvature of the potential becomes dominant again (but this time with a negative sign). The time $N_2$, at which \mbox{Stage 2} ends, is determined numerically from the condition
\begin{align}
-\frac{2}{3 }\frac{m_{\varphi}^2(N_2)}{H^2(N_2)}S(N_2)=\frac{H^2(N_2)}{4\pi^2}.
\end{align}
The initial conditions for \mbox{Stage 3} are 
\begin{align}
\left.\varphi\right|^{\mathrm{s3}}_{N_2}={}&S^{1/2}(N_2)&\left.\frac{\mathrm{d}\varphi}{\mathrm{d}N}\right|^{\mathrm{s3}}_{N_2}={}&\left.\frac{1}{2\sqrt{S}}\frac{\mathrm{d}S}{\mathrm{d}N}\right|_{N_2},\\
\left.\hat{\chi}\right|^{\mathrm{s3}}_{N_2}={}&\left.\hat{\chi}\right|^{\mathrm{s2}}_{N_2}&
\left.\frac{\mathrm{d}\hat{\chi}}{\mathrm{d}N}\right|^{\mathrm{s3}}_{N_2}={}&\left.\frac{\mathrm{d}\hat{\chi}}{\mathrm{d}N}\right|^{\mathrm{s2}}_{N_2}.
\end{align}
The numerical analysis confirms the analytic estimate \eqref{DeltaN} that the second stage typically lasts for less than one efold and that the values acquired by $\varphi$ at the beginning of \mbox{Stage 3} are very small. This a posteriori justifies the assumptions of slow-roll along $\varphi$ in \eqref{slp}, the Taylor expansion of the potential in \eqref{Drift}, and the Gaussian solution \eqref{GaussianGuess} to the Fokker-Planck equation \eqref{FockerPlanck}. 

The numerical solutions for $\varphi(N)$, $\hat{\chi}(N)$ and $a(N)$ are then used in the equations for the perturbations \eqref{DynEQPertPhi}, which are solved numerically with Bunch-Davis initial conditions imposed in the deep subhorizon regime. Finally, the power spectra \eqref{PowerR}-\eqref{Powerh} are computed numerically for the pivot scale $k_{*}=0.002\,\mathrm{Mpc}^{-1}$ crossing the horizon at $N=60$.
We discuss the numerical results separately for \mbox{Scenario I} and \mbox{Scenario II}. 
%
%
\subsection{\mbox{Scenario I}}

For the parameter values \eqref{ScenarioI}, for which \mbox{Scenario I} is realized, the effective single-field stage with the Starobinsky potential \eqref{StarobinskyPotential} along $\varphi_0$ will be inevitably attained, independently of the initial value $\varphi_{\mathrm{i}}$.

A successful phase of Starobinsky inflation during \mbox{Stage 1} requires $m_0^2/M_{\mathrm{P}}^2=1.18\times 10^{-5}$ to satisfy the COBE normalization \eqref{AsCMB}. The parameters are further constrained by demanding that PBHs are produced within a given PBH mass window (constraining the peak location in the power spectrum) and that these PBHs lead to a significant amount of CDM as observed today (constraining the amplitude of the peak). For instance, as explained in the context of the peak formation mechanism in Sect.~\ref{PeakFormation}, demanding that the power spectrum peaks on specific scales constraints the ratio $\xi/\zeta$ that, in turn, determines $\hat{\chi}_{\mathrm{c}}$ and, hence, also the efolding number around which the sourcing takes place. Altogether, these constraints fix three out of the four parameters $m_0$, $\zeta$, $\xi$, and $\lambda$.

\begin{figure}[!ht]
	\centering
	\begin{tabular}{cc}
		\includegraphics[width=0.45\linewidth]{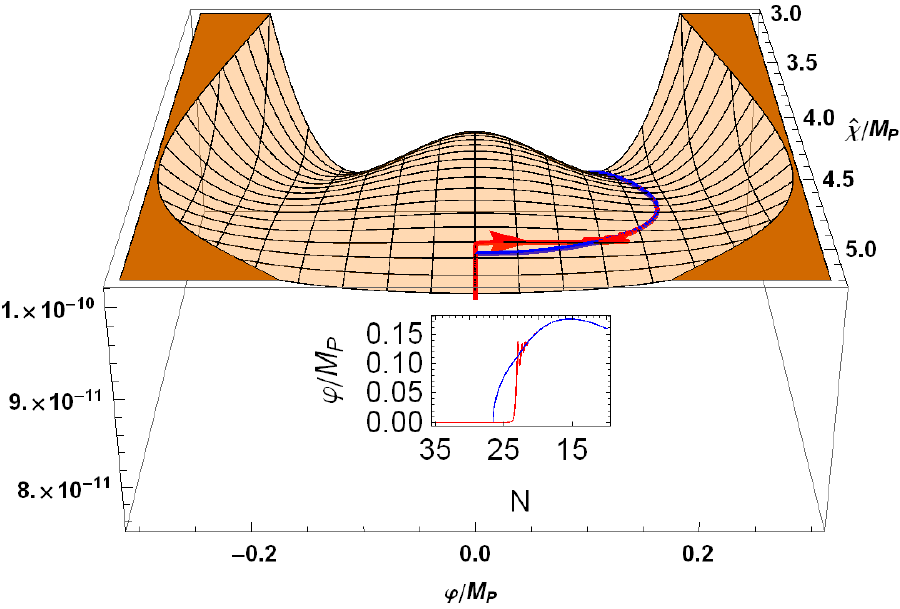}&
		\includegraphics[width=0.4\linewidth]{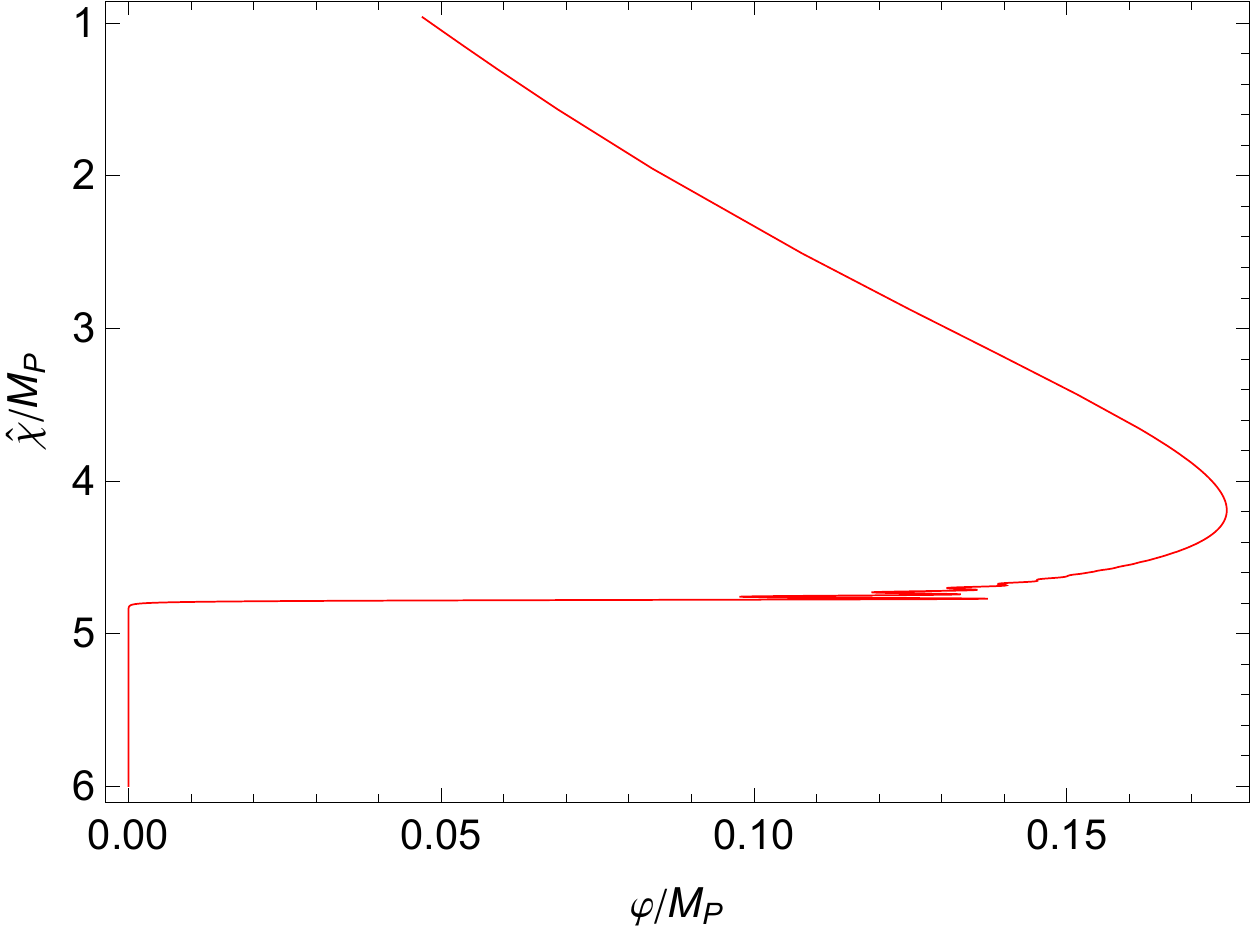}\\
		\includegraphics[width=0.45\linewidth]{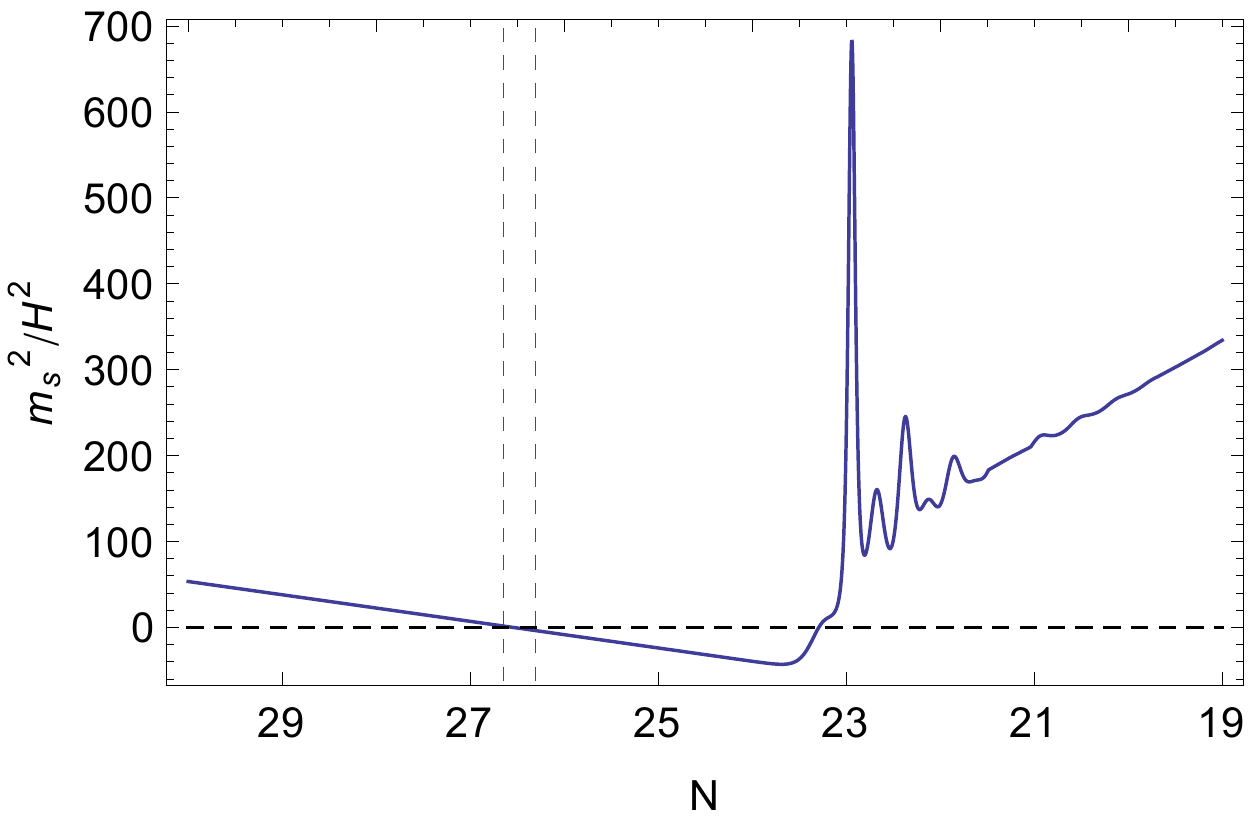}&
		\includegraphics[width=0.45\linewidth]{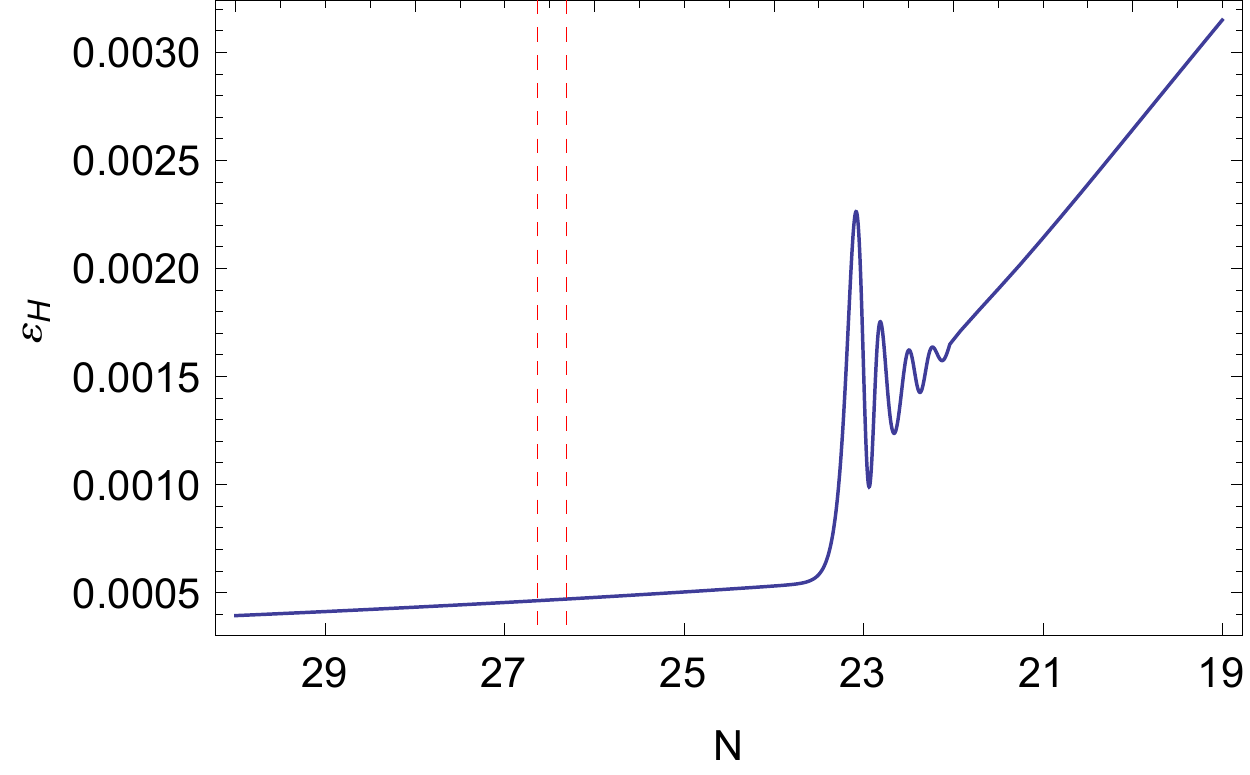}
	\end{tabular}
	\caption{Top left: the two-field potential \eqref{TwoFieldPotential} in the vicinity of the critical point \eqref{ChiCrit} superimposed by the background trajectory (red) and the $\varphi_{\mathrm{v}}^{+}$ solution (blue). Top right: the parametric plot showing $\varphi(\hat{\chi})$. Bottom left: the effective isocurvature mass as a function of $N$. Bottom right: the slow roll parameter $\varepsilon_{\mathrm{H}}$ as a function of $N$. The vertical red dashed lines in the bottom row mark the beginning and end of \mbox{Stage 2} which lasts for less than one efold. The actual fall into the valley leading to the spike in the slow roll parameter and the isocurvature mass occurs well inside \mbox{Stage 3}.}
	\label{Scenario1Fig1}
\end{figure}

\begin{figure}[!ht]
	\centering
	\begin{tabular}{cc}
		\includegraphics[width=0.45\linewidth]{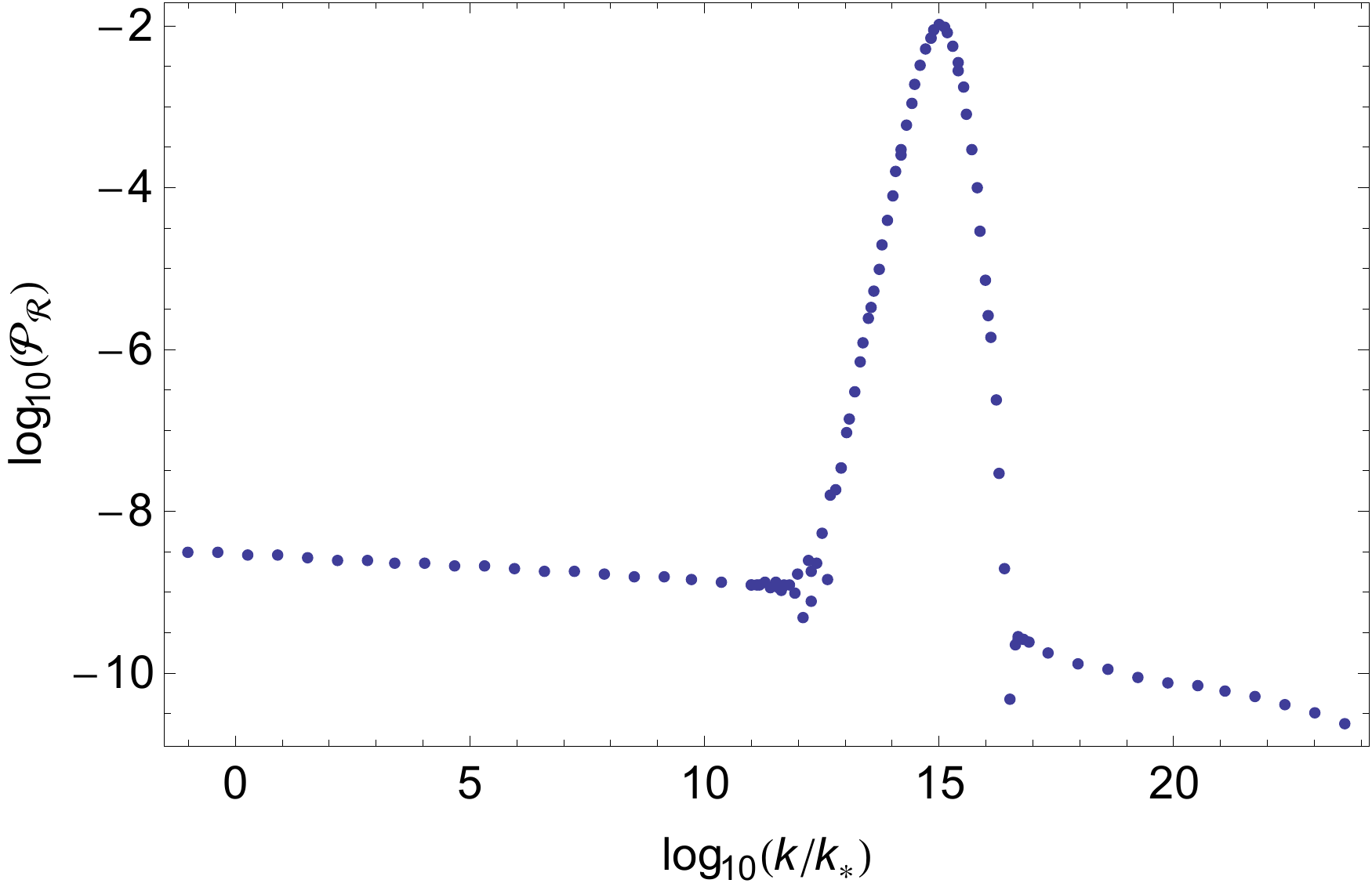}&
		\includegraphics[width=0.48\linewidth]{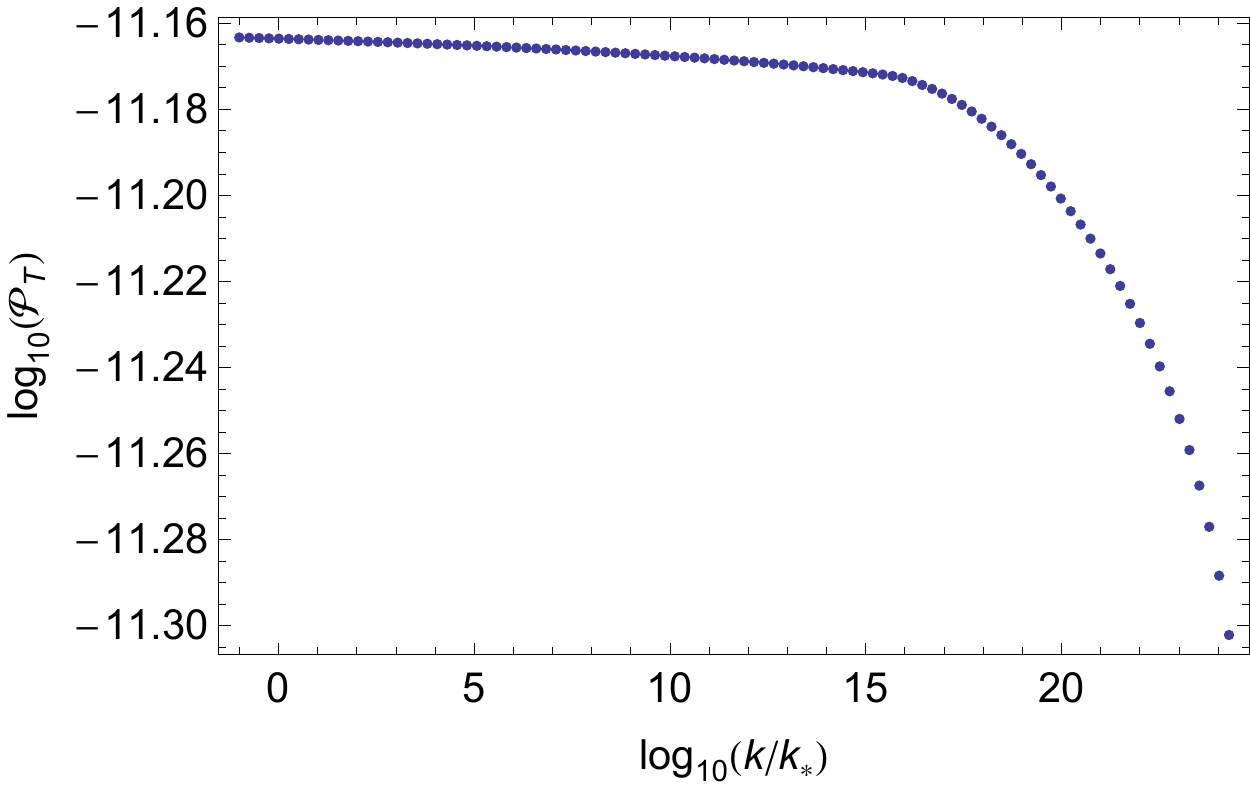}
	\end{tabular}
	\caption{The log-log plots of the power spectra $\mathcal{P}_{\mathcal{R}}$ (left) and $\mathcal{P}_h$ (right) evaluated at the end of inflation $N=0$ as a function of the wave number $k$. }
	\label{Scenario1Fig2}
\end{figure}

The numerical results in Fig.~\ref{Scenario1Fig1} and Fig.~\ref{Scenario1Fig2} are based on the parameter combination ${m_0/M_{\mathrm{P}}=1.18\times 10^{-5}}$, ${\zeta=2.6\times 10^{-10}}$, ${\xi=50}$, and ${\lambda=10^{-5}}$.

\noindent The inlay in the top left plot of Fig. \ref{Scenario1Fig1} shows $\varphi$ as a function of $N$ for both trajectories and illustrates that the inflationary trajectory (red) does not immediately turn/fall into the $\varphi_{\mathrm{v}}^{+}$ valley (blue) which re-emerges at $\hat{\chi}_{\mathrm{c}}$. Due to its inertia, it stays on the unstable hill along $\varphi_0$ for a short period. The exact moment, at which the trajectory (red) turns/falls before it catches up with the valley (blue), depends on the diffusive quantum kicks it experiences and requires the stochastic formalism discussed in Sect.~\ref{Stochastic}.

The top right plot in Fig.~\ref{Scenario1Fig1} shows $\varphi$ as a function of $\hat{\chi}$. It illustrates how the inflationary trajectory runs along $\varphi_0$, performs a sharp turn/fall and, after several mild oscillations, tracks the valley solution $\varphi_{\mathrm{v}}^{+}$.

The bottom left plot in Fig.~\ref{Scenario1Fig1} shows the ratio of the effective isocurvature mass \eqref{IsoMassDef} and the squared Hubble parameter as a function of $N$. It becomes tachyonic when it first crosses zero at $\hat{\chi}_\mathrm{c}$, corresponding to the moment at which the curvature of the potential in $\varphi$ direction $m_{\mathrm{s}}^2\propto\hat{W}_{,\varphi\varphi}$ changes sign. 
It turns positive again as the trajectory is pushed away from $\varphi_0$ and approaches the  $\varphi^{+}_{\mathrm{v}}$ valley where $m^2_{\mathrm{s}}$ turns positive again. The oscillations in $m^2_{\mathrm{s}}$ disappear after the trajectory settles down completely in the $\varphi^{+}_{\mathrm{v}}$ valley.

The bottom right plot in Fig.~\ref{Scenario1Fig1} shows the slow-roll parameter $\varepsilon_{\mathrm{H}}$ as a function of $N$. It remains well within the slow-roll regime $\varepsilon_{\mathrm{H}}\ll1$ during all three stages of the inflationary background evolution and at no point enters a regime of ultra slow-roll (in particular not during the turn/fall). This illustrates again that the multi-field amplification mechanism described in Sect.~\ref{PeakFormation} is essentially different from that based on an intermediate phase of ultra slow-roll in single-field models of inflation. 
After mild oscillations, $\varepsilon_{H}$ settles down to a higher value in $\varphi^{+}_{\mathrm{v}}$ compared to its value along $\varphi_0$. This implies that the inflationary dynamics in \mbox{Stage 3} is slightly faster than that in \mbox{Stage 1} due to the steeper slope of $\hat{W}(\hat{\chi},\varphi^{+}_{\mathrm{v}}(\hat{\chi}))$ in $\hat{\chi}$ direction compared to $\hat{W}(\hat{\chi},\varphi_0)$.

The formation of a peak in $\mathcal{P}_{\mathcal{R}}$ and the absence of any feature in $\mathcal{P}_h$ is shown in Fig.~\ref{Scenario1Fig2}.

\noindent The left plot of Fig.~\ref{Scenario1Fig2} shows the weak logarithmic $k$ dependence of the power spectrum $\mathcal{P}_{\mathcal{R}}$ for large wavelengths (small $k$) during the first slow-roll phase along $\varphi_0$ in \mbox{Stage 1} with the amplitude $\mathcal{P}_{R}\approx10^{-9}$ required for the consistency with CMB measurements. At smaller wavelengths (larger $k$), $\mathcal{P}_{\mathcal{R}}$ experiences a strong amplification leading to a peak centered around $k_{ \mathrm p}/k_{*}\approx 10^{15}$ with amplitude $\mathcal{P}_{\mathcal{R}}\approx 10^{-2}$. This peak corresponds to the modes which cross the horizon during the turn/fall of the inflationary trajectory. For modes that cross the horizon during the slow-roll phase in \mbox{Stage 3}, the amplitude of $\mathcal{P}_{\mathrm{R}}\approx10^{-10}$ is slightly smaller than that for the modes that cross the horizon during \mbox{Stage 1}.

The amplification in $\mathcal{P}_{\mathcal{R}}$ is entirely due to the sourcing of scalar perturbations. This can be seen from the right plot of Fig.~\ref{Scenario1Fig2}, which shows that the tensor power spectrum $\mathcal{P}_h$ remains constant for lower $k$ in \mbox{Stage 1}, and drops as $H$ decreases when the inflationary trajectory settles in the $\varphi_{\mathrm{v}}^{+}$ valley. No amplification or any other feature is visible, which is again explained by the fact that the isocurvature amplification mechanism only affects the scalar power spectrum $\mathcal{P}_{\mathcal{R}}$, cf.~\eqref{EQPertQsig}.
%
%
\subsection{\mbox{Scenario II}}

While the Starobinsky phase along $\varphi_0$ during \mbox{Stage 1} in \mbox{Scenario I} is similar to that in \mbox{Scenario II}, the transition to \mbox{Stage 3} is more violent in \mbox{Scenario II}, as the inflationary background trajectory falls much deeper from the central $\varphi_0$ valley into one of the $\varphi_{\mathrm{v}}^{\pm}$ valleys.    
All plots in this Section are obtained for ${m_0/M_{\mathrm{P}}=1.18\times 10^{-5}}$, ${\zeta=6\times 10^{-10}}$, ${\lambda=10^{-5}}$ and ${\xi=200}$.
\begin{figure}[!ht]
	\centering
	\begin{tabular}{cc}
		\includegraphics[width=0.45\linewidth]{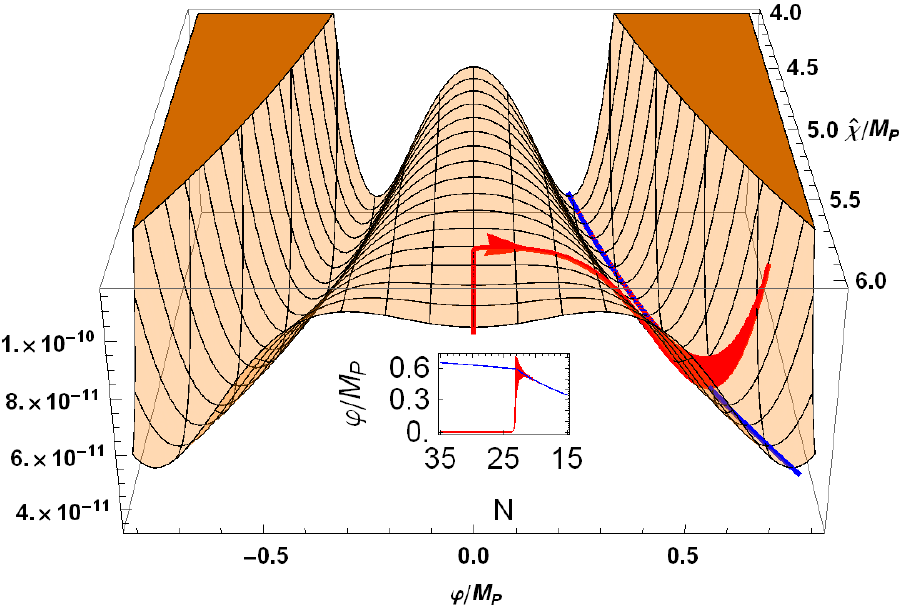}&
		\includegraphics[width=0.4\linewidth]{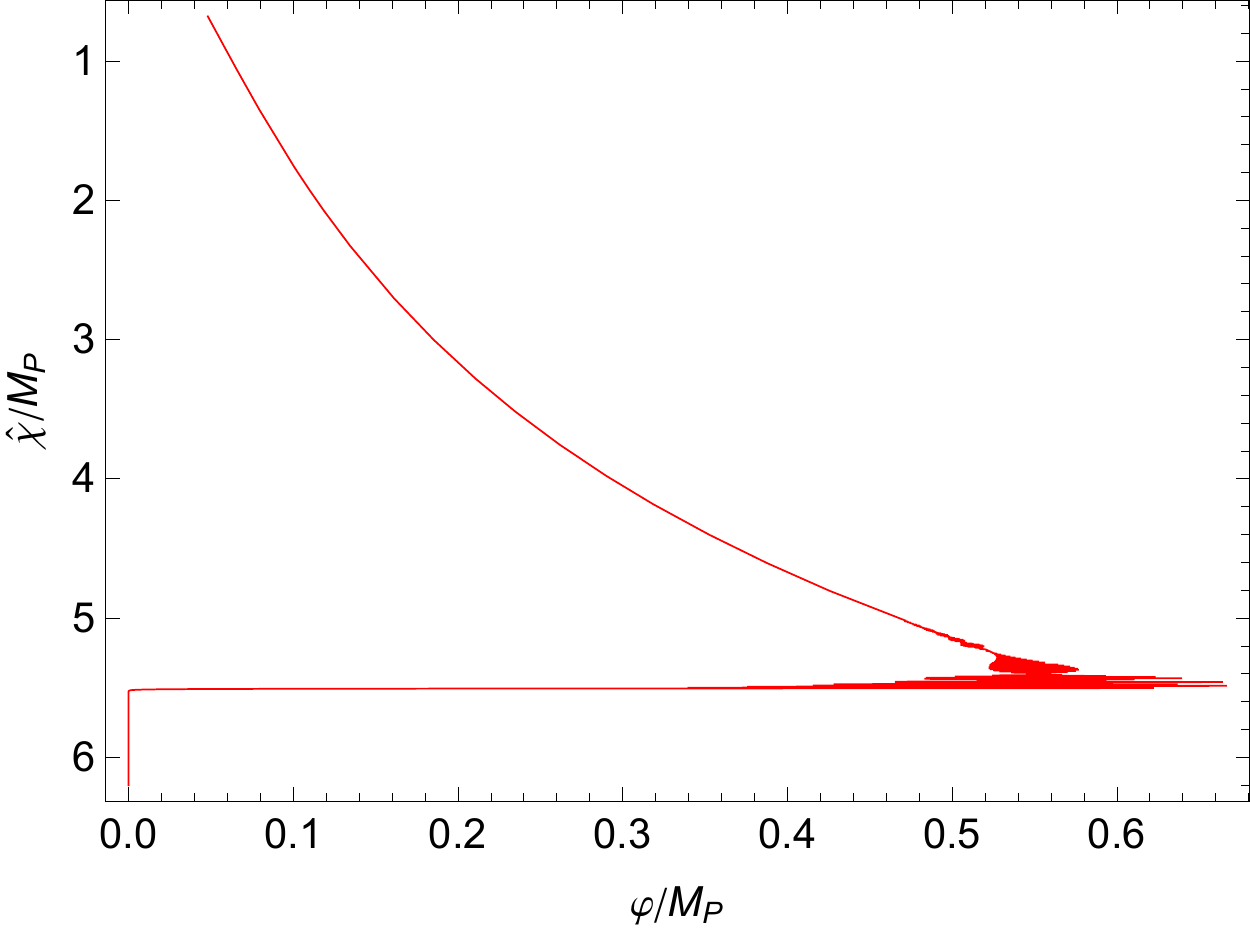}\\
		\includegraphics[width=0.45\linewidth]{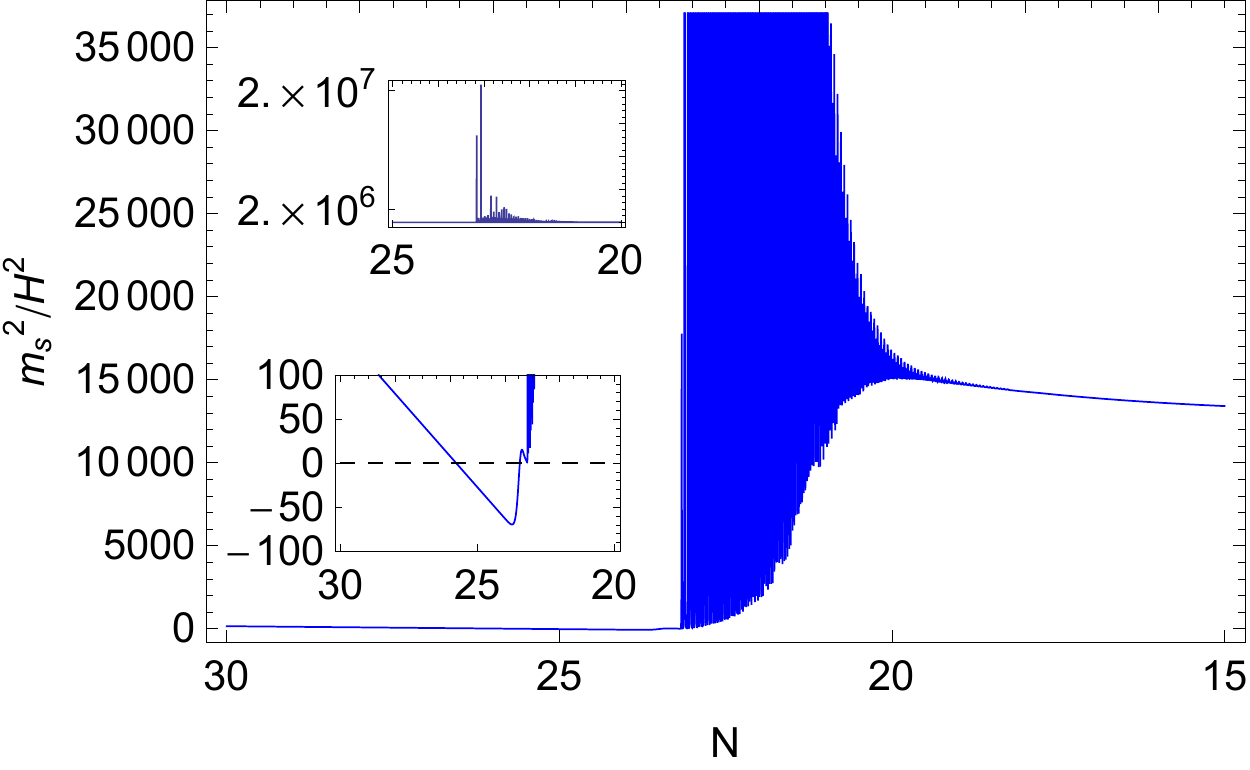}&
		\includegraphics[width=0.45\linewidth]{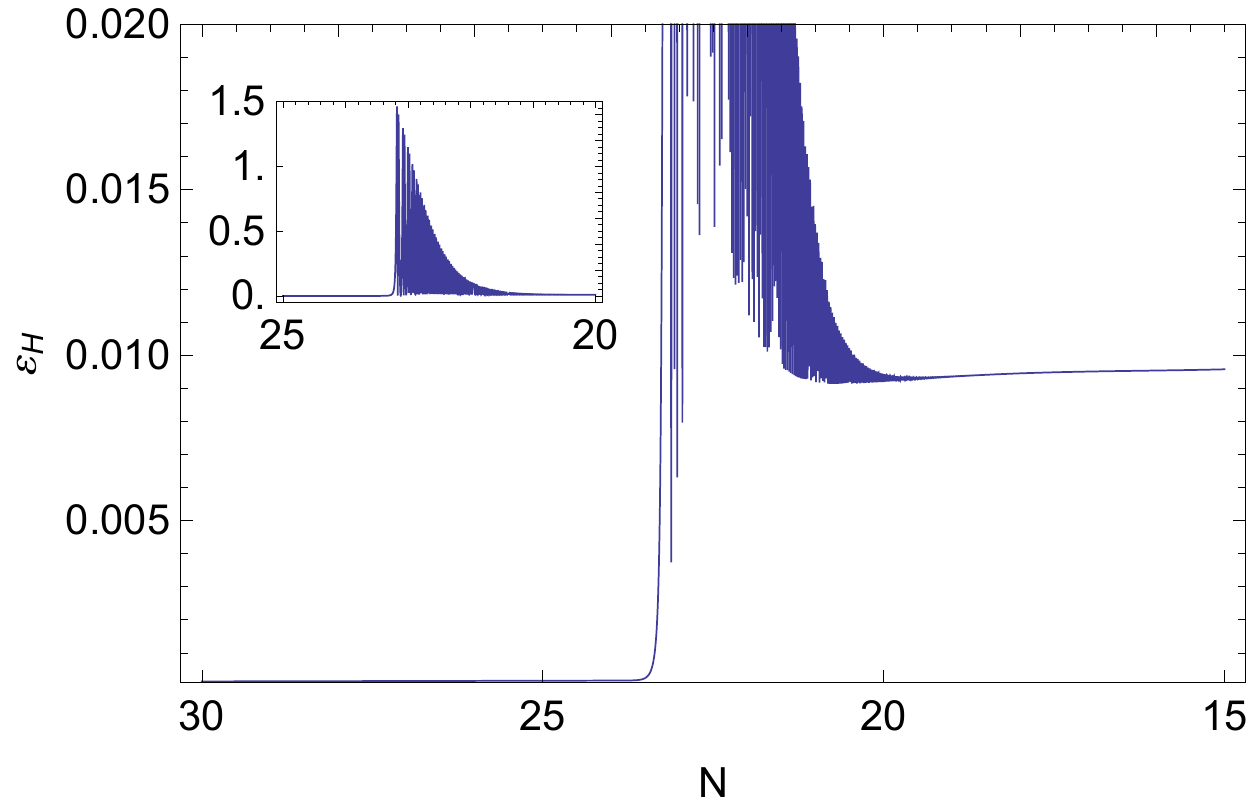}
	\end{tabular}
	\caption{Top left: The two-field potential \eqref{TwoFieldPotential} in the vicinity of the critical point \eqref{ChiCrit} superimposed by the background trajectory (red) and the $\varphi_{\mathrm{v}}^{+}$ valley solution (blue). Top right: Parametric plot $\varphi(\hat{\chi})$. Bottom left: The effective isocurvature mass as a function of $N$. Bottom right: the slow roll parameter $\varepsilon_{\mathrm{H}}$ as a function of $N$. As in Fig.~\ref{Scenario1Fig1} the spikes in the slow-roll parameter and the isocurvature mass occur well inside \mbox{Stage 3} (due to the inlay plots we refrain from displaying the vertical dashed lines indicating the beginning and end of \mbox{Stage 2}).}
\label{Scenario2Fig1}
\end{figure}

\noindent The top left plot of Fig.~\ref{Scenario2Fig1} shows the two-field potential superimposed by the numerically calculated exact inflationary trajectory (red) and the analytic valley equation $\varphi_{\mathrm{v}}^{+}$ (blue). The inlay plot shows that the trajectory sharply bends and falls into the $\varphi_{\mathrm{v}}^{+}$ valley. 

The top right plot of Fig.~\ref{Scenario2Fig1} shows the parametric dependence of $\varphi$ on $\hat{\chi}$ and the oscillations in $\varphi$ direction before settling in the $\varphi_{\mathrm{v}}^{+}$ valley.

The effective isocurvature mass $m_{\mathrm{s}}^2/H^2$ shown in the bottom left plot of Fig.~\ref{Scenario2Fig1} remains positive during the slow-roll phase in \mbox{Stage 1}, crosses zero at the critical point $\hat{\chi}_{\mathrm{c}}$ (shown by the lower inlay plot), grows negative while the trajectory is still on the unstable hill $\varphi_0$, crosses zero again, peaks during the steep fall into the  $\varphi_{\mathrm{v}}^{+}$ valley, undergoes heavy subsequent oscillations until the trajectory settles in the $\varphi_{\mathrm{v}}^{+}$ valley, and acquires a large positive value in the second slow-roll phase along $\varphi_{\mathrm{v}}^{+}$ well inside \mbox{Stage 3}. The upper inlay plot shows the peak amplitude of the oscillations in the effective isocurvature mass.

The bottom right plot of Fig.~\ref{Scenario2Fig1} shows the slow-roll parameter $\varepsilon_{\mathrm{H}}(N)$, that remains almost constant during \mbox{Stage 1}, grows rapidly during the fall and performs subsequent oscillations until the trajectory has settled in the $\varphi_{\mathrm{v}}^{+}$ valley. During \mbox{Stage 3}, it remains constant but is larger than in \mbox{Stage 1} due to the steeper slope of $\hat{W}(\hat{\chi},\varphi^{+}_{\mathrm{v}}(\hat{\chi}))$. While overall $\varepsilon_{\mathrm{H}}(N)$ is well within the slow-roll regime, in contrast to \mbox{Scenario I}, the slow-roll condition is temporarily violated during \mbox{Stage 3} when the trajectory falls into the valley, as shown by the inlay plot. This again emphasizes the difference between the isocurvature pumping mechanism and the ultra slow-roll mechanism for peak formation.
\begin{figure}[!ht]
	\centering
	\begin{tabular}{cc}
		\includegraphics[width=0.45\linewidth]{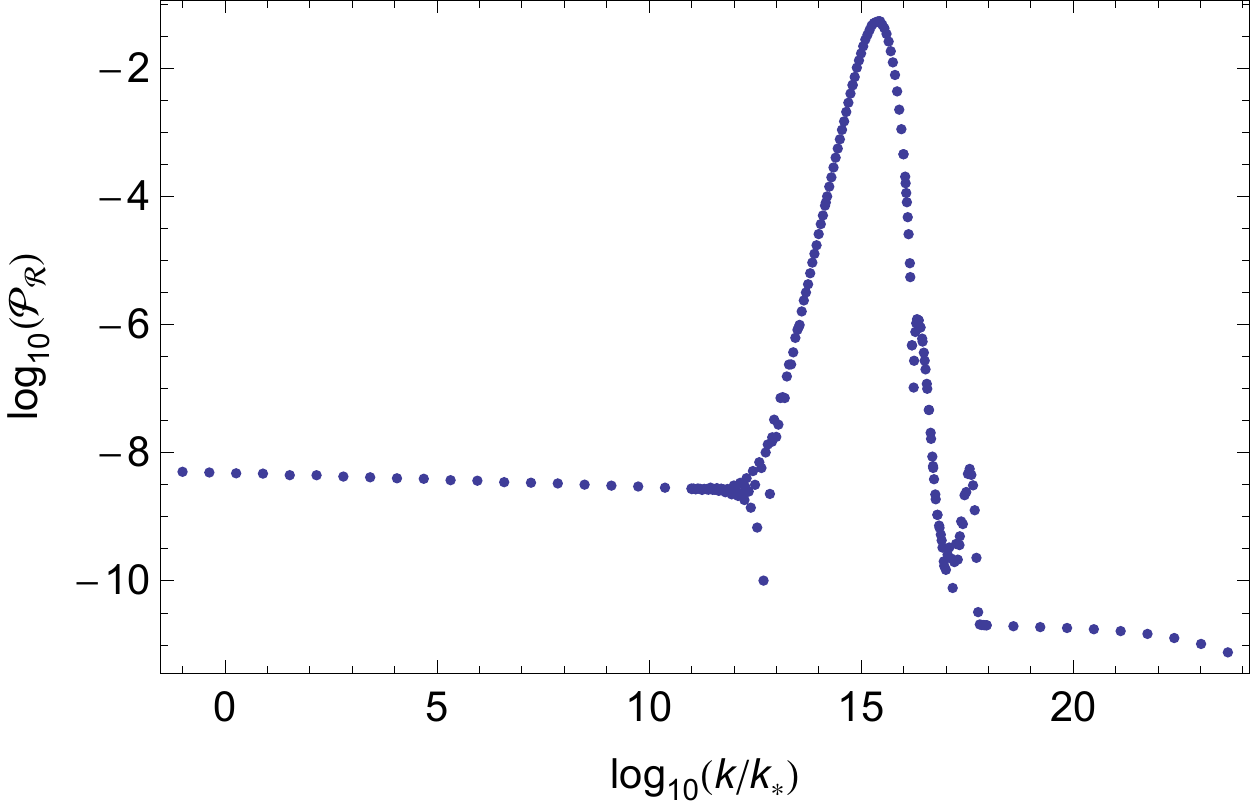}&
		\includegraphics[width=0.46\linewidth]{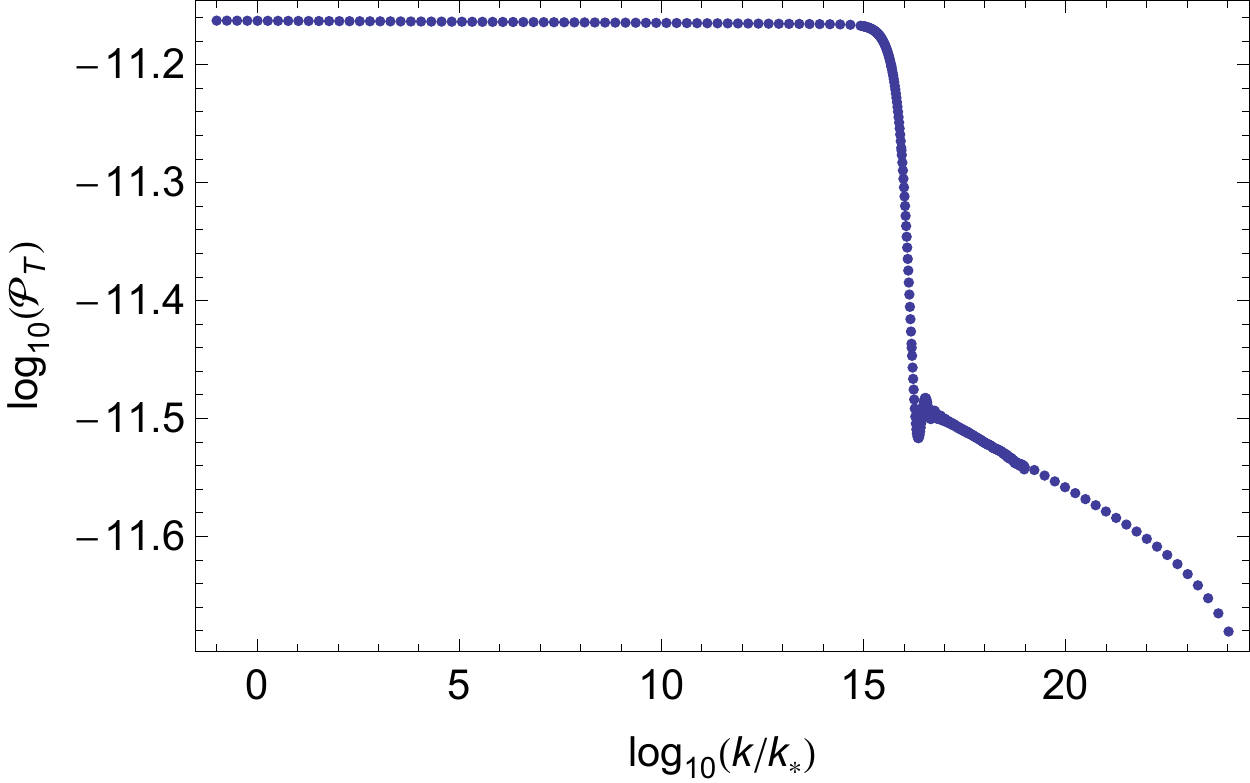}
	\end{tabular}
	\caption{The log-log plots of the adiabatic power spectrum $\mathcal{P}_{\mathcal{R}}$ (left) and the tensorial power spectrum $\mathcal{P}_h$ as a function of the wave number $k$. }
	\label{Scenario2Fig2}
\end{figure}

\noindent
The log-log plot of $\mathcal{P}_{\mathcal{R}}$ in Fig.~\ref{Scenario2Fig2} shows the weak logarithmic $k$ dependence for large wavelengths in \mbox{Stage 1} with amplitude $\mathcal{P}_{R}\approx10^{-9}$ consistent with CMB measurements. The strong amplification leads to a peak centered around $k_{ \mathrm p}/k_{*}\approx 10^{15}$ with amplitude $\mathcal{P}_{R}\approx10^{-2}$. 
Well inside \mbox{Stage 3}, the trajectory settles into the $\varphi_{\mathrm{v}}^{+}$ valley in which the value of the potential is considerably smaller than during \mbox{Stage 1} and the overall power $\mathcal{P}_{\mathcal{R}}\approx10^{-11}$ along $\varphi_{\mathrm{v}}^{+}$ is about two orders of magnitude lower compared to that in \mbox{Stage 1}.

The log-log plot of $\mathcal{P}_{h}$ in Fig.~\ref{Scenario2Fig2} shows that $\mathcal{P}_h$ remains constant for large  wavelengths during \mbox{Stage 1} and sharply drops when the inflationary trajectory falls into the $\varphi_{\mathrm{v}}^{+}$ valley. It shortly oscillates while the trajectory settles in the $\varphi_{\mathrm{v}}^{+}$ valley and continuously decreases as the trajectory approaches the end of inflation along  $\varphi_{\mathrm{v}}^{+}$.
%
%
\section{PBH dark matter from inflation}
\label{PBHDM}

The formation of PBHs in the early Universe was proposed more than 50 years ago \cite{Zeldovich1966,Carr:1974nx}. Since PBHs do not form by the gravitational collapse of a star or the merger of two neutron stars, PBH masses have a much wider mass spectrum than permitted by the Chandrasekhar mass bound.

Black holes emit Hawking radiation with the temperature ${T(M)\propto1/M}$, and have a lifetime ${\tau(M)\propto M^3}$ \cite{Hawking:1974sw}. Therefore, lighter black holes decay earlier. Since lighter black holes with Schwarzschild radius $r_{\mathrm{S}}(M)=2G_{\mathrm{N}}M$ are smaller, they have a higher density ${\rho=M/V\propto1/M^2}$, where we have used the volume $V(M)=4\pi r_{\mathrm{S}}^3(M)/3$ of a Schwarzschild black hole with mass $M$. 
Since high-density black holes could be formed naturally in the early stages of the radiation dominated Universe, they are referred to as PBHs.~\footnote{Even though PBHs could have also formed during inflation, their number density would have been strongly diluted due to the rapid expansion of the Universe.}

Despite the wide range of theoretically permitted PBH masses, there are strong observational constraints on the PBH mass spectrum which is probed by many different sources on a huge range of scales. The absence of any strong gamma ray burst that would have been emitted by the black hole evaporation \cite{Page:1976wx}, provides a direct lower bound on the allowed PBH mass. Tighter constraints come from Hawking radiation of decaying PBHs that would have heated the Universe and thereby delayed the formation of chemical elements during the Big Bang Nucleosynthesis (BBN). In addition, decaying PBHs might leave relics of Planck mass.\footnote{A consistent description of the final stage of black hole evaporation may require a more fundamental quantum theory of gravity, see e.g.~\cite{Steinwachs:2020jkj} for a review about various covariant approaches to quantum gravity.} Since the total energy density of such hypothetical relics cannot exceed the present CDM density, this also leads to constraints on the PBH mass spectrum. Further constraints unrelated to PBH evaporation come from gravitational lensing experiments, pulsar timing experiments, as well as CMB and LSS measurements. A recent overview of observational constraints can be found in \cite{Carr:2020gox}.
%
%
\subsection{PBH formation}

There are various formation mechanisms for PBHs, such as a sudden pressure reduction during phase transitions \cite{Kodama:1982sf}, cosmic strings \cite{Polnarev:1988dh} and bubble collisions \cite{Barrow:1992hq}. The most relevant formation mechanism, however, is provided by the collapse of large density perturbations. The probability of such perturbations with a large amplitude can be enhanced by peaks in the inflationary power spectrum of curvature perturbation \cite{Silk:1986vc,Ivanov:1994pa, Green:1997sz,Bringmann:2001yp}. Since CMB measurements only cover large wavelengths \eqref{CMBModes}, PBHs offer a unique opportunity to probe inflationary models on smaller wavelengths \cite{Aghanim:2018eyx}. 
A PBH forms whenever the overdensity 
\begin{align}
\delta(t,\mathbf{x}):=\frac{\rho(t,\mathbf{x})-\bar{\rho}(t)}{\bar{\rho}(t)}
\end{align}
exceeds a critical value $\delta_{\mathrm{c}}$ in a given (spherical) Hubble volume $V_{\mathrm{H}}(t):=4\pi r_{\mathrm{H}}^3(t)/3$. The Hubble radius is defined as $r_{\mathrm{H}}(t):=1/H(t)$ and the background density in a flat FLRW universe is given by $\bar{\rho}(t)=3H^2(t)/(8\pi G_{\mathrm{N}})$. 
In the simplest picture, a PBH with $r_{\mathrm{S}}=r_{\mathrm{H}}$ immediately forms once a perturbation $\delta>\delta_{\mathrm{c}}$ with wavelength $\lambda_{\delta}$ enters the Hubble horizon $\lambda_{\delta}=r_{\mathrm{H}}$.\footnote{A more detailed study of the formation time and its impact on the PBH abundance can be found in \cite{Musco:2008hv,Musco:2012au,Germani:2018jgr,DeLuca:2019qsy,Young:2019osy}.} In addition, the PBH mass $M_{\mathrm{PBH}}$ is assumed to be directly proportional to the horizon mass $M_{\mathrm{H}}$ at the time of formation $t_{\mathrm{f}}$,
\begin{align}
M_{\mathrm{PBH}}(t_{\mathrm{f}})={}&KM_{\mathrm{H}}(t_{\mathrm{f}}),\\
M_{\mathrm{H}}(t_{\mathrm{f}})={}&V_{\mathrm{H}}(t_{\mathrm{f}})\bar{\rho}(t_{\mathrm{f}})=\frac{r_{\mathrm{H}}(t_{\mathrm{f}})}{2G_{\mathrm{N}}}.\label{HM}
\end{align}
The critical density $\delta_{\mathrm{c}}$ is determined by the Jeans-length criterion applied in an expanding FLRW universe. The critical density $\delta_{\mathrm{c}}$, as well as the proportionality factor $K$, depend only on the FLRW background dynamics. For an equation of state ${p=\omega\rho}$, with constant $\omega$ determined by the dominant energy density fraction at the time of formation, $\delta_{\mathrm{c}}\approx\omega$ and $K\approx \omega^{3/2}$  \cite{Carr:1974nx}. In the radiation dominated phase $\omega=1/3$ this leads to $\delta_{\mathrm{c}}\approx0.33$ and $K\approx0.19$.

In this simplified picture, the PBH mass only depends on the horizon mass and, hence, on the time of formation but not on the amplitude $\delta$. However, numerical collapse simulations show that the PBH mass satisfies a more complicated critical scaling relation and depends both on the time of formation $t_{\mathrm{f}}$ as well as on the amplitude of the overdensity $\delta$ \cite{Niemeyer:1997mt,Niemeyer:1999ak},
\begin{align}\label{CritMPBH}
M_{\mathrm{PBH}}(\delta,t_{\mathrm{f}})=K\,M_{\mathrm{H}}(t_{\mathrm{f}})(\delta-\delta_{\mathrm{c}})^{\gamma}.
\end{align}
The parameters $K$, $\delta_{c}$ and $\gamma$ in \eqref{CritMPBH} are determined numerically \cite{Jedamzik:1999am,Shibata:1999zs,Musco:2004ak,Musco:2018rwt,Escriva:2019phb}. An analytic approach to determine the threshold $\delta_{\mathrm{c}}$ was proposed recently in \cite{Musco:2020jjb}.

All our numerical results are obtained by taking full account of the critical scaling relation \eqref{CritMPBH}.  Following the arguments of \cite{Gow:2020bzo} we fix the parameters at $K=10$, $\delta_{c}=0.25$ and $\gamma=0.36$ entering \eqref{CritMPBH} consistent with the choice of the window function as described in the subsequent section.

%
%
\subsection{PBH abundance}\label{PBHAb}

In order to calculate the PBH abundance, it is useful to define the fraction of the mass in the universe, which collapsed into PBHs at the time of formation,
\begin{align}
\beta(t_{\mathrm{f}}):=\frac{\rho_{\mathrm{PBH}}(t_{\mathrm{f}})}{\bar{\rho}(t_{\mathrm{f}})}.
\end{align}  
In the Press-Schechter formalism\footnote{For alternative methods based on peaks theory, see e.g.~\cite{Bardeen:1985tr,Green:2004wb,Young:2014ana,Young:2020xmk}.} $\beta$ is calculated as \cite{Press:1973iz}
\begin{align}\label{BetaMasterEq}
\beta(t_{\mathrm{f}})=2\int_{\delta_c}^{\infty}\mathrm{d}\delta\,\frac{M_{\mathrm{PBH}}(\delta,t_{\mathrm{f}})}{M_{\mathrm{H}}(t_{\mathrm{f}})}P(\delta,t_{\mathrm{f}}).
\end{align} 
Here $P(\delta, t_{\mathrm{f}})$ is the PDF of generating an overdensity with amplitude $\delta$ at the moment of formation $t_{\mathrm{f}}$. Assuming that the perturbations $\delta$ are independent random variables, they follow Gaussian statistics.\footnote{For a discussion of non-Gaussian effects, see e.g.~\cite{Byrnes:2012yx,Young:2013oia,Franciolini:2018vbk,Atal:2018neu,DeLuca:2019qsy,Young:2019yug,Panagopoulos:2019ail,Yoo:2019pma,Atal:2019erb,Ezquiaga:2019ftu,Germani:2019zez}.}  The lower integration bound in \eqref{BetaMasterEq} is determined by the critical collapse density $\delta_{\mathrm{c}}$.\footnote{Even though the upper integration bound may be constrained by demanding $\delta$ should still be within the linear regime $\delta\leq1$, it can safely be taken to infinity in view of the exponential suppression factor arising from the Gaussian PDF \eqref{ProbDist} for $\delta\gg1$.} 
The probability density of having an overdensity with amplitude $\delta$ is given by
\begin{align}\label{ProbDist}
P(\delta, t_{\mathrm{f}})=\frac{1}{\sqrt{2\pi\sigma^2(t_{\mathrm{f}})}}\exp{\left(-\frac{1}{2}\frac{\delta^2}{\sigma^2(t_{\mathrm{f}})}\right)}.
\end{align} 
Hence those perturbations forming PBHs are very rare and lie in the tail of the Gaussian PDF \eqref{ProbDist}.
Calculating $\beta$ from \eqref{BetaMasterEq} requires calculating the variance ${\sigma^2(t_{\mathrm{f}})=\langle\delta^2(t_{\mathrm{f}})\rangle}$ in \eqref{ProbDist}. The Fourier transform of the density contrast $\delta(t,\mathbf{x})$ is given by
\begin{align}
\delta(t,\mathbf{x})={}&\int\frac{\mathrm{d}^3\mathbf{k}}{(2\pi)^{3/2}}\,e^{i\mathbf{k}\mathbf{x}}\delta_{\mathbf{k}}(t).
\end{align}
In a homogeneous and isotropic FLRW universe the variance $\sigma^2(t)$ is completely determined by the power spectrum $\mathcal{P}_{\delta}(t,k)$ via the two-point correlation function
\begin{align}
\sigma^2(t)={}&\int_{0}^{\infty}\mathrm{d}(\ln k)\,\mathcal{P}_{\delta}(t,k),\label{var2}\\
\langle\delta_{\mathbf{k}}(t)\delta^{*}_{\mathbf{k}'}(t)\rangle={}&\frac{2\pi^2}{k^3}\mathcal{P}_{\delta}(t,k)\delta^3(\mathbf{k}-\mathbf{k}').
\end{align}
Assuming that the overdensities $\delta_{\mathbf{k}}(t)$ arise from the comoving curvature perturbations $\mathcal{R}_{\mathbf{k}}(t)$ amplified during inflation, we need to relate $\mathcal{P}_{\delta}(t,k)$ with the inflationary power spectrum $\mathcal{P}_{\mathcal{R}}(t,k)$ that is defined by the two-point correlation function
\begin{align}
\langle\mathcal{R}_{\mathbf{k}}(t),\mathcal{R}^{*}_{\mathbf{k}'}(t)\rangle={}&\frac{2\pi^2}{k^3}\mathcal{P}_{\mathcal{R}}(t,k)\delta^3(\mathbf{k}-\mathbf{k}').
\end{align}
The linear relation between $\delta_{\mathrm{k}}$ in the radiation dominated era and $\mathcal{R}_{\mathbf{k}}$ is given by \footnote{For a discussion taking into account the effects of the more general non-linear relation between the curvature perturbations and the density contrast see \cite{Germani:2019zez,Young:2019yug}.} 
\begin{align}
\delta_{\mathbf{k}}(t)=\frac{4}{9}\left(\frac{k}{aH}\right)^2T(t,k)\mathcal{R}_{\mathbf{k}}(t).
\end{align}
The transfer function $T(t,k)$ describes the sub-horizon dynamics ${k>aH}$ of $\delta_{\mathbf{k}}(t)$ after horizon re-entry,
while $T(t,k)=1$ for superhorizon scales ${k<aH}$. 
Thus, the variance \eqref{var2} is obtained as
\begin{align}
\sigma^2(t)=\int_{0}^{\infty}\mathrm{d}(\ln k)\,\frac{16}{81}\left(\frac{k}{a(t)H(t)}\right)^4\,T^2(t,k)\,\mathcal{P}_{\mathcal{R}}(t,k).\label{var3}
\end{align}
The integral \eqref{var3} diverges at the upper integration bound for small wavelengths $\lambda=1/k$. This is avoided by smoothing $\delta(t,\mathbf{x})$ with a unit normalized window function $W(\mathbf{x}-\mathbf{y},R)$ at a smoothing scale $R$,
\begin{align}
\delta_R(t,\mathbf{x})=\int\mathrm{d}^3y\, W(\mathbf{x}-\mathbf{y},R)\delta(t,\mathbf{y}).\label{SMD}
\end{align}
Physically, the coarse graining induced by the smoothing means that at every point $\mathbf{x}$, the smoothed overdensity $\delta_{R}(t,\mathbf{x})$ represents the average of $\delta(t,\mathbf{x})$ over a spherical region of radius $R$ centered at $\mathbf{x}$, i.e.~the substructures in the overdensity $\delta(t,\mathbf{x})$ below the resolution scale $R$ are smoothed out in $\delta_R(t,\mathbf{x})$ by the averaging procedure.
We choose a modified Gaussian window function $W_{\mathrm{G}}$ in \eqref{SMD}. Following the conventions in \cite{Young:2019osy}, the window function in Fourier space reads\footnote{Note the additional factor of $1/2$ in the argument of the exponential in \eqref{WindowG}. For a comparison of the impact of different window function see e.g.~\cite{Gow:2020bzo}.}
\begin{align}
W_{\mathrm{G}}(kR)={}&\exp\left[ -\frac{(kR)^2}{4}\right].\label{WindowG}
\end{align}
The window function \eqref{WindowG} strongly damps out contributions from modes much larger than the ``smoothing mode'' $k_{R}=1/R$. 
Since we assume that a PBH forms when the modes $\delta_{\mathbf{k}}(t)$ re-enter the horizon at $t=t_{\mathrm{f}}$, the smoothing mode should be identified with the comoving Hubble radius at formation 
\begin{align}
k_{R}\equiv a(t_{\mathrm{f}})H(t_{\mathrm{f}}).\label{SS}
\end{align}
With the choice of the smoothing scale \eqref{SS}, the subhorizon modes $k\gg a(t_{\mathrm{f}})H(t_{\mathrm{f}})$ are strongly suppressed by the window function such that the transfer function in \eqref{var3} effectively becomes $T(t,k)=1$ for the superhorizon modes $k\ll a(t_{\mathrm{f}})H(t_{\mathrm{f}})$.
The variance \eqref{var3} at $t_{\mathrm{f}}$, smoothed at the horizon scale, acquires the form 
\begin{align}
\sigma^{2}_{R}(t_{\mathrm{f}})=\int_{0}^{\infty}\mathrm{d}(\ln k)\,\frac{16}{81}\left(\frac{k}{k_R}\right)^4\,W^2(k/k_R)\,\mathcal{P}_{\mathcal{R}}(t_{\mathrm{f}},k).\label{var4}
\end{align}
In order to have a sizeable mass fraction \eqref{BetaMasterEq}, the smoothed variance \eqref{var4} must be sufficiently large. 
Since $\sigma_{R}(t_{\mathrm{f}})$ is strongly damped by the $(k/k_R)^4$ factor for modes $k\ll k_{R}$ and by the window function $W^2(k/k_{R})$ for modes $k\gg k_{R}$, a sufficiently large $\sigma_{R}(t_{\mathrm{f}})$ can only be realized for a power spectrum which features a strong amplification at $k\approx k_R$.\footnote{Note that smoothing has a negligible effect on sharp peaks in $\mathcal{P}_{\mathcal{R}}$, see e.g.~\cite{Young:2019yug} for more details.}
Thus, in case the inflationary power spectrum
$\mathcal{P}_\mathcal{R}(t_{\mathrm{f},k})$ features a sharp peak at $k_{\mathrm{p}}$, this peak scale should be arranged to be close to $k_{R}$ by tuning the parameters
of the inflationary model.
Moreover, the horizon mass $M_{\mathrm{H}}(t_{\mathrm{f}})$ at the time of formation is related to the peak scale $k_p$ by \cite{Nakama:2016gzw},
\begin{align}
M_{\mathrm{H}}(k_{\mathrm{p}} )={}&M_{\mathrm{H}}(t_{\mathrm{eq}})\left(\frac{g(t_{\mathrm{f}})}{g(t_{\mathrm{eq}})}\right)^{1/6}\left(\frac{k_{\mathrm{p}}}{k_{\mathrm{eq}}}\right)^{-2}\nonumber\\
={}&3.3\times10^{17} M_{\odot} \left({\frac{k_{\mathrm{p}}}{0.07 h^2 \Omega_m \mathrm{Mpc}^{-1}}}\right)^{-2}.
\label{HMpeak}
\end{align}
Here $g$  is the effective number of relativistic degrees of freedom, and $t_{\mathrm{eq}}$ and $k_{\mathrm{eq}}$ are the time of matter-radiation equality and the mode which crosses the horizon at that time, respectively.
A simple way to obtain a rough analytic estimate of the relationship for the PBH mass as a function of the peak scale $k_{\mathrm{p}}$ is to assume that the  horizon mass, \eqref{HMpeak}, is directly proportional to the PBH mass leading to
\begin{align}
M_{\mathrm{PBH}}(k_{\mathrm{p}})\approx 6.3\times 10^{12}M_{\odot} \left(\frac{k_{\mathrm{p}}}{\mathrm{Mpc}^{-1}}\right)^{-2}.\label{MPBHEst}
\end{align}	
In the transition from \eqref{HMpeak} to the estimate \eqref{MPBHEst}, we have used the explicit values ${M_{\mathrm{PBH}}=0.19 M_{\mathrm{H}}}$, $g(t_{\mathrm{f}})=106.75$, $g(t_{\mathrm{eq}})= 3.36$, ${h=0.674}$ and $\Omega_m=0.315$ \cite{Aghanim:2018eyx}. Conversely, if we are interested in a particular PBH mass range, the relation \eqref{MPBHEst} gives an estimate for the peak location in $\mathcal{P}_{\mathcal{R}}(t,k)$. For example, if we demand that all CDM we observe today is made of PBHs, only very small windows for the PBH mass are observationally allowed, and determine the scale $k_{\mathrm{p}} $ at which $\mathcal{P}_{\mathcal{R}}(t_{\mathrm{f}},k)$ must be strongly amplified. We emphasize that all our numerical results are based on the critical scaling relation \eqref{CritMPBH} and not on the simplified estimate \eqref{MPBHEst}.
%
%
\subsection{PBHs as CDM}

In the case when sufficiently large number of PBHs formed in the radiation dominated era, they could make up a large fraction of the presently observed CDM content in the Universe \cite{Hawking:1971ei,Chapline:1975ojl,Blais:2002nd,Carr:2016drx,Carr:2020xqk,Green:2020jor}. Particularly interesting is the PBH mass window accessible to gravitational wave experiments performed by LIGO \cite{Abbott:2016blz,Bird:2016dcv,Garcia-Bellido:2017aan,Sasaki:2018dmp}. The PBH mass distribution $f(M_{\mathrm{PBH}})$ derived in \eqref{fpbhfinal} measures the fraction of the presently observed CDM contributed by PBHs with mass $M_{\mathrm{PBH}}$. In our numerical analyses the PBH mass distribution is obtained by numerically integrating  $\eqref{fpbhfinal}$, by taking into account \eqref{CritMPBH} with the numerical values $K=10$, $\delta_{c}=0.25$ and $\gamma=0.36$. 

In case CDM is entirely made of PBHs, the total fraction $F_{\mathrm{PBH}}$ defined in \eqref{FMPBH} equals one. In case only a fraction of CDM is made of PBHs ${F_{\mathrm{PBH}}<1}$. Thus, the trivial constraint on the PBH abundance is given by ${F_{\mathrm{PBH}}\leq1}$ because the total PBH density cannot exceed the present CDM density. Since the mass spectrum of PBHs is already considerably constrained on a broad range of scales, there are only a few PBH mass windows in which ${F_{\mathrm{PBH}}=1}$ can be realized \cite{Carr:2020gox, Carr:2020xqk}:
\begin{align}
10^{-17}M_{\odot}\lesssim M_{\mathrm{PBH}}^{I}\lesssim 10^{-16}M_{\odot}\label{MI1},\\
10^{-13}M_{\odot}\lesssim M_{\mathrm{PBH}}^{II}\lesssim 10^{-9}M_{\odot}\label{MI2}.
\end{align}
In addition to $M_{\mathrm{PBH}}^{I}$ and $M_{\mathrm{PBH}}^{II}$,  the detection of binary black hole merges at 
LIGO/Virgo has renewed interest in the possibility of a primordial origin of CDM for PBHs in the mass range
\begin{align}
10 M_{\odot}\lesssim M_{\mathrm{PBH}}^{III}\lesssim 10^2 M_{\odot}\label{MI3}.
\end{align}
Even if the possibility of explaining all the observed CDM by PBHs in the mass window $M_{\mathrm{PBH}}^{III}$ seems to be ruled out observationally, as $F_{\mathrm{PBH}} \lesssim 10^{-2}\div10^{-3}$ \cite{Carr:2020gox, Carr:2020xqk, Sasaki:2018dmp}, a peak leading to the production of PBHs in the mass range $M_{\mathrm{PBH}}^{III}$ would provide an inflationary explanation for the observed merger events. While there are interesting scenarios connected to a mixed contribution of PBHs and additional CDM particles such as WIMPs (see~\cite{Carr:2020xqk} for an overview), we primarily focus on the possibility to explain all of the observed CDM content by PBHs with $F_{\mathrm{PBH}}=1$ in the mass windows $M_{\mathrm{PBH}}^{I}$ and $M_{\mathrm{PBH}}^{II}$. Nevertheless, in Sect.~\ref{NumResults}, we demonstrate that there are the parameter combinations that permit a generation of the observationally allowed distribution $f(M_{\mathrm{PBH}})$ in all mass windows \eqref{MI1}-\eqref{MI3}, i.e.~$F_{\mathrm{PBH}}\approx 10^{-1}\div1$ in the mass windows $M^I_{\mathrm{PBH}}$ and $M^{II}_{\mathrm{PBH}}$, and $F_{\mathrm{PBH}}\approx10^{-3}\div10^{-2}$ in the mass window $M^{III}_{\mathrm{PBH}}$.

The mass intervals \eqref{MI1}-\eqref{MI3} directly translate into $k$ intervals in which the peak featured in $\mathcal{P}_{\mathcal{R}}$, centered at $k_{\mathrm{p}}$, must lie:
\begin{align}
k^{I}_{\mathrm{p}}\approx 10^{15}\mathrm{Mpc}^{-1}\label{kpI},\\
10^{13}\mathrm{Mpc}^{-1}\gtrsim k^{II}_{\mathrm{p}}\gtrsim 10^{11}\mathrm{Mpc}^{-1}\label{kpII},\\
k^{III}_{\mathrm{p}}\approx 10^{6}\mathrm{Mpc}^{-1}\label{kpIII}.
\end{align}
Conversely, once a particular PBH mass $M_{\mathrm{PBH}}$ has been chosen, the constraint $F_{\mathrm{PBH}}=1$ directly translates into a constraint on the amplitude $A_{\mathrm{p}}$ of the peak in $\mathcal{P}_{\mathcal{R}}$ at $k_{\mathrm{p}} $. 
For a given PBH mass $M_{\mathrm{PBH}}$, a rough analytical estimate of the peak amplitude $A_{\mathrm{p}}$ leading to $F_{\mathrm{PBH}}=1$ is provided in Appendix \ref{Analytic}.
%
%
\section{Numerical Results}\label{NumResults}

We present our numerical results for the PBH mass distribution \eqref{fpbhfinal} in the three mass windows \eqref{MI1}-\eqref{MI3}. We show that for all three mass windows, there are parameter combinations which lead to $F_{\mathrm{PBH}}=1$. Moreover, in all cases it is also possible to find parameters such that $F_{\mathrm{PBH}}<1$, which is particularly relevant for the observationally most interesting LIGO mass window \eqref{MI3}. Therefore, we first present our results for the LIGO mass range, show that the log-normal function \eqref{SimpPLogNormal} provides an excellent fit to the peak of the numerically generated power spectrum, and finally compare the peak amplitude $A_{\mathrm{p}}$ to the results obtained in \cite{Gow:2020bzo}. For the mass windows \eqref{MI1}-\eqref{MI3}, we illustrate that the numerical value for $F_{\mathrm{PBH}}$ is highly sensitive to the parameter $\xi$.   
%
%
\subsection{LIGO mass window $M^{III}_{\mathrm{PBH}}$}\label{subsec:LigoMassWindow}

If the observed LIGO black hole merger events are of inflationary origin, the merger rates may be used to constrain the power spectrum via the PBH mass distribution $f(M_{\mathrm{PBH}})$. Observationally, in the LIGO mass window \eqref{MI3}, the total fraction $F_{\mathrm{PBH}}$ is most likely constraint to lie between $10^{-3}\div10^{-2}$, see e.g.~\cite{Ballesteros:2018swv, Carr:2020xqk}. While recent works \cite{Chen:2018czv,Raidal:2018bbj,Gow:2019pok,DeLuca:2020qqa} suggest that $F_{\mathrm{PBH}}$ is closer to $10^{-3}$, other works indicate that  $F_{\mathrm{PBH}}$ could attain much higher values \cite{Jedamzik:2020ypm,Young:2020scc,Jedamzik:2020omx,Clesse:2020ghq}.
\begin{figure}[!ht]
	\centering
	\begin{tabular}{cc}
		\includegraphics[width=0.46\linewidth]{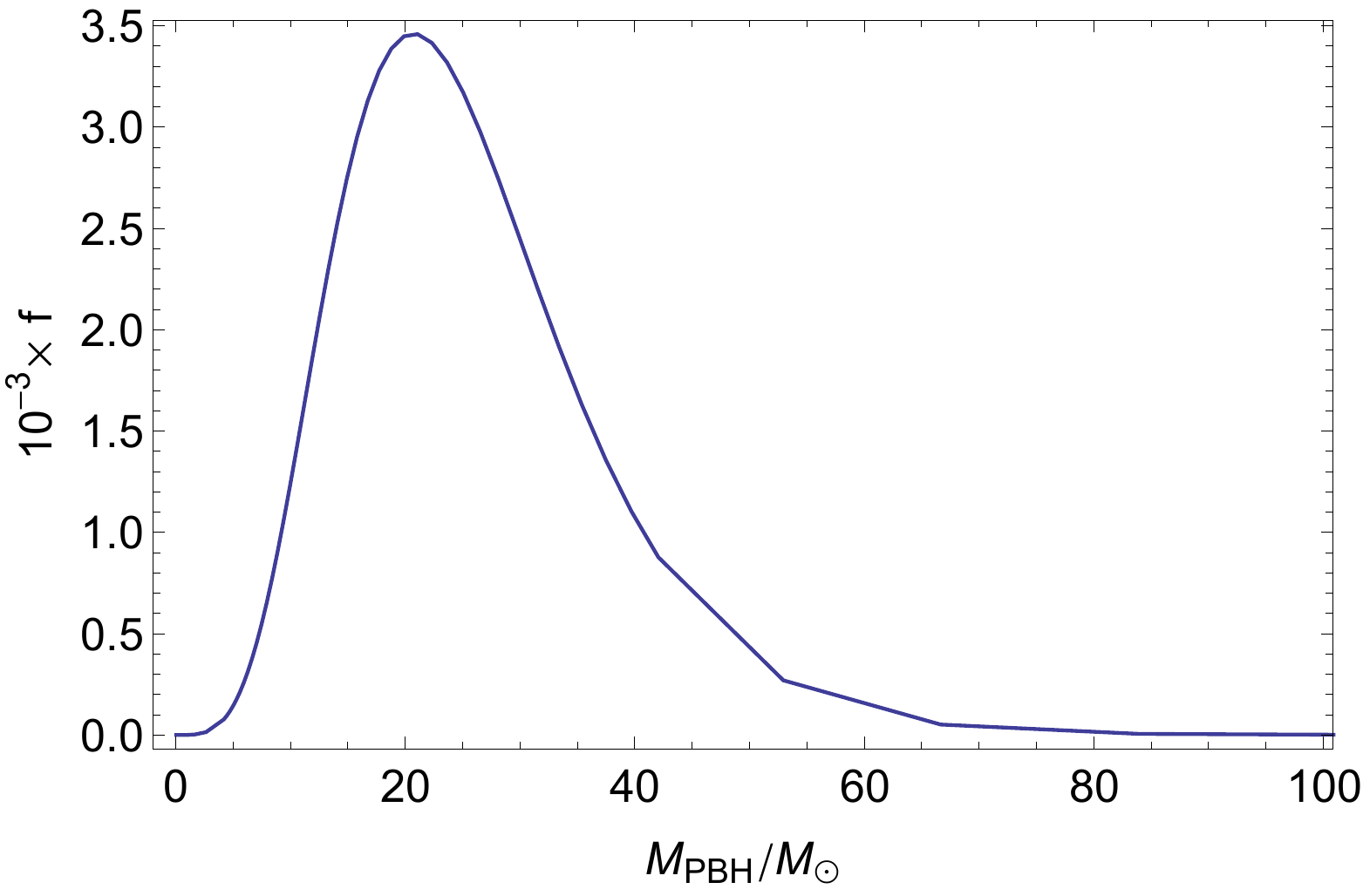}&
		\includegraphics[width=0.45\linewidth]{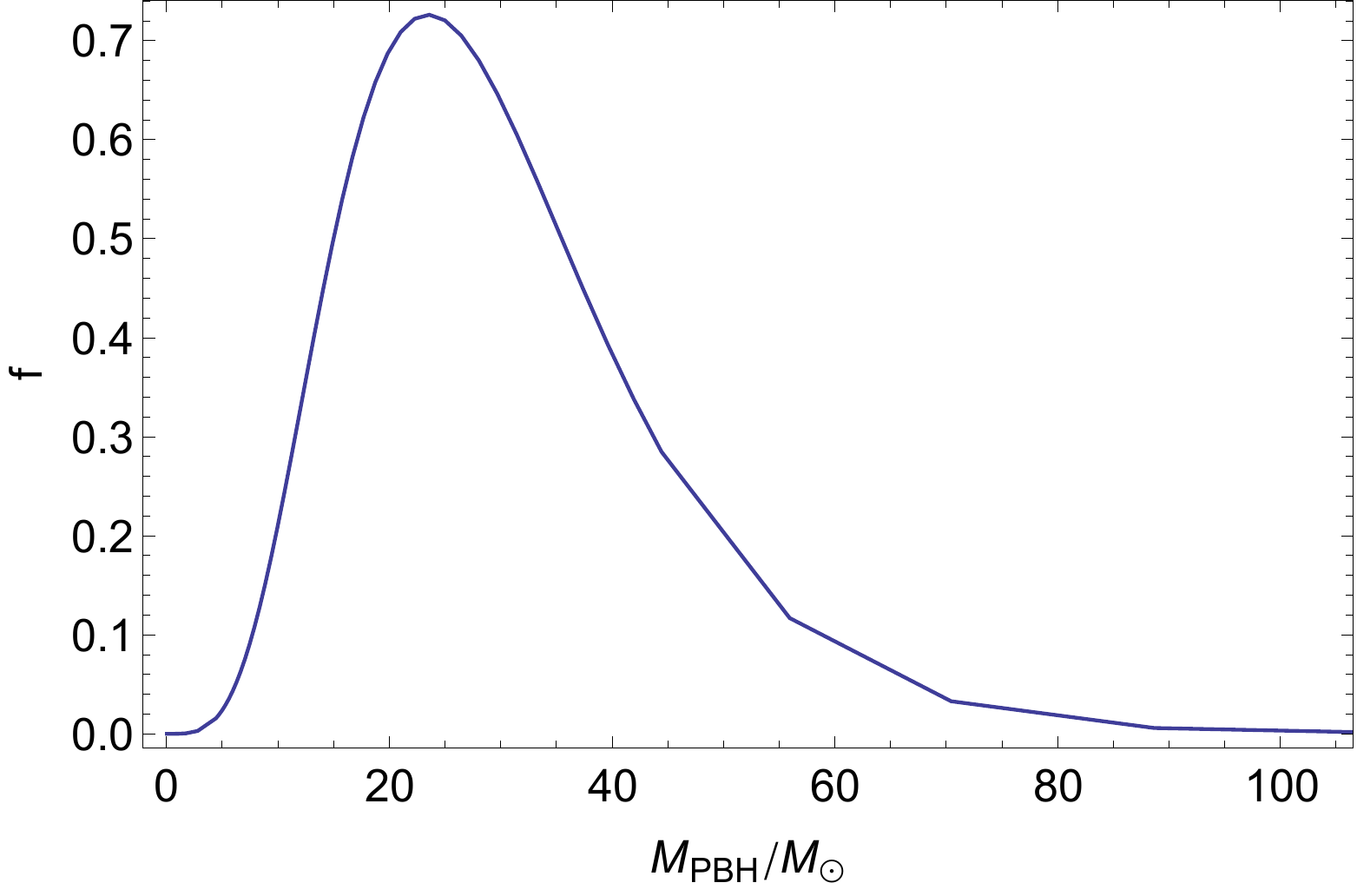}
	\end{tabular}
	\caption{Left: $f(M_{\mathrm{PBH}})$ obtained for the LIGO mass window $M_{\mathrm{PBH}}\approx10 M_{\odot}$ for parameter values $\lambda=10^{-5}$, $\xi=34$ and $\zeta=1.345\times 10^{-10}$ and $m_0$ as in \eqref{starm}. Right: $f(M_{\mathrm{PBH}})$ in the LIGO mass window $M_{\mathrm{PBH}}\approx 10 M_{\odot}$ for parameter values $\lambda=10^{-5}$, $\xi=36.70$ and $\zeta=1.396\times 10^{-10}$ and $m_0$ as in \eqref{starm}. In both plots $f(M_{\mathrm{PBH}})$ is evaluated for $g=10.75$.}
	\label{Fig:LigoMassWindow}
\end{figure}

The left plot in \noindent Fig.~\ref{Fig:LigoMassWindow} shows that the chosen model parameters generate a distribution $f(M_{\mathrm{PBH}})$ consistent with observational constraints. Hence, the two-field extended Starobinsky model \eqref{fg} with the potential \eqref{TwoFieldPotential} can explain the origin of the black holes involved in the LIGO observed merger events as PBHs formed due to the enhancement of the inflationary power spectrum on wavelengths smaller
than the ones probed by CMB. The fit-parameters, obtained by fitting the mass distribution $f(M_{\mathrm{PBH}})$ in the left plot of Fig.~\ref{Fig:LigoMassWindow} to a log-normal distribution, are close to the values reported in \cite{Wong:2020yig} for mass distributions that would give rise to the events in the recently released GWTC-2 event catalog of the LIGO-Virgo Collaboration \cite{Abbott:2020niy} and can easily be made compatible by a slight modification of the model parameters $\zeta$, $\xi$ and $\lambda$.

The total fraction $F_{\mathrm{PBH}}$ is highly sensitive to the model parameters, in particular to $\xi$, as can be seen by comparing the two plots in Fig~\ref{Fig:LigoMassWindow} where $F_{\mathrm{PBH}}=4.1\times10^{-3}$ for the left plot and $F_{\mathrm{PBH}}=0.9$ for the right plot. Therefore, by fine tuning the model parameters any numerical value $F_{\mathrm{PBH}}\leq1$ can be obtained.

\begin{figure}[!ht]
	\centering
	\begin{tabular}{cc}
		\includegraphics[width=0.46\linewidth]{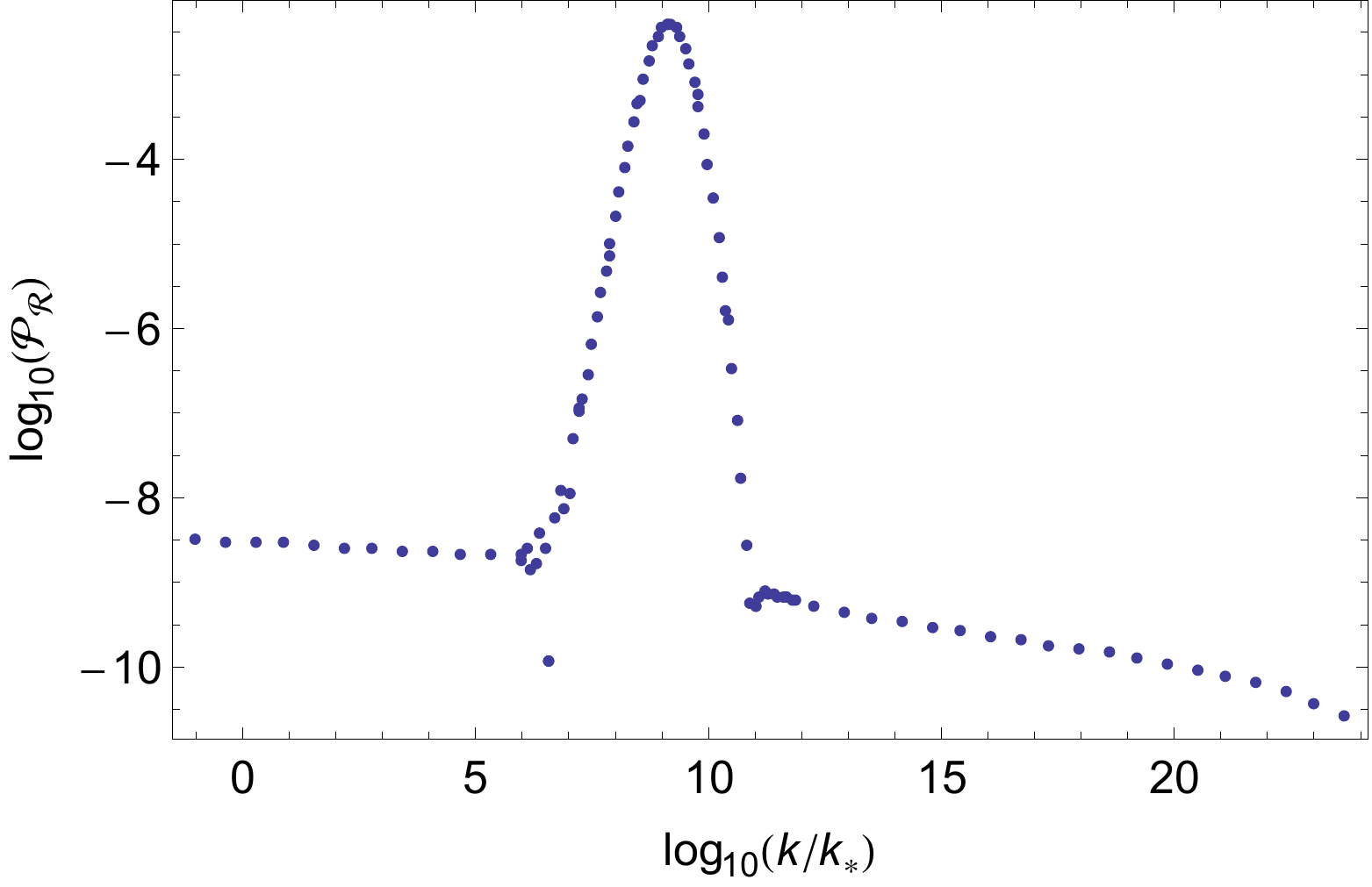}&
		\includegraphics[width=0.445\linewidth]{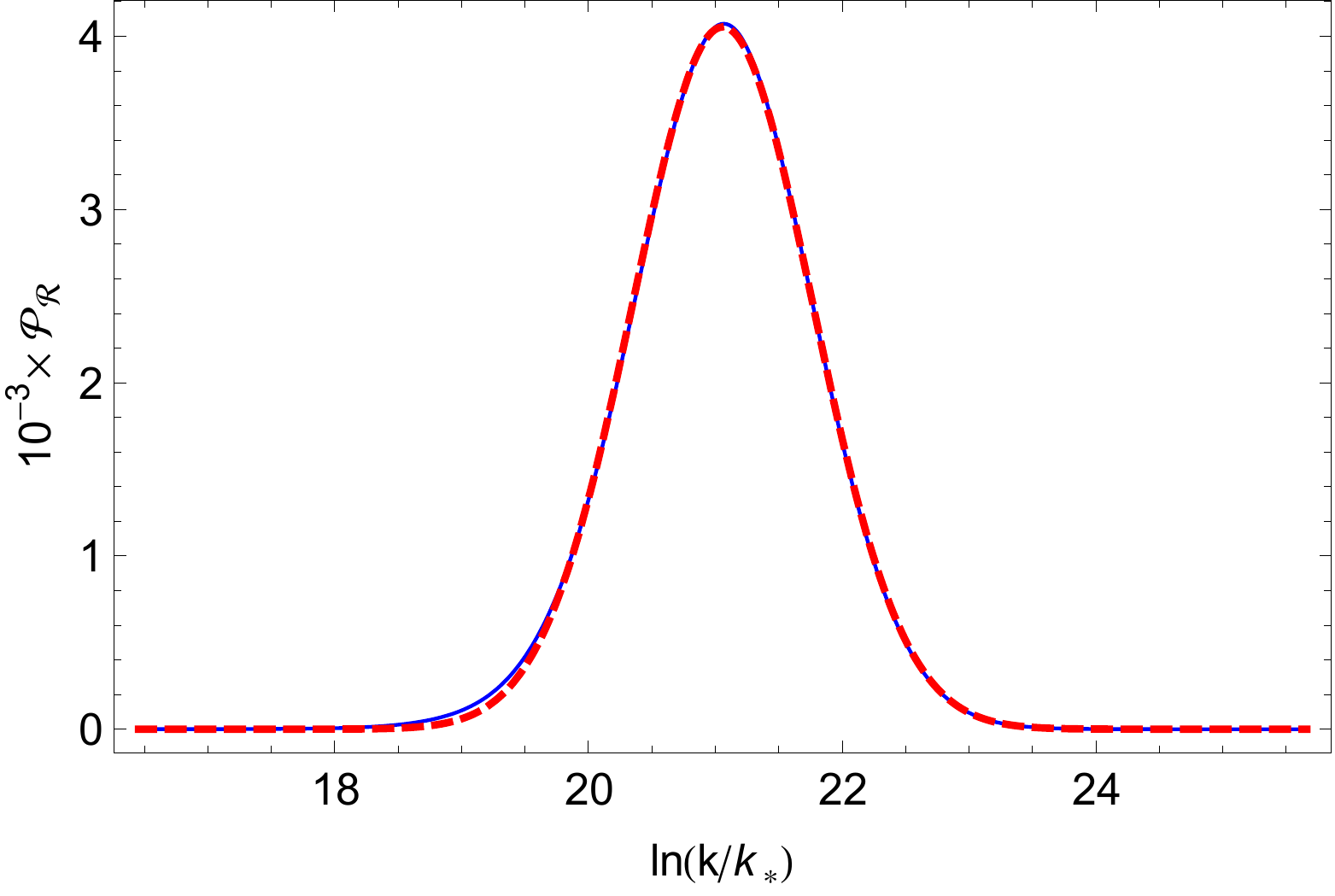}
	\end{tabular}
	\caption{Left: The log $\mathcal{P}_{\mathcal{R}}$ vs. log $k$ plot obtained for the parameter values ${\lambda=10^{-5}}$, ${\xi = 34}$ and ${\zeta = 1.345\times 10^{-10}}$. Right: The red dashed line shows the log-normal fit \eqref{SimpPLogNormal} to the peak of the numerically generated $\mathcal{P}_{\mathcal{R}}$ at ${k_{\mathrm{p}}=2.80\times10^{6}\mathrm{Mpc}^{-1}}$ with amplitude $A_{\mathrm{p}}=0.0072$ and width $\Delta_{\mathrm{p}}=0.709$ for the same parameter values as in the left plot. The blue solid line shows the \texttt{Mathematica} interpolating function of the numerically obtained $\mathcal{P}_{\mathcal{R}}$ in the peak region for the same parameters.}	
	\label{Fig:LigoMassWindowInterpolation}
\end{figure}

As illustrated in Fig.~\ref{Fig:LigoMassWindowInterpolation}, the log-normal function \eqref{SimpPLogNormal}
closely fits the peak of the exact numerically obtained power spectrum, implying that the peak is well characterized by three parameters: the peak scale $k_{\mathrm{p}}$, the peak amplitude $A_{\mathrm{p}}$ and the peak width $\Delta_{\mathrm{p}}$. In addition, by using the log-normal fit, the subsequent numerical integrations can be performed much more efficiently. 

\noindent Finally, the log-normal fit to the peak allows a direct comparison with \cite{Gow:2020bzo}. In Fig.~1 of \cite{Gow:2020bzo}, $f(M_{\mathrm{PBH}})$ for the LIGO mass window was also computed using the Press-Schechter formalism with a modified Gaussian window function and by assuming a log-normal shape for the peak in the power spectrum.
In \cite{Gow:2020bzo} the value of the total fraction was fixed to $F_{\mathrm{PBH}}=2\times10^{-3}$, which is of the same order as $F_{\mathrm{PBH}}=4.1\times10^{-3}$ obtained in left plot in Fig.~\ref{Fig:LigoMassWindow}. 

Moreover, a comparison with the results tabulated in Table I of \cite{Gow:2020bzo} shows (first column) that $A_{\mathrm{p}}$ must lie between ${4.14\times 10^{-3}\leq A_{\mathrm{p}}\leq 8.92\times 10^{-3}}$ in order to have ${F_{\mathrm{PBH}}=2\times10^{-3}}$ for $0.3\leq\Delta_{\mathrm{p}}\leq 1$. The values ${A_{\mathrm{p}}=7.2\times10^{-3}}$ and $\Delta_{\mathrm{p}}=0.709$ obtained in Fig.~\ref{Fig:LigoMassWindow} leading to $F_{\mathrm{PBH}}=4.1\times10^{-3}$ therefore provide an additional consistency check for the numerical evaluation of $f(M_{\mathrm{PBH}})$.
%
%
\subsection{Mass windows $M^{I}_{\mathrm{PBH}}$ and $M^{II}_{\mathrm{PBH}}$}\label{subsec:MassWindowII}

Current observations suggest that ${F_{\mathrm{PBH}}=1}$ cannot be realized within the entire mass window
\begin{align}
10^{-17}M_{\odot}\lesssim M_{\mathrm{PBH}}\lesssim 10^{-9}M_{\odot},\label{M12}
\end{align}
but only within the smaller mass windows $M^{I}_{\mathrm{PBH}}$ and $M^{II}_{\mathrm{PBH}}$ resulting from splitting \eqref{M12}, see \cite{Carr:2020xqk}. We explicitly show that for appropriate parameter values,  mass distributions $f(M_{\mathrm{PBH}})$ with $F_{\mathrm{PBH}}=1$ can be realized in both mass windows $M_{\mathrm{PBH}}^{I}$ and  $M^{II}_{\mathrm{PBH}}$. Note that recent data from the NANOGrav Collaboration \cite{Arzoumanian:2020vkk} lends further support to the proposal that PBHs may constitute a large part (or the whole) of CDM with the dominant contribution to the mass function in the range $10^{-15}M_{\odot}\div 10^{-11} M_{\odot}$ \cite{DeLuca:2020agl}. Thus, in light of the observational ambiguity regarding the strict upper bound on $F_{\mathrm{PBH}}$ in the respective mass windows $M^{I}_{\mathrm{PBH}}$ and $M^{II}_{\mathrm{PBH}}$, our main objective is not to derive stringent constraints on the model parameters for all the mass windows but to demonstrate that there are parameter values that lead to observationally viable mass distributions $f(M_{\mathrm{PBH}})$.

\begin{figure}[!ht]
	\centering
	\begin{tabular}{cc}
		\includegraphics[width=0.45\linewidth]{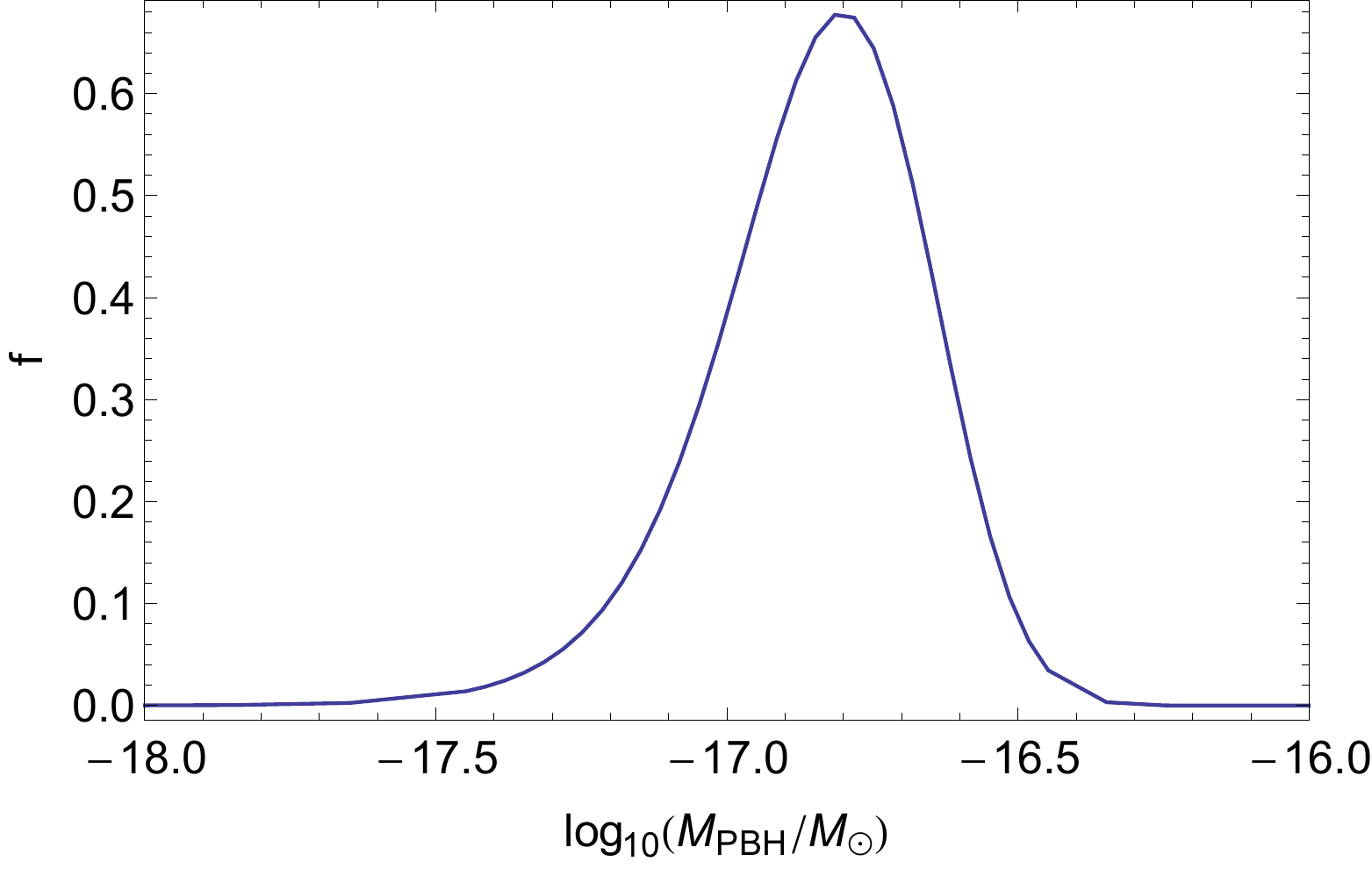}&
		\includegraphics[width=0.45\linewidth]{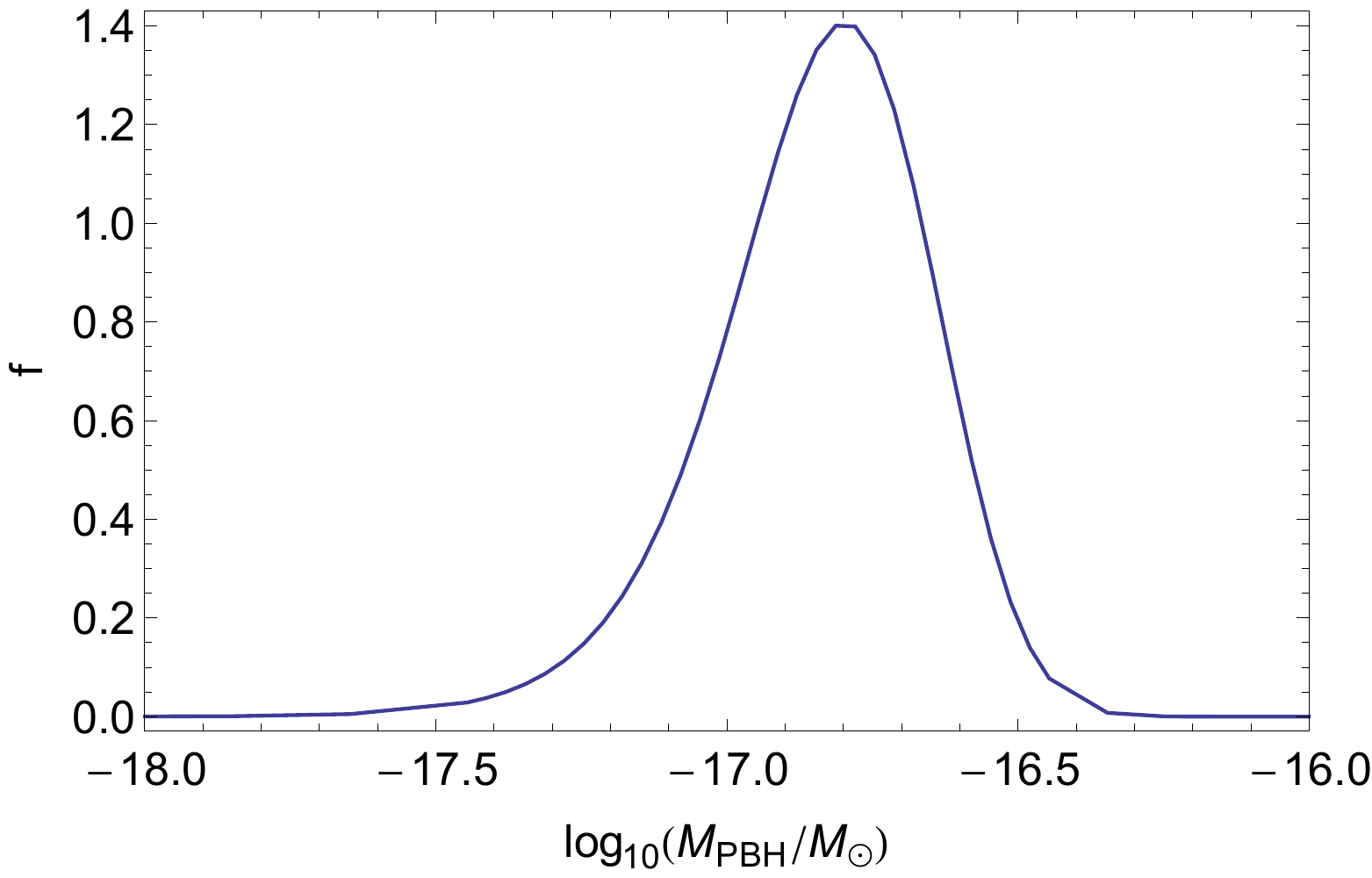}
	\end{tabular}
	\caption{Left: ${f(M_{\mathrm{PBH}})}$ for ${\lambda = 10^{-5}}$, ${\xi = 38}$, ${\zeta = 3.30\times 10^{-10}}$ leading to the log-normal power spectrum fit with ${A_{\mathrm{p}}=0.003835}$, ${\Delta_{\mathrm{p}}=0.592179}$ and ${k_{\mathrm{p}}=3.04\times 10^{15}\mathrm{Mpc}^{-1}}$. Right: $f(M_{\mathrm{PBH}})$ for ${\lambda = 10^{-5}}$, ${\xi = 38.2}$, ${\zeta = 3.31\times 10^{-10}}$ leading to the log-normal power spectrum fit with ${A_{\mathrm{p}}=0.00391}$, ${\Delta_{\mathrm{p}}=0.5919}$ and ${k_{\mathrm{p}}=3.03\times 10^{15}\mathrm{Mpc}^{-1}}$. In both plots $f(M_{\mathrm{PBH}})$ is evaluated for $g=106.75$.}
	\label{Fig:MassWindowOne}
\end{figure}

In Fig.~\ref{Fig:MassWindowOne}, we show an exemplary parameter combination for which $\mathcal{P}_{\mathcal{R}}$ peaks at ${k_{\mathrm{p}}\approx10^{15}\mathrm{Mpc}^{-1}}$ and generates a significant amount of CDM in mass window $M_{\mathrm{PBH}}^{I}$. The left plot in Fig.~\ref{Fig:MassWindowOne} shows a mass distribution $f(M_{\mathrm{PBH}})$ which leads to $F_{\mathrm{PBH}}=0.69$ for $\xi=38$.
The right plot in Fig.~\ref{Fig:MassWindowOne} illustrates the sensitivity of $f(M_{\mathrm{PBH}})$ on $\xi$. For a $\xi$ larger by only $0.5\%$, the amplification mechanism is already too strong and leads to the observationally unacceptable large value of $F_{\mathrm{PBH}}=1.4$.

\begin{figure}[!ht]
	\centering
	\begin{tabular}{cc}
		\includegraphics[width=0.45\linewidth]{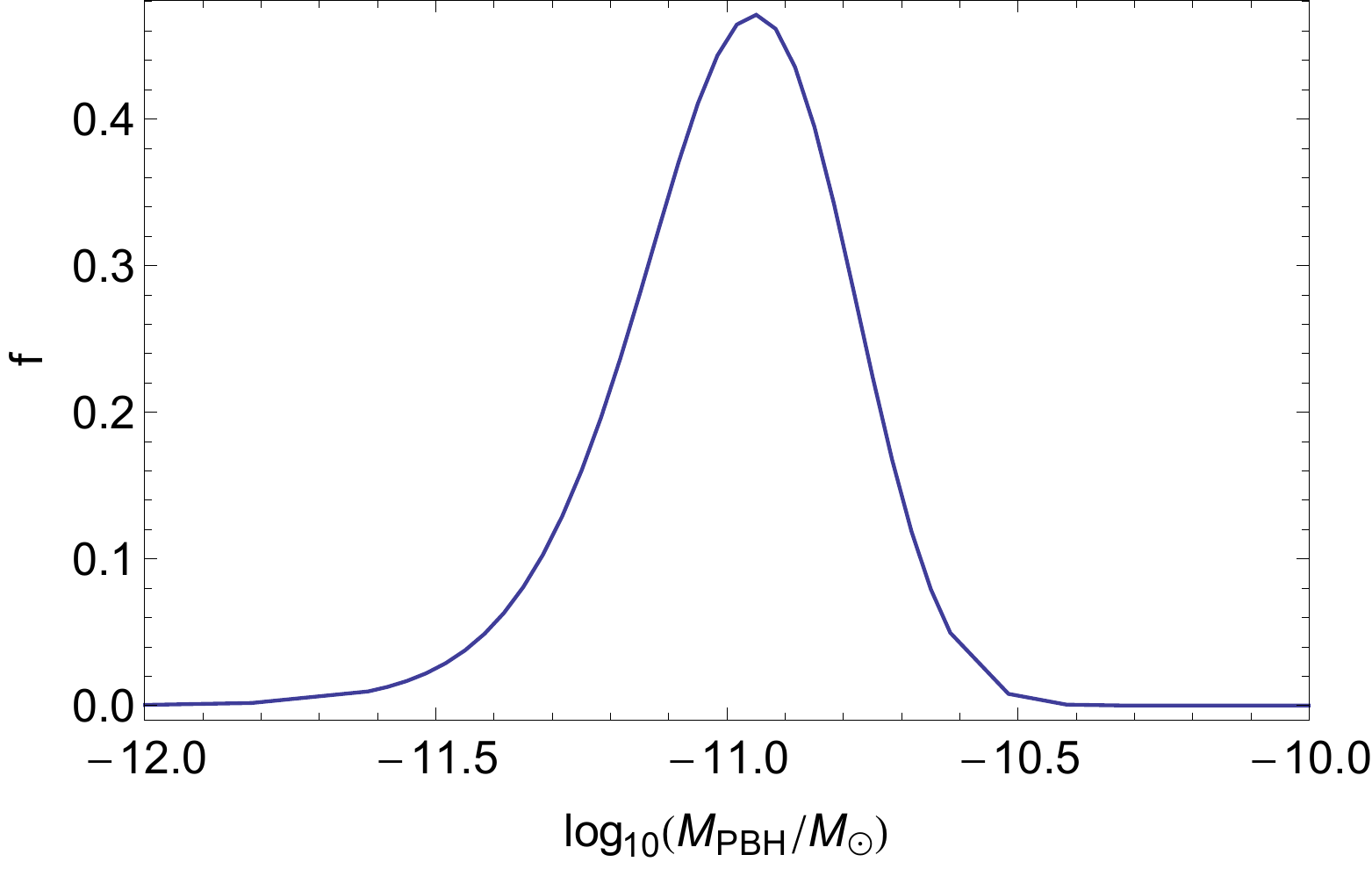}&
		\includegraphics[width=0.45\linewidth]{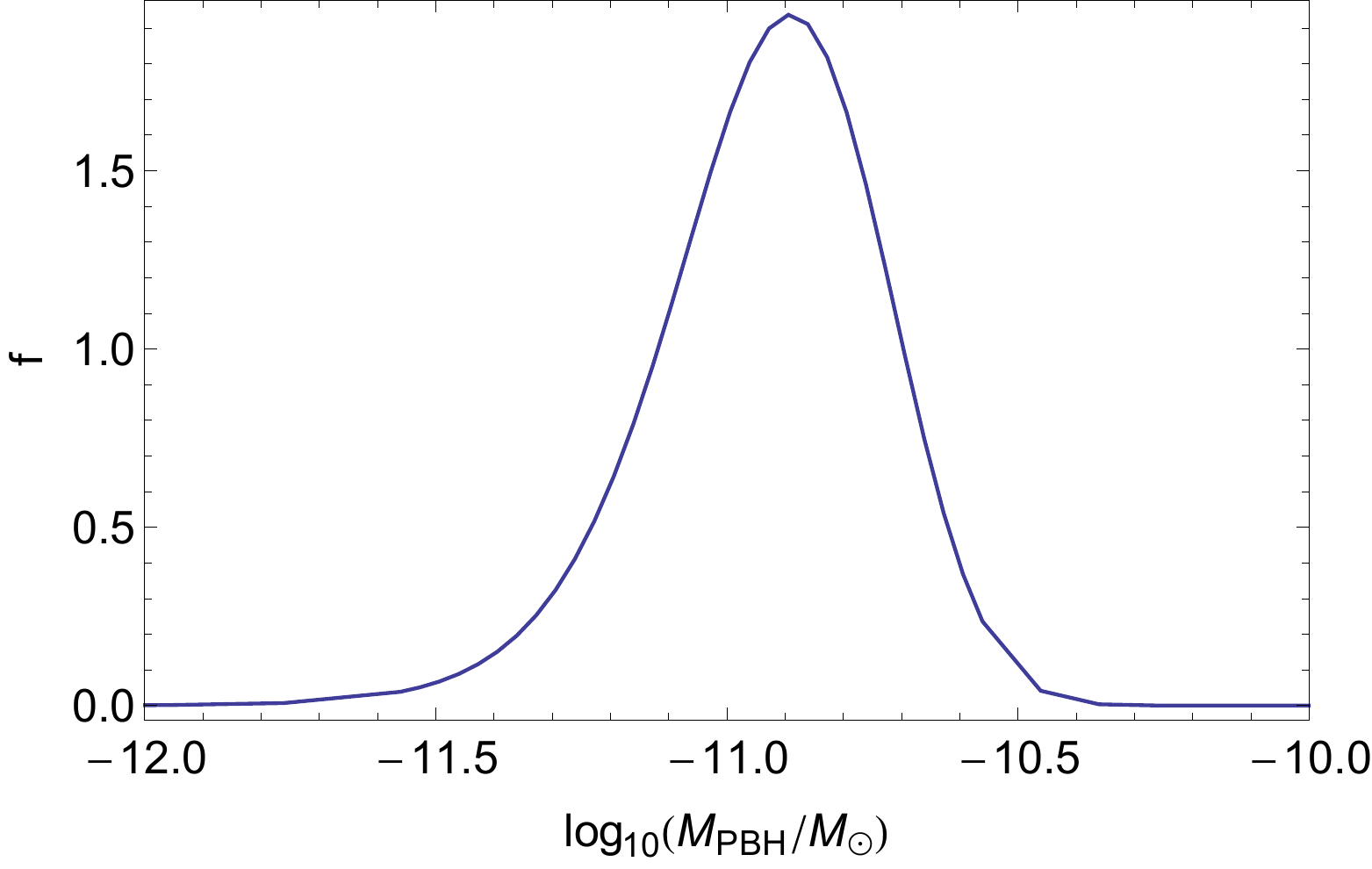}
	\end{tabular}
	\caption{Left: $f(M_{\mathrm{PBH}})$ for ${\lambda = 10^{-5}}$, ${\xi = 36}$, ${\zeta = 2.29\times 10^{-10}}$ leading to the log-normal power spectrum fit with ${A_{\mathrm{p}}=0.00485}$, ${\Delta_{\mathrm{p}}=0.6360}$ and ${k_{\mathrm{p}}=3.69\times 10^{12}\mathrm{Mpc}^{-1}}$. Right: $f(M_{\mathrm{PBH}})$ for ${\lambda = 10^{-5}}$, ${\xi = 36.5}$, ${\zeta = 2.31\times 10^{-10}}$ leading to the log-normal power spectrum fit with ${A_{\mathrm{p}}=0.0051}$, ${\Delta_{\mathrm{p}}=0.6348}$ and ${k_{\mathrm{p}}=3.46\times 10^{12}\mathrm{Mpc}^{-1}}$. In both plots $f(M_{\mathrm{PBH}})$ is evaluated for $g=106.75$.}
	\label{Fig:MassWindowTwo}
\end{figure}

\noindent Similarly, Fig.~\ref{Fig:MassWindowTwo} shows $f(M_{\mathrm{PBH}})$ for two parameter combinations in the mass window  $M_{\mathrm{PBH}}^{II}$. The left plot in Fig.~\ref{Fig:MassWindowTwo} shows an observationally viable scenario with $F_{\mathrm{PBH}}=0.5$, while the right plot leads to a an unacceptable value $F_{\mathrm{PBH}}=2.1$.
Summarizing, the discussion presented in this section illustrates that the model parameters can be adjusted such that a significant fraction of CDM (including all CDM) is made of PBHs in all the three mass windows \eqref{MI1}-\eqref{MI3}.
It is interesting to see whether the isocurvature pumping mechanism underlying
our two-field model can be distinguished from the ultra slow-roll enhancement mechanism in single-field models
of inflation on purely phenomenological grounds. A possible
indicator which would allow to discriminate between
these mechanisms is the growth rate of the power spectrum
at small scales. In single-field models the growth rate was found to be bounded by a $k^4$ enhancement in \cite{Byrnes:2018txb} (the refined analysis in \cite{Carrilho:2019oqg} lead to a weaker bound $k^5(\ln k)^2$). In all our numerical results the enhancement is well fitted
by the log-normal distribution \eqref{SimpPLogNormal}. Consequently, the growth rate
can be quantified by the (scale dependent) spectral index
\begin{align}
n_{\mathrm{s}}(k)-1 = \frac{\partial \ln\mathcal{P}_{\mathrm{PBH}}}{\partial \ln(k)}.
\end{align}
At $\ln(k_{\mathrm{max}}/k_{\mathrm{p}}) = -\Delta_{\mathrm{p}}$
the growth rate of the power spectrum has a maximum.
This implies that $n_{\mathrm{s}}(k_{\mathrm{max}})-1 = 1/\Delta_{\mathrm{p}}$, and hence $\Delta_{\mathrm{p}}\leq0.25$ to overcome the single-field bound $n_{\mathrm{s}}\leq5$. In
all our numerical results the value of $\Delta_{\mathrm{p}}$ does not fall below $\Delta_{\mathrm{p}} = 0.592$ (cf. Fig.~\ref{Fig:MassWindowOne}). Hence, for the scenarios considered in this work, we
cannot discriminate between the isocurvature pumping
mechanism in our model and the ultra slow-roll single-field mechanism by just comparing the growth rate of the power spectra.

%
%
\section{Conclusions}
\label{Conclusions}

The two-field dilaton extension of Starobinsky's inflationary model predicts a successful phase of inflation in perfect agreement with recent Planck measurements. At the same time, it is capable of predicting the presently observed dark matter content in our Universe in the form of PBHs. The generation of gravitational waves and CDM from PBHs is also possible in the Starobinsky supergravity theory with two-field double inflation  \cite{Aldabergenov:2020bpt,Aldabergenov:2020yok}.

It is interesting to compare our model to the model of scalaron-Higgs inflation \cite{Gundhi:2018wyz}.
First, we identify the non-minimally coupled scalar field extending Starobinsky's geometric model with a dilaton field, whereas in the scalaron-Higgs model this scalar field is identified with the Standard Model Higgs field. Second, we assume a dilaton-dependent scalaron mass $M^2(\varphi)$ in \eqref{smodf}, which effectively introduces an additional parameter $\zeta$.
 
For positive values of $\zeta$ the stable inflationary dynamics along $\varphi_0$ reduces to that of an effective single-field model with the same predictions as Starobinsky's model at the scales probed by the CMB.
However, the inflationary trajectory remains stable only up to the critical point $\hat{\chi}_{\mathrm{c}}$, at which $\varphi_0$ turns into an unstable hill. The trajectory subsequently falls into one of the outer $\varphi_{\mathrm{v}}^{\pm}$ valleys. As explained in Sect.~\ref{PeakFormation},
this feature of the multi-field dynamics leads to an amplification of the adiabatic power spectrum at small wavelengths. In contrast, the two-field potential in scalaron-Higgs inflation features an unstable hill-top along $\varphi_0$ for all values of $\hat{\chi}$ and, therefore, leads to an immediate fall of the inflationary background trajectory already at the very onset of inflation.

Even though the scalaron-Higgs model predicts the same values for the inflationary observables as Starobinsky's model at large wavelengths, it does not support the isocurvature sourcing mechanism for smaller wavelengths required for significant PBH production.
The detailed conditions required for a successful realization of the multi-field amplification mechanism include the growth of isocurvature perturbations resulting from an intermediate phase in which the effective isocurvature mass becomes tachyonic and the sourcing of adiabatic modes resulting from a curved trajectory in the scalar field space geometry.
This  "isocurvature pumping" mechanism, already discussed in \cite{Gundhi:2018wyz}, is a genuine multi-field effect and is essentially different from other amplification mechanisms in the single-field models of inflation. 

We emphasized the necessity of the stochastic treatment within the short (less than one efold) transition phase, where quantum diffusive effects dominate the inflationary dynamics, and set the initial conditions for the subsequent classical dynamics in which the peak in the adiabatic power spectrum is generated. Thus, the stochastic treatment is crucial for the production of PBHs and for a precise prediction of their contribution to the presently observed CDM. The importance of the stochastic formalism in the context of PBHs produced in multi-field models of inflation is also discussed in \cite{Mollerach:1990zf,Assadullahi:2016gkk,Vennin:2020kng}.

All numerical results presented in Sect.~\ref{NumResults} are obtained for the parameter values satisfying the condition \eqref{ScenarioI}, and, therefore, are realized in \mbox{Scenario I}. In \mbox{Scenario I}, the valley along $\varphi_0$ is a global attractor, so that all its predictions are independent of the initial conditions for the inflationary background trajectory.       

Even if the exact inflationary power spectrum is known numerically, the calculation of the PBH mass distribution depends on the details of the formalism used.
For generic power spectra with multiple peaks of different amplitudes and shapes, a choice of the window functions and the formalism (Press-Schechter \cite{Press:1973iz} vs. peaks theory \cite{Bardeen:1985tr,Green:2004wb,Young:2020xmk}) do, in general, also affect the result for the PBH mass distribution \cite{Young:2019osy}. 
The PBH collapse process does not only depend on the time at which the PBHs form, but also on the amplitude of the density perturbations. This is taken into account by the critical scaling relation for the PBH mass \cite{Jedamzik:1999am}.
In the present work, we assumed that PBHs form immediately once an overcritical density fluctuation enters the horizon, took into account the critical scaling relation \eqref{CritMPBH} for the PBH mass, worked with the Press-Schechter formalism \eqref{BetaMasterEq}, and used a Gaussian window function \eqref{WindowG}. However, since the dilaton-extended Starobinsky model predicts an almost scale invariant power spectrum with a single sharp peak, we neither expect our results to strongly depend on the specifics of the Press-Schechter formalism nor on the choice of the Gaussian window function. This expectation is further supported by the comparison with the results of \cite{Gow:2020bzo}.  

We investigated three different PBH mass windows \eqref{MI1}-\eqref{MI3} where the observational constraints permit a sizeable contribution of PBHs to the presently observed CDM content of our universe. This also includes the possibility to explain all CDM by PBHs in the mass windows \eqref{MI1} and \eqref{MI2}, which is supported by the recent data from the NANOGrav Collaboration \cite{Arzoumanian:2020vkk}.

Of particular interest is the mass window \eqref{MI3}, as the sources of the binary merger events observed by the LIGO  Collaboration \cite{Abbott:2016blz} may be identified with PBHs and thereby provide an additional observational window to the inflationary dynamics. 
While the maximum value of the fraction $F_{\mathrm{PBH}}$ in the LIGO mass window \eqref{MI3} is controversially discussed \cite{Chen:2018czv,Raidal:2018bbj,Gow:2019pok,DeLuca:2020qqa,Jedamzik:2020ypm,Young:2020scc,Jedamzik:2020omx,Clesse:2020ghq}, the extended dilaton two-field Starobinsky model can account for any observationally viable $F_{\mathrm{PBH}}$
by a suitable combination of parameters.

Our model allows for several future applications. It would be interesting to study the effect of non-Gaussianities, both primordial ones and those arising from the non-linear relation between the comoving curvature perturbation and the density perturbation.
Moreover, the formation of primordial black holes inevitably leads to the production of gravitational waves \cite{Nakama:2016gzw,Garcia-Bellido:2017aan}, which may be detected by the space-based gravitational interferometer LISA \cite{Cai:2018dig,Bartolo:2018evs,Braglia:2020eai} and could provide additional constraints on the model.
Finally, it may be possible to realize the same mechanism of PBH production with the Standard Model Higgs field instead of the dilaton in a similar extension of the scalaron-Higgs model \cite{Gundhi:2020zvb}.
%
%
\section*{Acknowledgments}
The authors are grateful to Valerio De Luca, Gabriele Franciolini and Ilia Musco for helpful correspondence.  
AG thanks the University of Trieste and INFN for financial support and Angelo Bassi for supporting this collaboration.
SVK was supported by Tokyo Metropolitan University, the World Premier International Research Center Initiative (WPI), MEXT, Japan, and the Competitiveness Enhancement Program of Tomsk Polytechnic University in Russia.
%
%
\appendix
%
%
\section{Derivation of the PBH mass distribution}\label{Appendix1}

To derive the mass fraction in terms of quantities observed today, we first provide a number of basic equations that relate the relevant quantities at different times $t$ and temperatures $T$,
\begin{align}
\frac{\rho(t)}{\rho(t_{\mathrm{eq}})}={}&\frac{g(T)}{g(T_{\mathrm{eq}})}\left(\frac{T}{T_{\mathrm{eq}}}\right)^4\label{RhoTemp},\\
\frac{H(t)}{H(t_{\mathrm{eq}})}={}&\left[\frac{g(T)}{g(T_{\mathrm{eq}})}\right]^{1/2}\left(\frac{T}{T_{\mathrm{eq}}}\right)^2\label{HubbleTemp},\\
\frac{a(t)}{a(t_{\mathrm{eq}})}={}&\left[\frac{g(T)}{g(T_{\mathrm{eq}})}\right]^{-1/3}\left(\frac{T}{T_{\mathrm{eq}}}\right)^{-1}\label{ScaleTemp},\\
\frac{k}{k_{\mathrm{eq}}}={}&\frac{a(t)H(t)}{a(t_{\mathrm{eq}})H(t_{\mathrm{eq}})}=\left[\frac{g(T)}{g(T_{\mathrm{eq}})}\right]^{1/6}\left(\frac{T}{T_{\mathrm{eq}}}\right).\label{HorizonTemp}
\end{align}
Here $t_{\mathrm{eq}}$ and $T_{\mathrm{eq}}$ are the time and temperature at matter-radiation equality, and $g(T)$ counts the effective number of relativistic degrees of freedom at a given temperature.\footnote{The function $g$ only depends on the temperature because the energy density and the entropy density are equal during the radiation dominated era, see e.g.~\cite{Riotto:2002yw}.} 

Let $\rho_{\mathrm{tot}}(t)$  and $\rho_{\mathrm{PBH}} (t)$ be the total energy density and the energy density of the primordial black holes at some given time $t$, respectively. Then the mass fraction $\beta(t)$ at time $t$ is defined by
\begin{align}\label{BetaDef}
\beta(t):=\frac{\rho_{\mathrm{PBH}}(t)}{\rho_{\mathrm{tot}}(t)}.
\end{align}
Since we focus only on PBHs that form well within the radiation dominated era, we have
\begin{align}\label{Betaapprox}
\beta(t_{\mathrm{f}})\approx \frac{\rho_{\mathrm{PBH}}(t_{\mathrm{f}})}{\rho_{\mathrm{rad}}(t_{\mathrm{f}})},
\end{align}
with $\rho_{\mathrm{rad}}$ the energy density of the relativistic matter and $t_{\mathrm{f}}$ the time of PBH formation. We would like to relate $\beta(t_{\mathrm{f}})$ to present observations. First, we relate $\beta(t_{\mathrm{f}})$ to its value at matter-radiation equality $t_{\mathrm{eq}}$. We express $\rho_{\mathrm{PBH}}(t_{\mathrm{f}})$ in terms of its value at $t_{\mathrm{eq}}$ via
\begin{align}\label{rhoPBHeq}
\rho_{\mathrm{PBH}}(t_{\mathrm{f}})=\rho_{\mathrm{PBH}}(t_{\mathrm{eq}})\frac{a^3(t_{\mathrm{eq}})}{a^3(t_{\mathrm{f}})}.
\end{align}
Next, we relate $\rho_{\mathrm{rad}}(t_{\mathrm{f}})$ to its value at $t_{\mathrm{eq}}$ via \eqref{RhoTemp}. Using \eqref{ScaleTemp}, \eqref{rhoPBHeq} and \eqref{RhoTemp} in \eqref{Betaapprox} we obtain
\begin{align}\label{BetaEq}
\beta(t_{\mathrm{f}})=\frac{T_{\mathrm{eq}}}{T_{\mathrm{f}}}\frac{\rho_{\mathrm{PBH}}(t_{\mathrm{eq}})}{\rho_{\mathrm{rad}}(t_{\mathrm{eq}})},
\end{align}  
with $T_{\mathrm{f}}$ and $T_{\mathrm{eq}}$ denoting the temperatures at PBH formation and matter-radiation equality, respectively. 

Next, we express $\beta(t_{\mathrm{f}})$ in terms of the currently observed values of cosmological parameters. For this purpose, note that the non-relativistic matter energy density $\rho_{\mathrm{m}}(t)$ satisfies $\rho_{\mathrm{rad}}(t_{\mathrm{eq}})=\rho_{\mathrm{m}}(t_{\mathrm{eq}})$, while $\rho_{\mathrm{m}}(t_{\mathrm{eq}})$ is related to the total energy density $\rho_{\mathrm{tot}}(t_0)$ at $t_0$ today via  
\begin{align}\label{RhoRadToday}
\rho_{\mathrm{rad}}(t_{\mathrm{eq}})=\rho_{\mathrm{m}}(t_{\mathrm{eq}})= \Omega _{\mathrm{m}}\frac{\rho_{\mathrm{tot}}(t_0)}{a^3(t_{\mathrm{eq}})},
\end{align}
with the matter density parameter ${\Omega_m=0.315\pm 0.007}$ \cite{Aghanim:2018eyx}. Furthermore, 
$\rho_{\mathrm{PBH}}(t_{\mathrm{eq}})$ is related to its current value $\rho_{\mathrm{PBH}}(t_0)$ via
\begin{align}\label{RhoPBHToday}
\rho_{\mathrm{PBH}}(t_{\mathrm{eq}})=\frac{\rho_{\mathrm{PBH}}(t_0)}{a^3(t_{\mathrm{eq}})}.
\end{align}
Using \eqref{RhoRadToday} and \eqref{RhoPBHToday}, \eqref{BetaEq} becomes
\begin{align}
\beta(t_{\mathrm{f}})=\frac{T_{\mathrm{eq}}}{T_{\mathrm{f}}}\frac{\rho_{\mathrm{PBH}}(t_0)}{\Omega_{\mathrm{m}}\rho_{\mathrm{tot}}(t_0)}.\label{betaT}
\end{align}
The total energy density today is expressed in terms of the CDM density today $\rho_{\mathrm{CDM}}(t_0)$ via the CDM density parameter $\Omega_{\mathrm{CDM}}$ by the relation ${\rho_{\mathrm{CDM}}(t_0)=\Omega_{\mathrm{CDM}}\rho_{\mathrm{tot}}(t_0)}$. When inserted into \eqref{betaT}, this yields
\begin{align}
\beta(t_{\mathrm{f}})=\frac{\Omega_{\mathrm{CDM}}}{\Omega_{\mathrm{m}}}\frac{T_{\mathrm{eq}}}{T_{\mathrm{f}}}\frac{\rho_{\mathrm{PBH}}(t_0)}{\rho_{\mathrm{CDM}}(t_0)}.\label{Betafin}
\end{align}
Since the horizon mass at a given time reads  ${M_{\mathrm{H}}(t)=4\pi\rho(t)/H^3(t)}$, using \eqref{RhoTemp} and \eqref{HubbleTemp}, the ratio $T_{\mathrm{f}}/T_{\mathrm{eq}}$ in \eqref{Betafin} can be expressed in terms of the horizon masses $M_{\mathrm{H}}(t_{\mathrm{f}})$ and $M_{\mathrm{H}}(t_{\mathrm{eq}})$. Defining, in addition, the ratio $\rho_{\mathrm{PBH}}(t_0)/\rho_{\mathrm{CDM}}(t_0)$ of the PBH energy density and the CDM density as observed today as
\begin{align}
f:=\frac{\rho_{\mathrm{PBH}}(t_{0})}{\rho_{\mathrm{CDM}}(t_{0})},
\end{align}
the relation \eqref{Betafin} can be rewritten as
\begin{align}\label{fPBHtf}
\beta(t_{\mathrm{f}})=\frac{\Omega_{\mathrm{CDM}}}{\Omega_{\mathrm{m}}}\left(\frac{g(t_{\mathrm{f}})}{g(t_{\mathrm{eq}})}\right)^{1/4}\left(\frac{M_{\mathrm{H}}(t_{\mathrm{f}})}{M_{\mathrm{H}}(t_{\mathrm{eq}})}\right)^{1/2}f.
\end{align}
Inverting this relation gives $f$ as a function of $t_{\mathrm{f}}$ with  ${M_{\mathrm{H}}:=M_{\mathrm{H}}(t_{\mathrm{f}})}$, ${g(t_{\mathrm{eq}}):= g_{\mathrm{eq}}}$ and ${M_{\mathrm{H}}^{\mathrm{eq}}:= M_{\mathrm{H}}(t_{\mathrm{eq}})}$,
\begin{align}
f(t_{\mathrm{f}})=\frac{\Omega_{\mathrm{m}}}{\Omega_{\mathrm{CDM}}}\left(\frac{g(t_{\mathrm{f}})}{g_{\mathrm{eq}}}\right)^{-1/4}\left(\frac{M_{\mathrm{H}}(t_{\mathrm{f}})}{M_{\mathrm{H}}^{\mathrm{eq}}}\right)^{-1/2}\beta(t_{\mathrm{f}}).\label{ffinal}
\end{align}
Since $M_{\mathrm{H}}(t)$ grows monotonically with time $t$, we trade the $t_\mathrm{f}$ dependence for $M_{\mathrm{H}}$ and obtain the PBH mass distribution $f$ as a function of the horizon mass $M_{\mathrm{H}}$,
\begin{align}
f(M_{\mathrm{H}})=\frac{\Omega_{\mathrm{m}}}{\Omega_{\mathrm{CDM}}}\left(\frac{g(M_{\mathrm{H}})}{g_{\mathrm{eq}}}\right)^{-1/4}\left(\frac{M_{\mathrm{H}}}{M_{\mathrm{H}}^{\mathrm{eq}}}\right)^{-1/2}\beta(M_{\mathrm{H}}).\label{ffinalMh}
\end{align}
The total fraction of CDM made from PBHs is obtained by integrating \eqref{ffinalMh} over the logarithmic mass intervals (or, equivalently, by integrating \eqref{ffinal} over all epochs when PBHs could have formed) as
\begin{align}\label{Ftotal}
F_{\mathrm{PBH}}:=\int_{-\infty}^{\infty}f(M_{\mathrm{H}})\mathrm{d}\ln M_{\mathrm{H}}.
\end{align}
In order to obtain the distribution \eqref{ffinalMh} as a function of the PBH mass, we proceed as follows. 
The mass fraction $\beta(M_{\mathrm{H}})$ as a function of $M_{\mathrm{H}}$ is given by \cite{Niemeyer:1997mt},
\begin{align}\label{BetaMasterEqAPP}
\beta(M_{\mathrm{H}})=2\int_{\delta_c}^{\infty}\frac{M_{\mathrm{PBH}}(\delta,M_{\mathrm{H}})}{M_{\mathrm{H}}}P(\delta)d\delta,
\end{align} 
with the Gaussian PDF \eqref{ProbDist} and variance $\sigma^2_{R}(M_{\mathrm{H}})$, 
\begin{align}\label{ProbDistAPP}
P(\delta)=\frac{1}{\sqrt{2\pi\sigma_{R}^2(M_{\mathrm{H}})}}\exp{\left(-\frac{\delta^2}{2\sigma^2_{R}(M_{\mathrm{H}})}\right)}.
\end{align}
The explicit dependence of $M_{\mathrm{PBH}}$ on the horizon mass $M_{\mathrm{H}}$ and the amplitude of the energy density contrast $\delta $ is given by the critical scaling \eqref{CritMPBH},
\begin{align}\label{ScalingLaw}
M_{\mathrm{PBH}} = K M_{\mathrm{H}} (\delta-\delta_{\mathrm{c}})^{\gamma}.
\end{align}
For fixed $M_{\mathrm{H}}$, we solve \eqref{ScalingLaw} for $\delta$ as a function of $M_{\mathrm{PBH}}$,
\begin{align}\label{DeltaInvert}
\delta=\mu^{1/\gamma}+\delta_{\mathrm{c}}\,,\qquad \mu:= \frac{M_{\mathrm{PBH}}}{K M_{\mathrm{H}}}.
\end{align}
Hence, for fixed $M_{\mathrm{H}}$, the $\mathrm{d}\delta$ integral in \eqref{BetaMasterEqAPP} can be traded for the $\mathrm{d}\ln M_{\mathrm{PBH}}$ integral with ${\mathrm{d}\delta=(\mu^{1/\gamma}/\gamma)\mathrm{d}\ln M_{\mathrm{PBH}}}$. After doing that and using \eqref{ProbDistAPP} in \eqref{BetaMasterEqAPP} and \eqref{BetaMasterEqAPP} in \eqref{ffinalMh}, \eqref{Ftotal} takes the form
\begin{align}
F_{\mathrm{PBH}}=\int_{-\infty}^{\infty} f(M_{\mathrm{PBH}}) d\ln M_{\mathrm{PBH}},\label{FMPBH}
\end{align} 
with the PBH mass distribution $f(M_{\mathrm{PBH}})$ given by 
\onecolumngrid
\begin{align}
f(M_{\mathrm{PBH}})=2\frac{\Omega_{\mathrm{m}}}{\Omega_{\mathrm{CDM}}}\int_{-\infty}^{\infty} d(\ln M_{\mathrm{H}})\frac{M_{\mathrm{PBH}}}{M_{\mathrm{H}}}\left(\frac{g(M_{\mathrm{H}})}{g_{\mathrm{eq}}}\right)^{-1/4}\left(\frac{M_{\mathrm{H}}}{M^{\mathrm{eq}}_{\mathrm{H}}}\right)^{-1/2}\frac{\mu^{1/\gamma}}{\gamma\sqrt{2\pi\sigma_{R}^2(M_{\mathrm{H}})}}\exp\left(-\frac{\left(\mu^{1/\gamma}+\delta_{\mathrm{c}}\right)^2}{2\sigma_{R}^2(M_{\mathrm{H}})}\right).\label{fpbhfinal}
\end{align}
\twocolumngrid
%
%
\section{Analytic estimates}\label{Analytic}

Under various simplifying assumptions, we obtain a simple analytic estimate of $\beta(M_{\mathrm{H}})$. However, we emphasize that the results obtained in this appendix are only to illustrate the basic relationship between the mass fraction and the peak amplitude of the enhanced power spectrum.
First, we calculate the variance $\sigma_{R}^2$ by assuming that the inflationary power spectrum can be written as a sum of the constant CMB part $A_{\mathrm{CMB}}$ and the part $\mathcal{P}_{\mathrm{PBH}}(k)$ responsible for the PBH production\footnote{The weak logarithmic scale dependence of $\mathcal{P}_{\mathcal{R}}$ at large scales is neglected as it has a negligible effect on the peak analysis.},
\begin{align}
\mathcal{P}_{\mathcal{R}}(k)=A_{\mathrm{CMB}}+\mathcal{P}_{\mathrm{PBH}}(k).\label{SimpP}
\end{align}
Assuming further that the PBH part of the power spectrum has a single symmetric peak centered at the peak scale $k_{\mathrm{p}}$, we parametrize the shape of the peak by a log-normal distribution, i.e. by a Gaussian in $\ln(k/k_p)$ with  the standard deviation $\Delta_{\mathrm{p}}$ centered around $k_{\mathrm{p}}$,
\begin{align}
\mathcal{P}_{\mathrm{PBH}}(k)=\frac{A_{\mathrm{p}}}{\sqrt{2\pi}\Delta_{\mathrm{p}}}\exp\left\{-\frac{\left[\ln(k/k_{\mathrm{p}})\right]^2}{2\Delta_{\mathrm{p}}^2}\right\}.\label{SimpPLogNormal}
\end{align}
In the case where an exact numerical treatment features a single peak, the parameters $A_{\mathrm{p}}$, $\Delta_{\mathrm{p}}$, and $k_{\mathrm{p}}$ are extracted by fitting the log-normal distribution \eqref{SimpPLogNormal} to the numerically obtained power spectrum. Here we focus on a derivation of the simple analytic estimate of the peak value $A_{\mathrm{p}}$ required for $f(M_{\mathrm{PBH}})=1$. We assume the simple scaling $M_{\mathrm{PBH}}=KM_{\mathrm{H}}$ (i.e.~not the critical scaling \eqref{CritMPBH}) and use \eqref{ffinal} to related $\beta$ with $f$.
We further assume that the peak is sufficiently sharp and can be approximated such that we can consider the limit $\Delta_{\mathrm{p}}\rightarrow 0$ where  the peak is described by a Dirac delta function,~\footnote{See also \cite{Germani:2018jgr,Byrnes:2018txb,DeLuca:2020ioi,Gow:2020bzo} for related discussions as regards the impact and comparison of different peak widths.}
\begin{align}
\mathcal{P}_{\mathrm{PBH}}(k)=A_{\mathrm{p}}\delta\left(\ln k-\ln k_{\mathrm{p}}\right).
\end{align}
Since the generation of a significant number of PBHs requires $A_{\mathrm{p}}\gg A_{\mathrm{CMB}}$, we can safely neglect the constant amplitude $A_{\mathrm{CMB}}\approx10^{-9}$ for the derivation of the PBH abundance and obtain
\begin{align}\label{SimpPDelta}
\mathcal{P}_{\mathcal{R}}(k)\approx A_{\mathrm{p}}\delta\left(\ln k -\ln k_{\mathrm{p}} \right).
\end{align}
Using \eqref{SimpPDelta} in \eqref{var4} for a peak scale $k_{\mathrm{p}}=k_{\mathrm{R}}$ \footnote{As explained in Sect.~\ref{PBHAb}, a strong amplification with a sufficiently large $\sigma_{R}$ can only be realized if the peak scales $k_{\mathrm{p}}$ is not significantly different from the smoothing scale $k_{R}$. Since we also assume that the time of formation coincides with the time the density perturbation enters the horizon, $k_{R}$ is identified with the comoving Hubble radius at the time of formation as in \eqref{SS}.}, we obtain a simple relation for the variance 
\begin{align}
\sigma^2_{R}(M_{\mathrm{H}})=\frac{16}{81}e^{-\frac{1}{2}}\,A_{\mathrm{p}}.\label{sigsimp}
\end{align} 
Furthermore, we assume that the PBH mass is directly proportional to the horizon mass at formation
${M_{\mathrm{PBH}}=KM_{\mathrm{H}}}$, so that the integral for $\beta(M_{\mathrm{H}})$ in the Press-Schechter formalism can be solved analytically as
\begin{align}
\beta(M_{\mathrm{H}})=K\,\mathrm{erf}\left(\frac{\delta_c}{\sqrt{2\sigma^2_R(M_{\mathrm{H}})}}\right).\label{erf}
\end{align}
For $\delta_c/\sqrt{2\sigma_{R}^2}\gg1$, the error function \eqref{erf} has the asymptotic expansion
\begin{align}
\beta(M_{\mathrm{H}})=\sqrt{\frac{2\sigma^2_R}{\pi\delta_c^2}}\exp\left(-\frac{\delta_c^2}{2\sigma^2_R}\right).
\end{align}
A quick estimate of $A_{\mathrm{p}}$, required to obtain $f(M_{\mathrm{PBH}})=1$ for a given PBH mass $M_{\mathrm{PBH}}$, can now be obtained easily. For example, let us take the values for the proportionality factor and the critical density found in \cite{Carr:1974nx} for a radiation dominated universe  as $K=0.19$ and ${\delta_{c}=0.33}$, and choose a PBH mass $M_{\mathrm{PBH}}=5\times10^{-12}\,M_{\odot}$. Then, according to \eqref{ffinal}, the mass fraction should be  $\beta=1.69\times 10^{-14}$ in order to have $f(M_{\mathrm{PBH}})=1$. Using \eqref{erf} and \eqref{sigsimp}, the corresponding peak value for the power spectrum can be deduced to be $A_{\mathrm{p}}\approx1.7\times10^{-2}$. Compared to the value of $A_{\mathrm{p}}\approx5\times 10^{-3}$ required to obtain $F_{\mathrm{PBH}}\approx 1$ in Fig.~\ref{Fig:MassWindowTwo} (the peak values in $f(M_{\mathrm{PBH}})$ almost coincide with the total integrated fraction $F_{\mathrm{PBH}}$), the analytic estimate roughly coincides up to an order of magnitude, with the main difference resulting from the omission of the critical scaling relation \eqref{CritMPBH} in the analytic estimate.   

\bibliography{DilatonStaroBib}{}
\end{document}